\documentstyle[aps,eqsecnum,epsf,epsfig]{revtex}

\begin{document}

\baselineskip=17pt

\draft

\title{ \hspace{11cm} Preprint WSU-NP-2-1998 \\
Scenario for Ultrarelativistic Nuclear Collisions: \\
   Space--Time Picture of Quantum Fluctuations and the Birth of QGP. } 
\author{ A.  Makhlin and  E. Surdutovich}
\address{Department of Physics and Astronomy, Wayne State University, 
Detroit, MI 48202}
\date{March 16, 1998}
\maketitle
\begin{abstract}

We study the dynamics of quantum fluctuations which take place at the
earliest stage of high-energy processes and the conditions under which
the data from {\em e--p} deep-inelastic scattering may serve as an
input for computing the initial data for heavy-ion collisions at high
energies. Our method is essentially based on the space-time picture of
these seemingly different phenomena.  We prove that the ultra-violet
renormalization of the virtual loops does not bring any scale into the
problem. The scale appears only in connection with the collinear
cut-off in the evolution equations and is defined by the physical
properties of the final state. In heavy-ion collisions the basic
screening effect is due to the mass of the collective modes (plasmons)
in the dense non-equilibrium quark-gluon system, which is estimated.
We avoid the standard parton phenomenology and suggest a dedicated
class of evolution equations which describe the dynamics of quantum
fluctuations in heavy ion collisions.

\end{abstract}


\section{Introduction}
\label{sec:SNi}

In this study, we wish to approach the theory of high energy
collisions of heavy ions based on a minimal number of first principles
instead of using the elaborate technique of factorization-based
scattering theory \cite{HPQCD}.  Our ultimate goal is to provide a
description of the transition process which converts two initial-state
composite systems (the stable nuclei in the ``normal'' nonperturbative
vacuum) into the quark-gluon plasma (QGP), which is a dense system of
quarks and gluons with a perturbative vacuum as its ground state.
This scenario was suggested by Shuryak 20 years ago \cite{Shuryak} and
it is our basic assumption that this phenomenon does indeed take
place. Our main result is that this dense system can be formed only
{\em in a single quantum transition}. This transition is similar to
the process described by the evolution equations of deep-inelastic
scattering (DIS) and differs from it only by the scale parameter
inherent to the final state. The limit of resolution for the emission
process in DIS is connected with the minimal mass of a jet, while in
the case of the QGP final-state, the scale is much smaller and defined
by the screening properties of the QGP itself.

To pose this essentially quantum--mechanical problem properly, one
(ideally) needs an exact definition of two main elements: the initial
state of the system and the observables in the expected final state.

We know very little about the initial state, and thus must rely on the
following phenomenological input only: The nuclei are stable bound
states of QCD and therefore, their configuration is dominated by the
stationary quark and gluon fields. The nuclei are well shaped objects;
the uncertainty of their boundaries does not exceed the typical Yukawa
interaction range. In the laboratory frame, both nuclei are Lorentz
contracted to a longitudinal size $R_0/\gamma \sim 0.1 fm$.  The tail
of the Yukawa potential is contracted in the same proportion.  The
world lines of the nuclei are the two opposite generatrices of the
light-cone that has its vertex at the interaction point.  Therefore,
no interaction between the nuclei is possible before they overlap
geometrically.

The final state is defined more accurately. We believe that
single-particle distributions of quarks and gluons at some early
moment after the nuclei have intersected, describe it
sufficiently. Thus, we may rely on a reasonably well-defined quantum
observable. The corresponding operator should count the number of
final-state particles defined as the excitations above the
perturbative vacuum.  To develop the theory for this transition
process we have to cope with a binding feature that the ``final''
state has to be defined at a finite time. This may look disturbing for
readers well versed in scattering theory.

Is it possible to compute the quantum-mechanical average of the
operator that counts the final-state particles without precise
knowledge of the wave function of the initial state?  We prove that
this is indeed possible, provided we have the data from a much simpler
process, like {\em e--p} DIS.  Moreover, we argue that this is
possible even without constructing any model for the proton or
colliding nuclei like, {\em e.g.}, the parton model.  Such a
possibility is provided solely by the inclusive character of the
measurements in both cases. Indeed, the measurement of one-particle
distributions is as inclusive as the measurement of the distribution
of the final-state electron in DIS.

The key to our proof and to the suggested algorithm is the principle
of causality in the quantum-mechanical measurement which has two major
aspects.  First, any statement concerning the time ordering must be in
manifest agreement with the light cone boundaries.  This is always
guaranteed in the theory based on the relativistic wave
equations. Second, the process of measurement physically interrupts
the evolution of a quantum system, and any dynamical information about
the quantum-mechanical evolution reveals itself only after the wave
function is collapsed.  In order to implement these principles in a
practical design of a theory, one has to start with a space-time
description of the measurement, {\em i.e.}, the Heisenberg picture of
quantum mechanics.

\begin{figure}[htb]
\begin{center}
\mbox{ \psfig{file=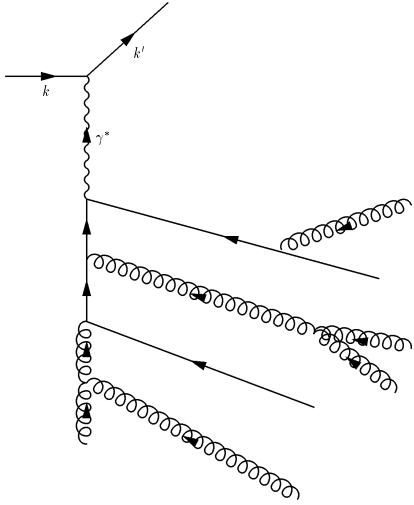,height=3.in,bb=70 370 400 745} }
\end{center}
\caption{An example of fluctuation resolved by the inclusive measurement
in DIS.}
\label{fig:fig1}
\end{figure}

To give a flavor of how the method works practically, let us start
with a qualitative description of the inclusive {\em e-p} DIS
measurement (for now, at the tree level without discussion of the
effects of interference). In this experiment, the only observable is
the number of electrons with a given momentum in the final state.
Something {\em in the past} has to create the electromagnetic field
that deflects the electron. {\em Before} this field is created, the
electromagnetic current, which is the source of this field, has to be
formed. Since the momentum transfer in the process is very high, the
current has to be sufficiently localized. This localization requires,
in its turn, that the electric charges which carry this current must
be dynamically decoupled from the bulk of the proton {\em before} the
scattering field is created (to prevent a recoil to the other parts of
the proton which could spread the emission domain). Such a dynamical
decoupling of a quark requires a proper rearrangement of the gluonic
component of the proton with the creation of short-wave components of
a gluon field. By causality, corresponding gluonic fluctuation must
happen before the {\em current} has decoupled, {\em etc.}  Thus we
arrive at the picture of the sequential-dynamical fluctuations which
create an electromagnetic field probed by the electron. It is very
important that the picture of the fluctuation is dynamical.  The
short-wave Fourier components of the quark and gluon fields are more
typical of free propagation with high momentum, than for a smooth
static design of a stable proton. The lifetimes of these fluctuations
are very short and they all {\em coherently} add up to form a stable
proton, unless the interaction of measurement breaks the proper
balance of phases. This intervention freezes some instantaneous
picture of the fluctuations, but with wrong ``initial velocities''
which results in a new wave function, and collapse of the old one.  An
example of such a fluctuation is depicted in Fig.~\ref{fig:fig1},
where the arrows point to the later moments of time.

This qualitative picture has been described many times and with many
variations in the literature, starting with the pioneering lecture by
Gribov \cite{Gribov}, and including a recent textbook \cite{BPQCD};
however, the sequential temporal ordering of the fluctuations has
never been a key issue. We derive this ordering as a consequence of
the Heisenberg equations of motion for the observables. The ordering
appears to be universal only {\em at the Born level}. In general, the
interference effects break the sequential temporal ordering, unless
the measurement prohibits the interference itself.  We eventually
arrive at a time-ordered picture which resembles the picture used by
Lepage and Brodsky \cite{Brodsky}. The difference is that they applied
time ordering exclusively in connection with the light-cone
Hamiltonian dynamics, while our picture holds in any dynamics as long
as the expansion in terms of two-point correlators makes sense.
Qualitative arguments supporting the strong time ordering in fast
hadrons were put forward earlier by Gribov \cite{Gribov}.  For our
practical goals, such an ordering is significant for several reasons.

First, it makes the whole picture of the evolution causal and clearly
indicates that the inclusively measured fluctuations are independent
of the nature of the inclusive probe.

Second, the causality allows one {\em to prove} important dispersion
relations. These relations provide a tool for the renormalization
which is also subjected to causality, {\em viz.}, there cannot be
radiative corrections to a stable proton (or nucleus)
configuration. This condition is physical and fully equivalent to the
on-mass-shell renormalization of the asymptotic states in a standard
scattering theory. Thus, we do not treat the intermediate quark and
gluon fields as fundamental fields that have certain states. Only
hadrons (in DIS) or collective modes (in dense matter) can be the
states of emission.

Third, the causality in the inclusive measurement requires that the
quark and gluon fields, which might exist after the measurement, {\em
are fully developed} as the virtual fluctuations {\em before the
measurement}. This is true both for freely-propagating hard quarks and
gluons and for their final-state hadronic equivalents
(jets). Therefore, the fluctuations have to satisfy the condition of
emission. This condition requires that (i) the field of emission
exists as a detectable state, and (ii) that two different states can
be resolved. Unlike QED, QCD has an intrinsic criterion of resolution
which requires that the states of emission in the physical vacuum must
be hadrons. Thus, {\em e.g.}, there cannot be a quark or gluon
emission which carries away an invariant mass $m_0$ less than several
hadronic masses. In other words, the modes with wave-length $\lambda >
m_{0}^{-1}\sim 1Gev^{-1}$ cannot propagate. The process of formation
of the final-state jets has to be fully accomplished in a few $~fm$.
This general rule, however, must be changed if a quark or gluon
becomes a part of a dense system, like the QGP, where the boundary of
the soft (and hopefully still perturbative) dynamics is determined by
a parameter similar to the plasmon mass (or the Debye screening
length) and not a hadronic scale \cite{Shuryak}. This condition
selects fluctuations that are materialized as the QGP after the
coherence of the nuclei is broken.

Fourth, our analysis, based on the causal picture of the evolution,
indicates that the only scale that may appear in the theory is
connected to the physical properties of the final states. This scale
penetrates into the theory only in connection with the infra-red
(collinear) problem but {\em not} in connection with ultra-violet
renormalization. This observation allows one to outline the scenario
of formation of the dense matter at the first 1{\em fm} of a nuclear
collision as follows: The nuclei probe each other locally at all
scales which are consistent with the spectrum of the allowed final
states. The most localized interactions create the final-state fields
with the highest $p_t$ which are the least sensitive to the collective
properties of the final state. (A similar process takes place in DIS
at sufficiently high $p^2\gg m^{2}_{hadr}>\Lambda^2_{QCD}$.) These
states are formed as (quasi-) particles by the time $\tau\sim
1/p_t$. In this domain, the structure functions and their rates of
evolution must be very similar for both nuclear collisions and
DIS. Therefore, the DIS data obtained with precise control of the
momentum transfer may serve as an input for the calculation of the
nuclei collision. The states with lower $p_t$ are formed at later
times.  This does not much affect the QCD evolution in DIS, where the
phase-space of the final states is almost unpopulated and only the
shape of the jets may lead to the power corrections.  In the case of
nuclei collisions this is not true, as the low--$p_t$ states are
formed in a space which is already populated with the earlier formed
(virtually decoupled) fast particles . These fast particles produce
screening effects. The process of the formation of the new states
saturates at $p_t\sim m_D$, where $m_D$ is an effective mass of
partons in the (highly non-equilibrium) collective system formed by
the time $\tau_D\sim 1/m_D$.  To quantify this scenario one has to
derive evolution equations that respect specific properties of the
final states in the QGP. A draft of such a derivation is given in the
last two chapters of the paper.

To conclude, the standard inclusive {\em e-p} DIS delivers information
about quantum fluctuations which may dynamically develop in the
proton.  We do not need to know the ``proton's wave function'' (since
the proton is a true fundamental mode of QCD, this notion is not even
well defined).  The theory can be built on two premises: causality,
and the condition of emission.  The latter is also known as the
principle of cluster decomposition which must hold in any reasonable
field theory. What the ``resolved clusters'' are is a very delicate
question. These states should be defined with an explicit reference as
to how they are detected.  Conventional detectors deal with hadrons
and allow one to hypothesize about jets. QGP turns out to be a kind of
 collective detector for quarks and gluons.
 
A vast search for a scale in ultra-relativistic nuclear collisions has been
undertaken  by McLerran {\em et al.} This study  started with the  formulation
of the {\em model} for a large nucleus as a system of valence quarks moving
at the speed of light (McLerran--Venugopalan model \cite{Larry}). The original
idea of this approach was that the scale is associated with the density of
valence quarks, that is, with the parameters of the initial state of the
nuclear collision. The model was gradually improved (see \cite{Kovner}, and
references therein) by the accounting for the radiative corrections with 
emphasis on the domain of small $x_F$ and a derivation of the BFKL equation,
which is known to have no scale parameter. 

The idea that screening effects should be taken into account, even at
the earliest moments of a collision of two nuclei has been articulated
earlier and with different motivations by Shuryak and Xiong
\cite{Xiong} and by Eskola, Muller and X.-N. Wang \cite{Eskola} .  In
Ref.\cite{Xiong}, the cut-off mass (expressed via the temperature) was
introduced to screen the singularities of the cross-section of the
purely gluonic process, $gg~\leftrightarrow~ng$, in the chemically
non-equilibrated plasma, $m^2\sim g^2n_{gluon}/T$. In
Ref.~\cite{Eskola}, the non-equilibrium Debye screening was introduced
in order to prevent the singular infrared behavior of the hard part of
the factorized cross-section of the mini-jet production.  The
expression for the screening mass in \cite{Eskola} is very similar to
our equation (\ref{eq:S1.14}).  However, in all of these papers, the
origin of the structure functions of the colliding nuclei is not
discussed, and there still remains an uncertainty connected with the
choice of the factorization scale.  We go one step further and prove
that the screening parameters of the collective system determine the
collinear cut-off self-consistently; these screening parameters are
{\em already in the evolution equations} for the structure functions
of individual nucleons (and therefore, in the nuclear structure
functions).

\section{Outline of calculations, main results, and conclusion}
\label{sec:SNo}

The QCD evolution equations are the main focus of this
study. Historically, their development started along two qualitatively
different lines which eventually merged into what is currently known
as the DGLAP equations. Despite their common name, there are still
certain concerns whether the marriage is happy. The one-to-one
correspondence between the DGL \cite{DGL} and the AP \cite{AP}
equations is evident only at the Born level (without the virtual
radiative corrections). A discussion of what happens in the higher
orders and particularly, what the argument of the running coupling is,
still continues (see, {\em e.g.}  Refs.~\cite{DMW,DSh}). The
variations appear due to {\em ab initio} different approaches. While
the AP equations re-interpret the equations of the operator product
expansion (OPE) based on the factorization and the renormalization
group (RG) for {\em e-p} DIS, the DGL equations are derived as a
probabilistic picture of the multiple process. The AP equations just
{\em accept} the running coupling of the RG for the exclusive Feynman
amplitudes, while in the DGL scheme one has to calculate the radiative
corrections as constituents of a real process explicitly and assemble
them into an effective quantity (the running coupling).

The first claim about a closer correspondence between the DGL and the AP
equations, at least in one loop beyond the Born approximation, was made in
Ref.~\cite{DDT} (widely known as DDT) almost 20 years ago. The ultimate goal
was to recover the running coupling in the natural environment of the QCD
evolution ladder by collecting various radiative corrections into an effective
form-factor instead of relying (as does the OPE method) on the RG for the
Feynman amplitudes of the S-matrix scattering theory. This program was a
success only for one subgroup of ladder diagrams, {\em vis.}, when the gluon is
an offspring of the fermion line and the dynamics is driven solely by the
splitting kernel $P_{qq}$.  This was a smart choice, and success was almost
predetermined by the well known fact that in physical gauges the
renormalization of the coupling is fully defined by the gluon propagator. 
Nevertheless, the DDT study did mighty battle with the double-logarithmic terms
which naturally appear in radiative corrections. It was a formidable task to
show that the running coupling depends on the highest transverse momentum in
the splitting point. Neither DDT, nor anybody else since that time, attempted
to do the same for other types of splittings, though it is not obvious if the
same result can be obtained for the coupling constant which accompanies the
kernel $P_{qg}$ (when a single gluon line carries the lowest of three
transverse momenta). We intend to bridge this gap and assemble the running
coupling constant of all other kernels.  The kernel $P_{gg}$ (when all three
lines are gluons) is of extreme interest since one may expect competition
between three gluon self-energies with all different momenta. The case of pure
glue-dynamics provides an ultimate test for the understanding of the dynamics
of QCD evolution. We address it first, and use it later as a touchstone for all
the tools we design.

For the sake of completeness, we begin in Sec.~\ref{sec:SNe} with the basics of
the so-called Keldysh-Schwinger technique~\cite{Keld}, in the form adjusted for
the calculations of the inclusive observables~~\cite{QFK}.  The net yield of
this section is summarized in Eq.~(\ref{eq:Q1.33}) for the fixed-order
two-point field correlators. This equation shows that the correlator ${\bf
D}(x,y)$ has two contributions: from the initial data, $D(\xi,\eta)$, and the
source, $\Pi(\xi,\eta)$, acting at an {\em earlier} time. In
Sec.~\ref{sbsec:SBe3} we discuss the connection between Eq.~(\ref{eq:Q1.33})
and the spectral representation of field correlators and explain why the
initial data can be dropped. In Sec.~\ref{sec:SNto} these equations are used to
derive the equations that lead to the picture in Fig.~\ref{fig:fig1} and prove
that the photon of {\em e-p} DIS must be treated as part of the proton.

We also examine (in Sec.~\ref{sec:SNto}) what types of fluctuation are
active in the {\em e-p} DIS which is inclusive with respect to a quark
jet (without observation of the final-state electron). In this
problem, the photon ``belongs'' to the initial-state electron and the
measurement is affected both by QED and QCD fluctuations which may
develop in the initial-state electron.

In Sec.~\ref{sec:SNE1} we derive the evolution equations of
glue-dynamics yet without the radiative corrections. Even at this
level, we discover the second {\em independent} element of the
evolution, {\em vis.}, the longitudinal fields, which are absent both
in the DGL and the AP equations.\footnote{It was Jianwei Qiu \cite{Tw4}, 
who first paid attention to 
the special role of the static part of the quark and gluon propagators
(he called them  {\em special} or {\em contact} propagators) in computing 
the twist-4 corrections to the DIS process. However, the main
emphasis of this study was on reproducing the high-twist effects in
the perturbative part of the OPE-expansion.}  The reason is that, in our
derivation, the evolution equation naturally appear in a tensor form
(\ref{eq:E1.3}). Projecting this equation onto the normal modes we
obtain {\em two} interconnected evolution equations, (\ref{eq:E1.10})
and (\ref{eq:E1.11}), for the transverse and longitudinal (static,
non-propagating) gluon modes.  This should not be considered a
surprise.  Indeed, it is well known that the photon on the top of the
evolution ladder (just proved to be a part of the proton) has
transverse and longitudinal polarizations which are probed by the
structure functions $F_2$ and $F_L$, respectively.  It is not clear,
{\em a priori}, why the gluon field should be only transverse. If the
longitudinal field is eliminated from Eqs.~(\ref{eq:E1.10}) and
(\ref{eq:E1.11}) by a fiat, the remaining single equation is
exactly the DGLAP evolution equation of glue-dynamics. Its kernel is
singular, and it is common to regulate it with the aid of the
$(+)$-prescription.  In Ref.\cite{Qui} this prescription which
incorporated self-energy corrections (in the fixed coupling
approximation) was derived.  For a similar singularity in
Eq.~(\ref{eq:E1.11}) this method does not work.  Moreover, it is not
clear what kind of coupling accompanies the transition between the
longitudinal and transverse fields in the evolution equations.
Potential consequence of this new-type of infrared divergence might be
a disaster since the standard remedies of scattering theory are not
designed to handle any problems related to the longitudinal fields. In
scattering theory the latter are never treated as observables.  The
specifics of heavy-ion collisions are different because the
observables must be defined at a finite time of evolution. Unlike
scattering theory, the longitudinal fields cannot be absorbed into the
definition of the final states.  Before looking for a cure, we must
make sure that this is not a false alarm, {\em e.g.} perhaps the
infrared divergences do not cancel between the different rungs of the
emission ladder.  Thus, we need to study the full set of radiative
corrections in the order $g^4$.

A full set of radiative corrections includes different types of
diagrams.  Two diagrams in Fig.~\ref{fig:fig4} are virtual vertex
corrections. Their vertex loops are formed by three $T$-ordered
propagators, $D_{00}$, in $V_{000}$ (diagram (a)) and three
$T^{\dag}$-ordered propagators, $D_{11}$, in $V_{111}$ (diagram
(b)). Even though all propagators which are connected to these
vertices are retarded, the vertex loops have {\em a priori} arbitrary
ordering of their space-time arguments. This may hurt the nice causal
picture of the lowest order approximation.  Our main observation is
that the net yield of the sum of these two diagrams does not change if
the vertex $V_{000}$ is replaced by the retarded vertex $V_{ret}$
(diagram (c)) and the vertex $V_{111}$ is replaced by the advanced
vertex $V_{adv}$ (diagram (d)). In this formulation, the picture of
the evolution remains totally causal even in the order $g^4$.

The retarded properties of the new vertex functions in coordinate
space can be translated into the analytic properties in momentum
space.  To comply with the light-cone dynamics and the light-cone
gauge $A^+=0$, we must treat the coordinate $x^+$ as the temporal
variable. The conjugated variable is the light-cone energy $p^-$. We
prove that the vertex function $V_{ret}(p,k-p,-k)$ is analytic in the
upper half-plane of complex $p^-$. The light-cone energy $p^-$
corresponds to the latest momentum $p^\mu$ resolved in the course of
the evolution (which has the lowest component $p^+$ of the light-cone
momentum and is, in fact, probed in the inclusive measurement).  These
results are obtained in Sec.~\ref{sbsec:SBE2b} and Appendix~1. Knowing
the analytical properties of the vertex function we are able to prove
the dispersion relations which allow one to find the real part of the
vertex via its imaginary part. It occurs that all singularities which
may affect the final answer are collinear and already appear in the
calculation of the imaginary part which is connected with real
processes. Thus, these singularities are physical and do not change in
the course of a calculation of the dispersion integrals. Besides, they
are always connected with emission into the final states and never
with the absorption from the initial states.

The dispersive method provides a tool for the UV-renormalization of
the vertex function, {\em viz.}, the subtractions in dispersion
relations.  These subtractions must be made at a value $p^-=-\Omega$
of the light-cone energy in the domain were the imaginary part is
zero. This domain occurs to be limited to a finite segment
(\ref{eq:E2.28}) of the real axis in the complex plane of $p^-$.  The
position $p^-= -\Omega$ of the subtraction point can be transformed
into the value of $p^2$,~{\em vis.}, $p^2=-\mu^2=-p^+\Omega -p_t^2$.
Under the assumption of strong ordering of the transverse momenta,
$p_t>>k_t$, the allowed values of $\mu^2$ are $ 0< \mu^2 < p_t^2$.
Even though this limitation is a direct consequence of causality, it
may seem abnormal from the RG point of view. Indeed, we cannot take an
asymptotically high renormalization point in $p^2$. However, in a
causal picture of gradually increasing resolution along the line of
evolution, this limitation is absolutely natural. {\em It is illegal
to renormalize what is not yet resolved!}

If we proceed with renormalization at an arbitrary scale $\mu^2$, then
the calculation of the vertex function leads to the terms
$~log^2(p_t^2/\mu^2)~$ along with $~log(p_t^2/\mu^2)~$.  The
double-logarithms find no counterparts among either virtual loops or
real emission processes. Therefore, we immediately undermine the
leading logarithmic approximation (LLA) in which the DGLAP equations
are derived (see Eq.~(\ref{eq:E3.6})).  Besides, in order to have
$~log(p_t^2/\mu^2)>>1~$ one should choose a very low renormalization
scale $\mu^2$ which would contradict all beliefs about perturbative
QCD. These problems compel us to look for a physically-motivated
condition for the renormalization.  Surprisingly, this condition is
readily found at the opposite end of the allowed interval. If we take
$\mu^2=p_t^2$ which corresponds to the subtraction point at $p^-=0$,
then all of the above problems miraculously disappear.  The vertex
function, as well as all retarded self-energies (also calculated via
dispersion relations), become free of logarithms, $log(p_t^2/\mu^2)$,
which originate from ultraviolet renormalization (where they are most
expected to come from). Technically, the toll is that we have to look
for the large logarithms in real processes.  However, from the
physical point of view, this is a gain rather than a loss, since we
get rid of the logarithms which are just an artifact of the
renormalization prescription. In fact, we can promote the status of
the running coupling; instead of being a formal attribute of the
renormalization group it may become a dynamical form-factor due to the
real processes. Before we search for these large logarithms, let us
try to understand why the subtraction point $p^-=0$ is a natural
physical point for the renormalization.

According to the principle of causality, the full picture of the QCD
evolution develops {\em before} the measurement. Thus, the system of
fields must be coherent. It represents, at most, virtual decomposition
of a hadron in terms of an unnatural set of modes which emerge as
propagating fields only after the interaction happens. Therefore, the
radiative corrections to the propagation of fields should not cause
any phase shifts along the line of hadron propagation and hence, the
real part of virtual loops should vanish for the state of free
propagation. This is exactly what is achieved by the on-mass-shell
renormalization of the asymptotic states in scattering theory.  If the
hadron is assumed to move along the light cone, the real parts of the
vertices and self-energies should vanish at $p^-=0$.

Contributions (\ref{eq:E3.9}) and (\ref{eq:E3.10}) of the retarded
gluon self-energy and vertex (both renormalized at the point $p^-=0$)
to the evolution equation are double-logarithmic with respect to the
collinear cut-off $~\epsilon$.  This cut-off already appears in the
imaginary parts of the self-energy and vertex. Therefore, it is
connected with real processes. Even if the divergent terms do cancel,
one still has to understand why. The physical mechanism of this
cancelation can only be due to interference, and it has to work even
at the classical level. The collinear divergence is both a classical
effect and an artifact of theoretical model with massless charged
particles. Therefore, it should not be treated as an artificial
phenomenon caused by spurious poles of a special gauge. Indeed, the
physical mechanism of the collinear divergence can be easily
translated into the language of space-time.

Consider a charged particle that instantaneously changes its velocity
from ${\bf v}_1$ to ${\bf v}_2$ at a time $t=0$. At some later time
$t$, the picture of the field is as follows. Outside the sphere of
radius $R=t$, there is an unchanged static proper field of the
initial-state particle. Inside this sphere, there is the newly created
proper field of the final-state particle.  The field of radiation
(which has the old and the new static fields as the boundary
conditions) is located on the sphere itself.  If $v_2<1$ then the
charge is inside the sphere and decoupled from the radiation field. At
$v_2=1$, the charge never decouples from the radiation field and this
leads to the divergence in the integration over the time of the
interaction.  Thus, assigning a finite mass $m$ to the field is a
test for true collinear divergence. This is also a cure, since then
$v_2<1$ and the problem disappears. Alternatively, one may provide at
least a single interaction with a third body which interrupts the
emission process after some time $\Delta t$.  In both cases, we
encounter a dimensional physical parameter which regulates the
time-length of interaction between the charge and the radiation
field. If $m\neq 0$, then the smallest of two quantities, $\Delta t$
and $m^{-1}$, is the physical cut-off for the emission process.  Let
us assume that some mass $m_0$ can be introduced as a collinear
cut-off.  If this cut-off cancels in the full assembly of diagrams in
any given order, then the theory has no single dimensional
parameter. This is exactly the case of pQCD where the parameter
$\Lambda_{QCD}$ establishes, at most, an absolute boundary of its
validity. All practical calculations are possible only for momenta
higher than some $Q_0\gg\Lambda_{QCD}$. All the divergences, including
soft and collinear, are usually removed by means of the dimensional
regularization with reference to the KLN theorem; however, this is not
the case we address.

In Sec.~\ref{sec:SNE4}, we calculate contributions of real emission
processes to the causal QCD evolution. The diagrams in
Figs.~\ref{fig:fig7}a and \ref{fig:fig7}b, with the unitary cut
through the vertex, and the interference diagram in
Fig.~\ref{fig:fig7}c, do not lead to terms with $~log(p_t^2)~$. They,
however, include many terms with the collinear cut-off which eliminates
a part of phase-space which does not satisfy the emission criterion.
In Sec.~\ref{sec:SNE5} we study the radiative corrections in the
LT-transition mode. They confirm the general trend.

Thus, we have proved that in the time-ordered picture of evolution
the UV-renormalization should be performed in such a way that
radiative corrections do not affect the initial state and bring no
scale into the problem. The scale is brought in by the emission process
and is defined by parameters of the final state that emerge after the
collision. The most important conclusion that follows from this
observation is that the evolution equations must be modified, {\em
viz.}, the universal running coupling that accompanies splitting kernels
in the DGLAP equations must be exchanged for the form-factors which depend
on the type and scale of the emission process. This is required by
causality of the inclusive measurement.

It is high time to understand why we need a theory with a
physical cut-off of the collinear emission and why we cannot rely on the
standard approach of pQCD. The latter was constructed as a theory of
the S-matrix at asymptotic energies. The concept of asymptotic freedom
is formulated precisely in this context. The running coupling is
always introduced as a certain assembly of radiative corrections
regardless of whether the perturbation series is divergent. In
pQCD, the way to assemble vertices and self-energies is defined by the
structure of Feynman amplitudes of exclusive processes, and it
is very important that the collinear (mass)
singularities are, at any price, removed from this assembly. 
The most popular technical
tool to do this is based on dimensional regularization of the Feynman
amplitudes. This method does not distinguish between UV-, IR-, and
mass-singularities, despite the different physical nature of these
divergences.
Sometimes, one may trace that cancelations of certain divergences, are
due to interference of various partial amplitudes.  In other
cases, one may observe that the soft amplitudes exponentiate and die
out in the quantities like total cross-sections. The underlying
calculations always implicitly rely on the structure of the space of
states, which includes {\em arbitrarily soft} states of emission (otherwise,
the exponentiation makes no sense). Such a space of states is
completely physical in QED but not in realistic QCD. Usually, the
collinear singularities are forced to cancel with the aid of the KLN
theorem which redefines the states of scattering by admixing a
probabilistic distribution of soft (collinear) states.  The proof of
the KLN theorem is intimately connected with the definition of the
states, {\em vis.}, it is sensitive to the parameter of resolution,
which is hidden 
in the definition of the degenerate states.  Once again, the theorem was
derived under the influence of QED where even very soft photons can be
resolved in the asymptotic states, provided a sufficiently long time is
allowed for the measurement. To approach this idealized regime in QCD,
one has to limit the theoretical analysis to exclusive processes
with very energetic final-state jets (insensitive to
several additional soft hadrons which might simulate the emission of
extra soft gluons). However, even in QED there are cases when the
recipe of KLN is not 
supposed to work. For instance, in the QED
plasma, the space of states is populated,
and we are not free to assign arbitrary statistical weights to the
collinear states.

In nuclear collisions, the space of states has a quite different
structure. Unlike a scattering problem, the states must be defined at
a finite time of evolution. Furthermore, the longitudinal fields
cannot be absorbed into the definition of stable charged particles (by
means of on-mass-shell renormalization) and are not geometrically
separated from the fields of emission. Nevertheless, the technical
problems are not frightening, since the normal modes in dense and exited (hot)
nuclear matter (to the first approximation) can be computed
perturbatively.  In Sec.~\ref{sec:SS1}, we compute the mass of a
transverse plasmon (a collective mode), which is now the final state for
the emission in the QCD evolution process.
The distribution of gluons that induce a plasmon mass is defined
by the process of multiple-gluon production, and the plasmon
mass is the function of the time of the distribution
measurement. The result of this calculation clarifies the correspondence
between the two approaches. It is summarized by Eq.~(\ref{eq:S1.12}):
The larger the transverse momentum $p_t$ of the emission is, the earlier
it is formed as the radiation field, the less it is affected by the
interaction with other partons, and the smaller its effective mass is.
This is exactly the limit when the OPE-based calculations of DIS
and the here advocated self-screened evolution should merge. The waves
with a very large $p_t$ just do not ``see'' the medium and have to
hadronize into jets as if there was no medium.  In order to quantify
this observation,
one has to estimate the corrections to the LLA on both
sides. These corrections lead to the broadening of jets in DIS and
radiation losses of fast partons in AA-collisions.

Formula (\ref{eq:S1.12}) was derived in three steps. We started
(in Sec.~\ref{sbsec:SBS1a}) with an attempt to compute the mass of the
transverse plasmon in the null-plane dynamics, which we essentially
relied on in the derivation of the evolution equations.  It turns
out that the dispersive part of the retarded self-energy is always
proportional to $p^2$ and may contribute only to the renormalization
of the propagator. At first glance, the non-dispersive part seems
to be capable of generating the effective mass of the
plasmon. However, even in the case of a finite
density of emitted gluons, the non-dispersive part remains
quasi-local because of the singular nature of longitudinal fields
in the null-plane dynamics, and  it vanishes in the course of
renormalization. We conclude that it is
impossible to generate an effective mass of a quasi-particle in the
null-plane dynamics.  

In order to smooth out the geometry of the longitudinal fields, we had
to choose another Hamiltonian dynamics which is described in
Sec.~\ref{sbsec:SBS1b}.  This new dynamics uses the proper time
$\tau$, $\tau^2=t^2-z^2$, as the time variable. For each slice in
rapidity, $\tau$ is just the local time of the co-moving reference
frame.  An advantage of this dynamics is that it naturally
incorporates the longitudinal velocity of a particle as one of its
quantum numbers and visualizes the process of the particle formation
by the time $\sim 1/p_t$. In its global formulation, the proper-time
dynamics uses the gauge $A^\tau=0$. The physical mechanisms of the
screening effects (like the generation of the plasmon mass) are known to
be localized in a finite space-time domain. This understanding allowed
us to use the gauge $A^\tau=0$ in a local fashion. In order to
estimate the plasmon mass we approximated this gauge by the
temporal-axial 
gauge (TAG) $A_\ast^0=0$ of the local (in rapidity) reference
frame.

An estimate of the plasmon mass is accomplished in
Sec.~\ref{sbsec:SBS1c}.  Once again, in local TAG only the
non-dispersive part of the retarded self-energy (in which the
propagation is mediated by the longitudinal field) and the tadpole
term (with the inertia-less contact interaction) lead to the plasmon
mass, which is given by Eq.~(\ref{eq:S1.12}). The low--$p_t$
mode of the radiation field acquires a finite effective mass as a
result of its forward scattering on the strongly localized (and formed
earlier) particles with $q_t\gg p_t$.  The smaller $p_t$ is, the larger
the plasmon mass is. Thus, the whole process, that converts the nuclei
into the dense system of (yet non-equilibrium) quarks and gluons, must
saturate. In this way, the scenario approaches the end of the
``earliest stage'', {\em i.e.}, the dense system is already created and the
collisions between the partons-plasmons start to take over the
dynamics of the quark-gluon system.

To derive the evolution equations with screening, we had to find
a space-time domain where the proper-time dynamics is close to the
null-plane dynamics. The limit is found at large negative rapidity,
{\em viz.}, in the photon fragmentation region.  In the last section
of the paper, we show that when the effective mass of the plasmon is
used as the pole mass of the radiation field in the evolution
equations, the evolution equations become regular.  The collinear
singularity of 
the splitting kernel $P_{gg}$ as well as a similar singularity in the
LT-transition mode are screened by the plasmon mass.  Thus, we obtain
a simple version of the evolution equations with screening,
which can be applied to the fluctuations in the photon
fragmentation region. They become the DGLAP equations when $s\to\infty$. In
order to obtain the evolution equations valid in the central-rapidity
region, we have to consider the whole problem in the scope of the 
``wedge dynamics'' \cite{WD1,WDG}. The last thing we estimate
within our ``local approximation'' is that the plasmon mass leads to
strong suppression of all radiative corrections in the order
$\alpha_s^2$.

The net yield of this paper is as follows. The interaction between the
two ultrarelativistic nuclei switches on almost instantaneously. This
interaction explores all possible quantum fluctuations which could have
developed by the moment of the collision and freezes (as the final
states) only the fluctuations compatible with the measured
observable. These snapshots cannot have an arbitrary structure, since the
emerging configurations must be consistent with all the interactions
which are effective on the time-scale of the emission process.  In
other words, the modes of the radiation field which are excited in the
course of the nuclear collision should be the collective excitations
of the dense quark-gluon system. This conclusion is the result of an
intensive search of the {\em scale} inherent in the process of a heavy-ion
collision. We proved that the scale is determined only by the
physical properties of the final state. Eventually, the scenario for
the ultra-relativistic nuclear collision promises to be more
perturbative than the standard pQCD.  It will be free from the
ambiguities of the standard factorization scheme inherent in the
cascade models.

This approach is novel for the field of ultrarelativistic nuclear
collisions, but the underlying physics is not new. Quantum transitions
in solids always excite collective modes, like photons in the
medium, phonons, or electrons with effective masses and
charges (polarons). Similar phenomena are known in particle and
atomic physics. For example:

{\em 1}. Let an electron-positron pair be created by two photons. If the
energy of the collision is large, then the electron and positron are
created in the states of freely propagating particles and the
cross-section of this process accurately agrees with the tree-level
perturbative calculation. However, if the energy of the collision is near
the threshold of the process, then the relative velocity of the
electron and positron is small, and they are likely to form
positronium.  It would be incredibly difficult to compute this case
using scattering theory.  Indeed, one has to account for the
multiple emission of soft photons which gradually builds up the
Coulomb field between the electron and positron and binds them into the
positronium. However, the problem is easily solved if we realize that
the bound state {\em is} the final state for the process. We can still 
use low-order perturbation theory to study the transition between
the two photons and the bound state of a pair \cite{Sakharov}.

{\em 2}. Let an excited atom be in a cavity with
ideally conducting walls. The system is characterized by three
parameters: the size $L$ of the cavity, the wave-length $\lambda\ll L$
of the emission, and the life-time $\Delta t=1/\Gamma$ of the excited
state. The questions are, in what case  will the emitted photon bounce
between the cavity walls, and when will the emission field be one of
the normal cavity modes. The answer is very simple. If $c\Delta t\ll
L$, the photon will behave like a bouncing ball.  When the line of
emission is very narrow, $c\Delta t\gg L$, the cavity mode will be
excited. It is perfectly clear that in the first case, the transition
current that emits the photon is localized in the atom. In the second
case it is not.  By the time of emission, the currents in the
conducting walls have to rearrange charges in such a way that the
emission field immediately satisfies the proper boundary
conditions. We thus have a collective transition in an extended system.

{}From a practical point of view, these two different problems are
united by the method of obtaining their solutions. A part of the
interaction (Coulomb interaction in the first case, and the
interaction of radiation with the cavity walls in the second case) is
attributed to the new ``bare'' Hamiltonian which is diagonalized by
the wave functions of the final state modes. This allows one to reach
the solution in a most economical way. The modes of the proper-time
dynamics are introduced with the same goal of optimizing the solution of
the nuclear collision problem by an explicit account of the collision
geometry (adequate choice of the quantum numbers) already at the level
of normal modes.

A list of known examples can continue. The example of nuclear
collisions is new in one respect only. Traditionally, the collective
modes are constructed against the existing ``heavy matter
background''. In nuclear collisions, the collective modes are created
simultaneously with the matter which supports them.

\section{The equations of relativistic quantum field kinetics}
\label{sec:SNe}

The equations of quantum field kinetics were derived in Ref.~\cite{QFK}  with
the goal of calculating the observables which cannot be reduced to the form of
the matrix elements of the S-matrix, {\em viz.}, the composite operators that
cannot be written as the $T$-ordered products of the fundamental fields.
The inclusive rates (or inclusive cross-sections)  which are really the 
number-of-particles operators, are quantities of this sort. In \cite{QFK}, we
concentrated on the phenomena associated with the dynamics of fermions. In this
study, the main emphasis is on the detail calculations in the gluon sector,
and thus, we use equations of QFK for vector gauge field.

\subsection{Basic definitions}
\label{subsec:SBe1} 

The calculation of observables such as inclusive cross sections
is based on work by Keldysh\cite{Keld}.  It incorporates
a specific set of exact (dressed) field correlators. These
correlators are products of Heisenberg operators, averaged with the
density matrix of the initial state.  For the gluon field they read

\begin{eqnarray}
{\bf D}_{10}^{ab;\alpha\beta}(x,y) & =-i\langle{\bf A}^{a;\alpha}(x)
{\bf A}^{b;\beta}(y)\rangle ,~~~~~
{\bf D}_{01}^{ab;\alpha\beta}(x,y) & = -i\langle {\bf  A}^{b;\beta}(y)
{\bf A}^{a;\alpha}(x)\rangle,\nonumber \\
{\bf D}_{00}^{ab;\alpha\beta}(x,y) & =-i\langle T({\bf A}(x)^{a;\alpha}
{\bf A}^{b;\beta}(y))\rangle ,~~
{\bf D}_{11}^{ab;\alpha\beta}(x,y) & =-i\langle T^{\dag}({\bf A}^{a;\alpha}(x)
{\bf A}^{b;\beta}(y))\rangle ,
\label{eq:Q1.1}
\end{eqnarray}
where $T$ and $T^{\dag}$  are the symbols of the time and anti-time ordering.
They may be rewritten in a unified form,
\begin{eqnarray}
{\bf D}_{AB}^{ab;\alpha\beta}(x,y)=-i\langle T_{c}({\bf A}^{a;\alpha}(x_{A})
    {\bf A}^{b;\beta}(y_{B}))\rangle,  
\label{eq:Q1.2}
\end{eqnarray}
in terms of a special ordering $T_{c}$ along a contour $C=C_{0}+C_{1}$ (the
doubled time axis) with $T$-ordering on $C_{0}$ and $T^{\dag}$-ordering on
$C_{1}$. The operators labeled by `1' are $T^{\dag}$-ordered, and stand
before the $T$-ordered operators labeled by `0'. Recalling that
\begin{eqnarray}
 {\bf A}(x)= S^{\dag}T(A(x)S)\equiv T^{\dag}(A(x)S^{\dag})S,
\label{eq:Q1.3}
\end{eqnarray}
we may introduce the formal operator $S_{c}=S^{\dag}S$,
and rewrite (\ref{eq:Q1.3}) using the operators of the $in$-interaction 
picture,
\begin{eqnarray}
{\bf D}_{AB}^{ab;\alpha\beta}(x,y)=-i\langle T_{c}(A^{a;\alpha}(x_{A})
    A^{b;\beta}(y_{B})S_{c})\rangle~.  
\label{eq:Q1.4}
\end{eqnarray}

Except for the matrix form, the Schwinger-Dyson equations for the Heisenberg
correlators remain the same as in any other technique. An elegant and
universal way to  derive them (that does not rely on the initial diagram
expansion) can be found in Ref.\cite{Bogol}. For the gluon field these
equations are of the form 
\begin{eqnarray} 
{\bf D}_{AB} =
D_{AB}+ \sum_{R}D_{AR}\circ (V{\cal A}_R)\circ {\bf D}_{RB}
+\sum_{RS} D_{AR}\circ \Pi_{RS}\circ {\bf D}_{SB},
\label{eq:Q1.5} \end{eqnarray} 
where the dot stands for both convolution in coordinate space, and the usual
product in  momentum space (providing the system can be  treated as homogeneous
in space and time). Indeed, the only tool used to derive these equations was
the Wick theorem for  the ordered products of the operators. (The type of
ordering is not essential for the proof of the Wick theorem  \cite{Bogol}). The
second term on the right corresponds to the interaction with the ``external'' 
field ${\cal A}$. 

The formal solution to this matrix equation (\ref{eq:Q1.5})
can be cast in the following symbolic form,
\begin{eqnarray} 
[{\bf D}^{-1}]_{AB} =[D^{-1}]_{AB}-\Pi_{AB},
\label{eq:Q1.6} \end{eqnarray}
Explicit expressions for  the self-energies $\Pi_{AB}$ emerge automatically
in the course of the derivation of  the  Schwinger-Dyson equations; we find, 
\begin{eqnarray}
 \Pi^{\alpha\beta,ab}_{AB}(x,y)=-{i\over 2} (-1)^{A+B}
 \sum_{R,S=0}^{1}(-1)^{R+S}
\int \! d \xi d \eta V^{\mu\alpha\lambda}_{caf}(\xi,x,\eta)
{\bf D}_{AR}^{ff',\lambda\nu}(\eta,\eta')
 {\bf V}^{\nu\beta\sigma}_{RBS;f'bc'}(\eta',y,\xi') 
{\bf D}_{SA}^{c'c,\sigma\mu}(\xi',\xi)\nonumber \\
- {i\over 2} (-1)^{A+B} \sum_{R,S=0}^{1}(-1)^{R+S}
\int \! d \xi d \eta {\bf V}^{\mu\alpha\beta\lambda}_{RABS;cabf}(\xi,x,y,\eta)
{\bf D}_{SR}^{fc,\lambda\mu}(\eta,\xi)~~.
\label{eq:Q1.7}
\end{eqnarray}
where the first and the second term correspond to the loop and tadpole
diagrams, respectively.
The 3-gluon vertex (in coordinate representation) is defined as
 \begin{eqnarray}
 {\bf V}^{\nu\beta\sigma}_{bcf,RSP}(x,y,z)=(-1)^{R+S+P}{
 {\delta [{\bf D}^{-1}(x,y)]^{bc;\nu\beta}_{RS} } \over
  {g ~\delta {\cal A}^{f}_{\sigma}(z_{P}) }  }~.
\label{eq:Q1.8}
\end{eqnarray}
The coordinate expression for the bare  3-gluon vertex is
 \begin{eqnarray}
V^{\alpha\beta\gamma}_{abc}(x)\equiv V^{\alpha\beta\gamma}_{abc}(x_1,x_2,x_3)=
-gf_{abc}[g^{\alpha\beta}(\partial_{1}^{\gamma}-\partial_{2}^{\gamma})+
g^{\beta\gamma}(\partial_{2}^{\alpha}-\partial_{3}^{\alpha})+
g^{\gamma\alpha}(\partial_{2}^{\beta}-\partial_{3}^{\beta})]~~,
\label{eq:Q1.9}
\end{eqnarray}
where one has to put $x_1=x_2=x_3=x$ at the end of the calculation.
In the momentum representation, it may be written  as
\begin{eqnarray}
  V^{\alpha\beta\gamma}_{ABC;abc}(p_1,p_2,p_3)=-ig~\delta_{AB}\delta_{AC}
f^{abc}[g^{\alpha\beta}(p_1 -p_2)^\gamma +g^{\beta\gamma}(p_2 -p_3)^\alpha
+g^{\gamma\alpha}(p_3-p_1)^\beta ]~ ~ ~.
\label{eq:Q1.10}
\end{eqnarray}
The next (third) order correction to the three-gluon vertex is obtained by means
of the Eq.~(\ref{eq:Q1.7}), after differentiating the second-order self-energy
with respect to the field ${\cal A}$,
\begin{eqnarray}
 ^{(3)}{\bf V}^{\nu\beta\sigma}_{bcf,RSB}(x,y,z)=(-1)^{1+R+S+B}
{ \delta [^{(2)}\Pi^{bc;\nu\beta}_{RS}(x,y)] \over
  g~ \delta {\cal A}^{f}_{\sigma}(z_{B})   }=\nonumber\\
 = -i g V(x_R)D_{RS}(x,y)V(y_S)D_{SB}(y,z)V(z_B)D_{BR}(z,x)~,
\label{eq:Q1.11}
\end{eqnarray}
where all Lorentz and color indices in the second equation are dropped for 
brevity. 

In this paper, the tadpole term in the gluon self-energy will not be considered
beyond the one-loop approximation. Hence, we need only the bare
four-gluon vertex,
\begin{eqnarray}
V^{\alpha\beta\rho\sigma}_{ABRS;abrs}= 
- g^2 \delta_{AB} \delta_{AR} \delta_{AS}
\big[ f_{nar} f_{nbs}( g^{\alpha\beta} g^{\rho\sigma}- 
                  g^{\alpha\sigma}g^{\beta\rho})+ \nonumber \\
+f_{nas} f_{nbr}(g^{\alpha\beta} g^{\rho\sigma}- 
                 g^{\alpha\rho\sigma}g^{\beta\sigma})
+f_{nab} f_{nrs}(g^{\alpha\rho} g^{\beta\sigma}- 
                 g^{\alpha\sigma}g^{\beta\rho}) \big]~.                    
\label{eq:Q1.10a}
\end{eqnarray}

The four types of operator ordering which enter Eqs.~(\ref{eq:Q1.1})
are not linearly independent, {\it i.e.}, there exists a set of  relations
between the field correlators and the self-energies:
\begin{eqnarray}
D_{00}+D_{11}=D_{10}+D_{01},\;\;\;\;\;\;
 \Pi_{00}+\Pi_{11}=-\Pi_{10}-\Pi_{01} ~ ~.
\label{eq:Q1.12}
\end{eqnarray}
These indicate that only three elements of the $2 \times 2$ matrices  $D,
\Pi$, {\it etc.} are independent. To remove the over-determination let us
introduce new functions
\begin{eqnarray}
 D_{ret}  =D_{00}-D_{01},\;\;\;\;\; D_{adv}  =D_{00}-D_{10},\;\;\;\;\;
 D_{1}  =D_{00}+D_{11};  \nonumber \\
 \Pi_{ret}  =\Pi_{00}+\Pi_{01},\;\;\;\;\;
 \Pi_{adv}  =\Pi_{00}+\Pi_{10}, \;\;\;\;\;
 \Pi_{1}  =\Pi_{00}+\Pi_{11},
\label{eq:Q1.13}
\end{eqnarray}
One possible way to exclude the extraneous quantities is to use the following 
unitary transformation \cite{Keld},
\begin{eqnarray}
\tilde{D}=R^{-1}DR, \;\;\;\;\; \tilde{\Pi}=R^{-1}\Pi R ,\;\;\;\;\;
R={{1}\over {\sqrt{2}}} \left| \begin{array}{rc}
                                            1 & 1 \\
                                           -1 & 1  \end{array} \right|.
\label{eq:Q1.14}
\end{eqnarray}
In this new representation, the matrices of the field correlators and
self-energies have a triangle form,
\begin{eqnarray}
  \tilde{D}= \left| \begin{array}{ll}
                                       0 & D_{adv} \\
                                 D_{ret} & D_{1}     \end{array} \right| ,
~~~~~~  \tilde{\Pi}= \left| \begin{array}{ll}
                                 \Pi_{1}    & \Pi_{ret}  \\
                                 \Pi_{adv}  & 0      \end{array} \right|~.
\label{eq:Q1.15}
\end{eqnarray}
Applying the transformation ~(\ref{eq:Q1.15}) to the matrix 
Schwinger-Dyson equations  (\ref{eq:Q1.5}) we may rewrite them  in the 
following form:
\begin{eqnarray}
 {\bf D}_{ret} = D_{ret}+ D_{ret}\circ \Pi_{ret}\circ {\bf D}_{ret}~,
\label{eq:Q1.16}\end{eqnarray}
\begin{eqnarray}
 {\bf D}_{adv} = D_{adv}+ D_{adv}\circ \Pi_{adv}\circ {\bf D}_{adv}~,
\label{eq:Q1.17}\end{eqnarray}
\begin{eqnarray}
 {\bf D}_{1} = D_{1}+ D_{ret}\circ \Pi_{ret}\circ {\bf D}_{1}+
D_{1}\circ \Pi_{adv}\circ {\bf D}_{adv}+
D_{ret}\circ \Pi_{1}\circ {\bf D}_{adv}~.
\label{eq:Q1.18}\end{eqnarray}
Using the definitions  (\ref{eq:Q1.7}), (\ref{eq:Q1.13}) and the identities,
\begin{eqnarray}
 D_{00}(x_1,x_2)D_{00}(y_1,y_2)-D_{01}(x_1,x_2)D_{10}(y_1,y_2)=
 D_{10}(x_1,x_2)D_{01}(y_1,y_2)-D_{11}(x_1,x_2)D_{11}(y_1,y_2)=\nonumber \\
= {1\over 2}[D_{ret}(x_1,x_2)D_{1}(y_1,y_2)+D_{1}(x_1,x_2)D_{adv}(y_1,y_2)]~,
\label{eq:Q1.19}
\end{eqnarray}
we may write the explicit expressions for the retarded and advanced
gluon self-energies in the approximation of the bare vertex,
\begin{eqnarray}
 \Pi^{\alpha\beta,ab}_{{ret\choose adv}}(x,y)= - {i\over 4} 
 [ V^{\mu\alpha\lambda}_{fac}(x)
{\bf D}_{{ret\choose adv}}^{cc',\mu\nu}(x,y)
 V^{\nu\beta\sigma}_{c'bf'}(y) 
{\bf D}_{1}^{f'f,\sigma\lambda}(y,x) \nonumber \\
+ V^{\mu\alpha\lambda}_{fac}(x)
{\bf D}_{1}^{cc',\mu\nu}(x,y)
 V^{\nu\beta\sigma}_{c'bf'}(y) 
{\bf D}_{{adv\choose ret}}^{f'f,\sigma\lambda}(y,x)]~ +
~\delta(x-y)\Pi^{\alpha\beta,ab}_{tdpl}(x,x).
\label{eq:Q1.20}
\end{eqnarray}
It is easy to see that one of the propagators in the loop keeps track on the
temporal ordering of the  arguments of these correlators. The second
propagator, $D_1$, is nothing but the density of physical states  with the
on-mass-shell momentum which mediates the propagation of the gluon.

\begin{figure}[htb]
\begin{center}
\mbox{ 
\psfig{file=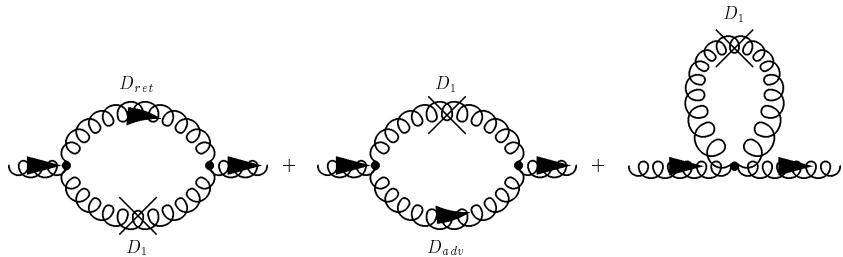,height=1.5in,bb=94 480 556 633} 
}
\end{center}
\caption{The retarded gluon self-energy. The arrows point to the latest temporal
argument. The crossed lines correspond to correlator $D_1$ which represents
the density of states of free propagation.}
\label{fig:figr}
\end{figure}

In order to find the tadpole part of the retarded gluon self-energy,
$\Pi_{tdpl}(x,x)$,
we have to go back to the Schwinger-Dyson equation which contains the following
fragment,
\begin{eqnarray}
{\bf D}_{ret}= ... -~{i\over 2}\sum_{R} (-1)^R 
\int dz [D^{\alpha\sigma}_{0R}(x,z)
V^{\sigma\lambda\mu\rho}{\bf D}^{\mu\lambda}_{RR}(z,z)
{\bf D}_{R1}^{\rho\beta}(z,y)
-D^{\alpha\sigma}_{0R}(x,z)
V^{\sigma\lambda\mu\rho}{\bf D}^{\mu\lambda}_{RR}(z,z)
{\bf D}_{R1}^{\rho\beta}(z,y)]~.\nonumber
\end{eqnarray}
The sum over the contour index $R$ involves two Green functions, $D_{00}(z,z)$
and $D_{11}(z,z)$, with coinciding arguments which are ill-defined since the
arguments are assumed to be ordered.  Only the sum 
$D_{00}(z,z)+D_{11}(z,z)=D_{1}(z,z)$ is unambiguous. Using the following
chain of transformations (which is very similar to (\ref{eq:Q1.19})), we obtain
\begin{eqnarray}
{\bf D}_{ret}= ...
+D_{0R}V{\bf D}_{RR}{\bf D}_{R0}-D_{0R}V{\bf D}_{RR}{\bf D}_{R1}=
D_{1R}V{\bf D}_{RR}{\bf D}_{R0}-D_{1R}V{\bf D}_{RR}{\bf D}_{R1}= \nonumber\\
={1\over 2}[D_{ret}V{\bf D}_1{\bf D}_{ret}+
D_1~V({\bf D}_{00}-{\bf D}_{11}){\bf D}_{ret}]~.
\label{eq:Q1.20a}
\end{eqnarray}
Since the correlator $D_1$ in the last equation is on mass shell, 
and ${\bf D}_{ret}$ on the left is not, the second term on the right drops out
and the tadpole term of the retarded self-energy becomes
\begin{eqnarray}
\Pi^{\alpha\beta,ab}_{tdpl}(z,z)= 
{-i\over 2}V^{\alpha\lambda\mu\beta}_{almb}{\bf D}^{\mu\lambda,lm}_{1}(z,z)~.
\label{eq:Q1.20b}
\end{eqnarray}
This term is entirely due to the instantaneous contact interaction of the 
propagating  field with on-mass-shell modes.

\subsection{The formal solution of the integral equations}
\label{sbsec:SBe2} 

The solution we shall look for now is equivalent to the rearrangement 
of the perturbation series for observables. This rearrangement is useful
when  specific features of the non-equilibrium system must be taken into 
account.  First, let us  introduce two new correlators:
\begin{eqnarray}
 D_{0}= D_{ret}-D_{adv}=D_{10}-D_{01} , 
 \label{eq:Q1.21}\end{eqnarray}
which coincides with a commutator of the gluon fields and thus
disappears outside the light cone, and 
\begin{eqnarray}
 \Pi_{0}= \Pi_{ret}-\Pi_{adv}=-\Pi_{10}-\Pi_{01}~,
\label{eq:Q1.22}\end{eqnarray}
which is the commutator of two charged currents and has the same causal
properties as (\ref{eq:Q1.21}).  The integral equation 
for $ D_{0}$ may be derived by subtracting  Eq.~(\ref{eq:Q1.16}) from
Eq.~(\ref{eq:Q1.17}),
\begin{eqnarray}
 {\bf D}_{0} = D_{0}+ D_{ret}\circ \Pi_{ret}\circ {\bf D}_{0}+
D_{0}\circ \Pi_{adv}\circ {\bf D}_{adv}+
D_{ret}\circ \Pi_{0}\circ {\bf D}_{adv}.
\label{eq:Q1.23}\end{eqnarray}
The sum and the difference of Eqs.~(\ref{eq:Q1.18}) and (\ref{eq:Q1.23})
give corresponding equations for the off-diagonal correlators 
$D_{10}$ and $D_{01}$,
\begin{eqnarray}
 {\bf D}_{{\stackrel{01}{\scriptscriptstyle 10}}} 
= D_{{\stackrel{01}{\scriptscriptstyle 10}}}+
 D_{ret}\circ \Pi_{ret}\circ {\bf D}_{01,10}+
D_{{\stackrel{01}{\scriptscriptstyle 10}}}\circ 
\Pi_{adv}\circ {\bf D}_{adv} -
D_{ret}\circ \Pi_{{\stackrel{01}{\scriptscriptstyle 10}}}\circ 
{\bf D}_{adv}  .
\label{eq:Q1.24}\end{eqnarray}
Since Eq.~(\ref{eq:Q1.16}) for the retarded propagator may be identically 
rewritten in the same form,
\begin{eqnarray}
 {\bf D}_{ret} = D_{ret}+ D_{ret}\circ \Pi_{adv}\circ {\bf D}_{adv}+
D_{ret}\circ \Pi_{ret}\circ {\bf D}_{ret} -
D_{ret}\circ \Pi_{adv}\circ {\bf D}_{adv} ,
\label{eq:Q1.25}\end{eqnarray}
we may use Eq.~(\ref{eq:Q1.24}) and derive the corresponding equations for 
the $T$- and  $T^{\dag}$-ordered propagators,
\begin{eqnarray}
 {\bf D}_{{\stackrel{00}{\scriptscriptstyle 11}}} = 
D_{{\stackrel{00}{\scriptscriptstyle 11}}}+ 
D_{ret}\circ \Pi_{ret}\circ 
{\bf D}_{{\stackrel{00}{\scriptscriptstyle 11}}}+
D_{{\stackrel{00}{\scriptscriptstyle 11}}}
\circ \Pi_{adv}\circ {\bf D}_{adv} + D_{ret}
\circ \Pi_{{\stackrel{11}{\scriptscriptstyle 00}}}\circ {\bf D}_{adv}.
\label{eq:Q1.26}\end{eqnarray}
This chain of routine transformations reduces the equations that make up all
elements of the matrix correlator $D_{AB}$ to a unified form.
On the one hand, this representation shows that linear relations
between the correlators (or, explicitly, different types of orderings)
hold even for the equations that the correlators obey. On the other hand,
this representation of the equations singles out the role of retarded and 
advanced propagators over 
all other correlators. In order to understand why their
role is special,  let us transform them further, and  begin by
rewriting of Eq.~(\ref{eq:Q1.18}) for the density of states
${\bf D}_{1}$ identically as
\begin{eqnarray}
 (1- D_{ret}\circ \Pi_{ret})\circ {\bf D}_{1}=
D_{1}\circ (1+\Pi_{adv}\circ {\bf D}_{adv}) +
D_{ret}\circ \Pi_{1}\circ {\bf D}_{adv}~ ~ ~.
\label{eq:Q1.27}\end{eqnarray}
Since ${\bf D}_{ret}\circ \Pi_{ret}\circ D_{ret}={\bf D}_{ret}-D_{ret}$,
it is easy to show that
\begin{eqnarray}
 (1+ {\bf D}_{ret}\circ \Pi_{ret}) (1- D_{ret}\circ \Pi_{ret})=1~~.
\label{eq:Q1.28}
\end{eqnarray}
Furthermore, we have the following two relations:
\begin{eqnarray}
 (1+\Pi_{adv}\circ {\bf D}_{adv}) ={\stackrel{\rightarrow}{D_{(0)}^{-1}}}
 \circ {\bf D}_{adv}~,\;\;\;\;\;
 (1+ {\bf D}_{ret}\circ \Pi_{ret})=   {\bf D}_{ret}\circ 
 {\stackrel{\leftarrow}{D_{(0)}^{-1}}}~ ,
\label{eq:Q1.29}
\end{eqnarray}
where ${\stackrel{\rightarrow}{D_{(0)}^{-1}}} (x) $ and
$ {\stackrel{\leftarrow}{D_{(0)}^{-1}}} (x)$
are the left and the right differential operators of the wave equation,
respectively. They explicitly depend on the type of the Hamiltonian
dynamics, including the gauge condition imposed on the field $A(x)$.
Multiplying Eq.~(\ref{eq:Q1.27}) 
by $(1+ {\bf D}_{ret}\circ \Pi_{ret}) $ 
from the left, we find the final form of the equation,
\begin{eqnarray}
 {\bf D}_{1} = {\bf D}_{ret}
 \circ {\stackrel{\leftarrow}{D_{(0)}^{-1}}} \circ D_{1}
 \circ {\stackrel{\rightarrow}{D_{(0)}^{-1}}} \circ {\bf D}_{adv}
+{\bf D}_{ret}\circ \Pi_{1}\circ {\bf D}_{adv}.
\label{eq:Q1.30}
\end{eqnarray}
Repeating these transformations for the other equations, we obtain the
corresponding forms that are most convenient for the subsequent analysis:
\begin{eqnarray}
 {\bf D}_{{\stackrel{10}{\scriptscriptstyle 01}}}  = 
{\bf D}_{ret}\circ  {\stackrel{\leftarrow}{D_{(0)}^{-1}}} 
\circ D_{{\stackrel{10}{\scriptscriptstyle 01}}}
 \circ {\stackrel{\rightarrow}{D_{(0)}^{-1}}} \circ {\bf D}_{adv}
-{\bf D}_{ret}\circ \Pi_{{\stackrel{10}{\scriptscriptstyle 01}}}
\circ {\bf D}_{adv},           
\label{eq:Q1.31}
\end{eqnarray}
\begin{eqnarray}
  {\bf D}_{{\stackrel{00}{\scriptscriptstyle 11}}}  = 
{\bf D}_{ret}\circ  {\stackrel{\leftarrow}{D_{(0)}^{-1}}} 
\circ D_{{\stackrel{00}{\scriptscriptstyle 11}}}
 \circ {\stackrel{\rightarrow}{D_{(0)}^{-1}}} \circ {\bf D}_{adv}
+{\bf D}_{ret}\circ \Pi_{{\stackrel{11}{\scriptscriptstyle 00}}}
\circ {\bf D}_{adv}.
\label{eq:Q1.32}
\end{eqnarray}
While equations ~(\ref{eq:Q1.16}) and (\ref{eq:Q1.17}) are the standard
integral equations which have an unknown function on both sides; however,
after transformation the unknown function appears only on the l.h.s.,
and we may consider Eqs.~(\ref{eq:Q1.30})--(\ref{eq:Q1.32}) to be the formal
representation of the required solution.

At this point, the first and the most naive idea is to ignore the arrows
indicating the direction the differential operators act in, and to rewrite
Eqs.~(\ref{eq:Q1.31})--(\ref{eq:Q1.32}) in the momentum representation. Then
the first term in each of Eqs.~(\ref{eq:Q1.31})  will contain the expression
like $p^{2}\delta(p^{2})$, which equals zero. 
This feature  reflects the simple fact that the
off-diagonal correlators $D_{10}$ and $D_{01}$ are solutions of the linearized
homogeneous Yang-Mills equations. However, this approach does not appear to be
sufficiently consistent: we  lose the identity between 
Eqs.~(\ref{eq:Q1.31})   and  (\ref{eq:Q1.23}), and make it impossible to
generate the standard perturbation expansion in powers of the  coupling
constant. Eqs.~(\ref{eq:Q1.32}) will be corrupted as well.

A more careful examination of Eqs.~(\ref{eq:Q1.31})-(\ref{eq:Q1.32})
shows that all four dressed correlators ${\bf D}_{AB}$ can be found as
the formal solution of the retarded Cauchy problem, with bare field correlators
being the initial data and the self-energies being the sources. Indeed, 
integrating the first term of each these equations twice by parts, we find
for all four elements of ${\bf D}_{AB}$ to be
\begin{eqnarray}
{\bf D}_{{\stackrel{10}{\scriptscriptstyle 01}}}(x,y)= 
\int d \Sigma_{\mu}^{(\xi)}
 d \Sigma_{\nu}^{(\eta)} {\bf
 D}_{ret}(x,\xi){\stackrel{\leftrightarrow} {\partial^{\mu}}}
D_{{\stackrel{10}{\scriptscriptstyle 01}}}(\xi.\eta)
{\stackrel{\leftrightarrow}{\partial^{\nu}}} 
{\bf D}_{adv}(\eta,y) 
-\int d^{4}\xi d^{4} \eta {\bf D}_{ret}(x,\xi)
\Pi_{{\stackrel{10}{\scriptscriptstyle 01}}}(\xi.\eta) 
{\bf D}_{adv}(\eta,y) ,
\label{eq:Q1.33}
\end{eqnarray} 
\begin{eqnarray}
{\bf D}_{{\stackrel{00}{\scriptscriptstyle 11}}}(x,y)= 
\!\!\int\!\! d\Pi_{\mu}^{(\xi)}
 d \Pi_{\nu}^{(\eta)} {\bf D}_{ret}(x,\xi)
{\stackrel{\leftrightarrow} {\partial^{\mu}}} 
D_{{\stackrel{00}{\scriptscriptstyle 11}}}(\xi,\eta)
{\stackrel{\leftrightarrow}{\partial^{\nu}}}  
{\bf D}_{adv}(\eta,y)\! 
-\! \int \!\! d^{4}\xi d^{4} \eta {\bf D}_{ret}(x,\xi)
[\pm D_{0}^{-1}+ \Pi_{{\stackrel{11}{\scriptscriptstyle 00}}}(\xi.\eta)] 
{\bf D}_{adv}(\eta,y)~.
\label{eq:Q1.34}
\end{eqnarray} 
     
The equations (\ref{eq:Q1.33}) and (\ref{eq:Q1.34}) are
the basic equations of relativistic quantum field kinetics.  They are
identical to the initial set of Schwinger-Dyson equations, but have
the advantage that the time direction is stated explicitly. 
There are two terms of different origin that contribute to any
correlator (and, consequently, to any observable).
The first term retains some memory of the initial data. 
The length of time for which this memory is kept depends on 
the retarded and advanced propagators.
The second term describes the current dynamics of the system. A comparison
of these two contributions allows one to judge if the system has 
various time scales.

\subsection{ Initial data and spectral densities.}
\label{sbsec:SBe3} 

We perform all calculations in the gauge $A^+=0$ which is complementary to the
null-plane dynamics with the the time variable $x^+$. In this gauge, the
on-shell correlators and the propagators of the perturbation theory are as
follows,
\begin{eqnarray} 
D^{\#ab,\mu \nu}_{{\stackrel{10}{\scriptscriptstyle01}}}(p)  
= -2 \pi i   \delta_{ab} d^{\mu\nu}(p) \theta (\pm p_{0})
\delta(p^{2})~, \nonumber \\
D^{\#ab,\mu \nu}_{{\stackrel{ret}{\scriptscriptstyle adv}}}(p) =
{d^{\mu\nu}(p) \over p^+(p^- \pm i0) -p_t^2}~,~~~~
d^{\mu\nu}(p)=-g^{\mu\nu}+{p^\mu n^\nu+n^\mu p^\nu \over p^+}~.   
\label{eq:Q1.35}\end{eqnarray} 
The polarization tensor $\Pi^{\mu\nu}$ appears  only between the
retarded and advanced propagators, {\em i.e.}, in the combination
$[D_{ret}(p)\Pi (p)D_{adv}(p)]^{\mu\nu}$. Both propagators contain the same
projector, $d^{\mu\nu}(p)$ which (by the gauge condition) is orthogonal to
the 4-vector $n^{\mu}$. So, of the general tensor form, only two terms
survive,
\begin{equation}
 \Pi^{\mu\nu}(p) =g^{\mu\nu}~p^2~w_1(p)+ p^\mu p^\nu w_2(p)+... ~~.
\label{eq:E1.4}\end{equation}  
The others, like $p^{\mu}n^{\nu}+ n^{\mu}p^{\nu}$  or $n^{\mu}n^{\nu}$ ,
cancel out.  
The invariants $w_1$ and $w_2$ can be found from the two contractions,
\begin{eqnarray}
-{\overline d}_{\mu\nu}(p) \Pi^{\mu\nu}(p) =2p^2~ w_1(p),~~~{\rm and}~~~
n_\mu n_\nu \Pi^{\mu\nu}(p)=(p^+)^2~ w_2(p)~,
\label{eq:E1.6}\end{eqnarray} 
independently of the other invariants accompanying the missing tensor 
structures. (The projector $ {\overline d}_{\mu\nu}(p)$ is defined below in
Eq.~(\ref{eq:E1.5}).)
The solutions of the Schwinger-Dyson equations (\ref{eq:Q1.16}) 
and (\ref{eq:Q1.17})
for the retarded and advanced gluon  propagators,
\begin{eqnarray}
{\bf D}_{{\stackrel{ret}{\scriptscriptstyle adv}}} 
= D_{{\stackrel{ret}{\scriptscriptstyle adv}}}
+  D_{{\stackrel{ret}{\scriptscriptstyle adv}}} 
\Pi_{{\stackrel{ret}{\scriptscriptstyle adv}}} 
{\bf D}_{{\stackrel{ret}{\scriptscriptstyle adv}}}~,\nonumber
\end{eqnarray} 
 can be cast in the form,
\begin{eqnarray}
{\bf D}^{\mu\nu}_{ret\choose adv}(p)=
{{\overline d}^{\mu\nu}(p)\over p^2 + p^2 w_{1}^{R\choose A}(p)}+
{1 \over (p^+)^2}~ {n^{\mu}n^{\nu} \over  1 - w_{2}^{R\choose A}(p)}. 
\label{eq:E1.8}\end{eqnarray} 
The first term is the propagator of the transverse field and 
has a normal pole corresponding to the causal propagation. 
It contains the polarization sum for the two transverse
modes,
\begin{equation}
{\overline d}^{\mu\nu}(p)\equiv - d^{\mu\rho}(p) d_{\rho}^{\nu}(p) 
=-g^{\mu\nu}+{p^\mu n^\nu+n^\mu p^\nu \over p^+ }
 -p^2{n^\mu n^\nu \over  (p^+)^2} ,
\label{eq:E1.5}\end{equation}
  which is orthogonal to both vectors $n^{\nu}$ and $p^{\mu}$.  The second term
in Eq.(\ref{eq:E1.8}) is the response function of the longitudinal field which
does not lead to causal propagation. When radiative corrections $w_1$ and $w_2$
are  dropped, the bare propagators are recovered with a clear separation of
the transverse and longitudinal parts. This separation tells us
that the longitudinal field in the null-plane dynamics has only the component
$A^-$ which is generated by the charge density $j^+$ of the source current
which enters Gauss' law but not the equations of motion.  If all other
components of the current vanish, then (by the current conservation) we have
$\partial_+j^+=0$ and, consequently, $j^+=j^+(x^-)= j^+(x^0-x^3)$. Therefore,
the purely static field of the null-plane dynamics is produced by the charge
traveling at the speed of light in the positive  $x^3$-direction. The radiation
field is everything else minus this static field. Physically, the  field is
called the radiation field if it geometrically decouples from the source. Therefore,
we can formulate a criterion for radiation in the null-plane dynamics. The
radiation field has to be slowed down with respect to the static source. There
is no problem with real massive sources, they are never static in the
afore-mentioned sense. However, if the source is put into the infinite momentum
frame, a paradox arises and the problem re-appears through the poles $1/p^+$
in the propagators. Fortunately, in the case of QCD the radiation is always 
massive, which resolves the paradox and  screens the light-cone singularity in
the emission process. However, since the screening is a non-local effect 
associated with the longitudinal fields, it cannot be derived in the singular
geometry of the null-plane dynamics (see Sec.\ref{sec:SS1}).

With the shorthand notation,
$~{\cal W}^{{R\choose A}}_{1}(p) = p^2 + p^2~w_{1}^{R\choose A}(p)$, and
$~{\cal W}^{R\choose A}_{2}(p)=1-w_{2}^{R\choose A}(p)$, we readily obtain,
\begin{eqnarray}
\left[{\bf D}_{ret}(k) \Pi_{01\choose 10}(k){\bf D}_{adv}(k)\right]^{\mu\nu}=
{-{\overline d}^{\mu\nu}(p)p^2 w_{1}^{01\choose 10}(p)\over 
{\cal W}^{R}_{1}(p){\cal W}^{A}_{1}(p)}+
{ w_{2}^{01\choose 10}(p)n^{\mu}n^{\nu}
\over (p^+)^2{\cal W}^{R}_{2}(p){\cal W}^{A}_{2}(p)}~,
\label{eq:E1.9}\end{eqnarray} 
which also exhibits a clear separation between the transverse and static
fields.   The first term on the right side is extracted by contraction with
the metric tensor $g^{\mu\nu}$; the second one is extracted with the aid of
the projector $p^{\mu}p^{\nu}/(p^{+})^2$.  The combination (\ref{eq:E1.9}) 
enters
the right side of Eq.~(\ref{eq:Q1.31}) and represents the density of states
into which the gluon field may decay. 

The first term  in Eq.~(\ref{eq:Q1.31}) corresponds to the initial data for the
field correlator. The role of this term is two-fold. According to the picture
qualitatively described in the Introduction, the final states in the inclusive
measurement are fully developed being the virtual fluctuations before the 
measurement. Therefore, we encounter two  types of initial fields.  The fields
of the first type (with the correlators labeled as $D^{\ast}$) are  real quark
and gluon fields of the proton or nuclei before the collision.  They always
have sources and no initial data which can be expressed in terms of the free
fields. Hence, the first term in Eq.~(\ref{eq:Q1.31}) for these  fields has to
be dropped. To understand  further consequences of this step,  let us  assume 
that the full retarded and advanced propagators in Eq.~(\ref{eq:Q1.31}) are
renormalized according to some condition (to be specified later). At the
renormalization point, the retarded and advanced self-energies vanish and,
therefore,    ${\bf D}_{ret}{\stackrel{\leftarrow}{D_{(0)}^{-1}}}=
{\stackrel{\rightarrow}{D_{(0)}^{-1}}} {\bf D}_{adv} =1$~. This leads to
\begin{eqnarray}
 {\bf D}^{\ast}_{10}  = Z~D^{\ast}_{10}-
 {\bf D}^{\ast}_{ret}\Pi^{\ast}_{10} {\bf D}^{\ast}_{adv},           
\label{eq:E1.1a}
\end{eqnarray}
where $Z$ is the residue of the pole of the renormalized propagator. Thus, the
absence of the initial data for the free gluon field is equivalent to the
requirement that its renormalization factor $Z$ equals  zero. Hence, this is
not a fundamental field and its spectral density is formed solely by 
multiparticle states into which it decays. (This option to make a distinction
between the elementary and composite fields was intensively discussed by
Weinberg \cite{Weinberg}.) 

The fields of the second type (with the correlators labeled as $D^{\#}$) are
used to decompose the colliding composite objects in terms of the  {\em true
normal modes} of the  final state which are excited only after the
collision. For example, in the case of the two nuclei colliding, the normal
modes are plasmons which have  effective masses $m$. To some approximation
(connected to the possible damping of excitations in the collective systems)
we have,  $D^{\#}_{10}\equiv D^{\#}_+\sim\delta(p^2-m^2)\theta(p^0)$, and we
may transform Eq.~(\ref{eq:Q1.31}) into  the known expression for the spectral
density,
\begin{eqnarray}
 {\bf D}_{+}^{\#\mu\nu}(p^2)  = 
 \int_{0}^{\infty} d m^2 \rho^{\mu\nu} (m^2)\delta(p^2-m^2)
 =Z~d^{\mu\nu}(p)\delta(p^2-m^2)\theta(p^0)+
 \int_{0}^{\infty} d m^2 \sigma^{\mu\nu} (m^2)\delta(p^2-m^2)\theta(p^0)~.       
\label{eq:E1.1b}\end{eqnarray}
Masses that enter this spectral representation  shield the abundant   collinear
singularities that may appear in the evolution equations.

\section{Temporal order in inclusive processes}
\label{sec:SNto} 

The program for computing the quark and gluon distributions in heavy-ion
collisions relies on the data obtained in seemingly more simple processes,
{\em e.g.}, $ep$-DIS. It turns out that differently triggered sets
of data may carry significantly different information. In this section,
we discuss two examples and show that even minor change in the way the
data are taken may strongly affect what is actually observed.

Let us consider a collision process where the  parameters  of only one final 
state particle  explicitly measured. Let  this  particle be an electron 
with momentum ${\bf k}'$ and spin $\sigma'$.  Deep inelastic 
electron-proton scattering is an example of such an experiment.  All vectors 
of final states which are accepted into the data ensemble are of the
form  $ a^{\dag}_{\sigma'}({\bf k}')|X\rangle $  where the vectors
$|X\rangle $  form a complete set. The initial state consists of the electron
with momentum ${\bf k}$ and spin $\sigma$ and the proton
carrying quantum numbers $P$.  Thus, the initial state
vector is  $ a^{\dag}_{\sigma}({\bf k})|P\rangle $.  The 
inclusive transition amplitude reads as
$\langle X|a_{\sigma'}({\bf k}')~S~a^{\dag}_{\sigma}({\bf k})|P\rangle $  
and the inclusive momentum distribution of the final-state electron is the
sum of the squared moduli of these amplitudes over the full set of the
non-controlled states $|X\rangle$. This yields the following formula,
\begin{eqnarray}
{ d N_e\over d {\bf k}'}=
\langle P|a_{\sigma}({\bf k}) S^{\dag} a^{\dag}_{\sigma'}({\bf k}')
a_{\sigma'}({\bf k}') S  a^{\dag}_{\sigma}({\bf k}) |P\rangle ~,
\label{eq:T.1}    
\end{eqnarray}      
which is just an average of the Heisenberg operator of the number of 
the final-state electrons  over the initial state.  Since the state
$|P\rangle$ contains no electrons, one may commute electron creation and
annihilation operators with the $S$-matrix and its conjugate $S^{\dag}$
pulling the Fock operators $a$ and $a^{\dag}$  to the right and to the
left, respectively. Let $\psi_{{\bf k}\sigma}^{(+)}(x)$ be the one-particle 
wave function of the electron. This procedure results in
\begin{eqnarray}
{ d N_e\over d {\bf k}'}={1\over 2} \sum_{\sigma\sigma'}
\int dxdx'dydy'{\overline\psi}_{{\bf k}\sigma}^{(+)}(x)
{\overline\psi}_{{\bf k'}\sigma'}^{(+)}(x')
\langle P| {\delta^2 \over \delta{\overline\Psi}(x) \delta \Psi(y)}
\bigg( {\delta S^{\dag} \over \delta\Psi(y')}
{\delta S \over \delta {\overline\Psi}(x')}\bigg) |P\rangle 
\psi_{{\bf k}\sigma}^{(+)}(y)\psi_{{\bf k'}\sigma'}^{(+)}(y')~.
\label{eq:T.2}    
\end{eqnarray}      
Since the electron couples only to the
electromagnetic field, then, to the lowest order,
\begin{eqnarray}
{ d N_e\over d {\bf k}'}={1\over 2} \sum_{\sigma\sigma'}
 \int dxdy{\overline \psi}_{{\bf k'}\sigma'}^{(+)}(y)
{\overline \psi}_{{\bf k}\sigma}^{(+)}(x)
\langle P|\not\!\!{\bf A^{(\gamma)}}(x) \not\!\!{\bf A^{(\gamma)}}(y) 
|P\rangle 
\psi_{{\bf k}\sigma}^{(+)}(y)\psi_{{\bf k'}\sigma'}^{(+)}(x) ~~,
\label{eq:T.4}    
\end{eqnarray}    
where  ${\bf A^{(\gamma)}}(x)$ is the Heisenberg operator of the
electromagnetic field. Already at this very early stage of calculations, the
answer has a very clear physical interpretation. Since only the final-state
electron is measured, the probability of the electron scattering is entirely
defined by the electromagnetic field produced by the rest of the system
evolved from its initial state until the moment of interaction with the
electron. 

The correlator of the two electromagnetic fields in  Eq.~(\ref{eq:T.4}) is the
function $i{\bf D}^{(\gamma)}_{10}(x,y)$. Therefore, we may use
Eq.~(\ref{eq:Q1.33}) to express it via the electromagnetic polarization tensor 
of the proton. Summation over the spins of the electrons brings in the leptonic
tensor,  $~L_{\mu\nu}(k,k')$.~  If $~q=k-k'~$ is the space--like momentum
transfer, then the DIS cross-section is given by
\begin{equation}
{d N_e \over d{\bf k'}}= {e^2 L_{\mu\nu}(k,k')\over 4(2\pi)^6 EE'} 
{\bf D}^{(\gamma)}_{ret}(q)\pi_{10}^{\mu\nu}(q){\bf D}^{(\gamma)}_{adv}(q)~~, 
\label{eq:T.6}
\end{equation}   
where $\pi_{10}^{\mu\nu}(q)$ is the notation for the 
correlator of the two electromagnetic currents,
\begin{equation}
 \pi_{10}^{\mu\nu}(x,y)=
\langle P|{\bf j}^\mu(x){\bf j}^\nu(y)|P\rangle~~.
\label{eq:T.7}\end{equation} 
(Alternatively, we may obtain the answer by means of the Yang-Feldman equation
\cite{YF},  
\begin{eqnarray}
~{\bf A}(x)=\int d^4y {\bf D}^{(\gamma)}_{ret}(x,y){\bf j}(y)~,\nonumber
\end{eqnarray} 
where $~{\bf j}(y)~$ is the Heisenberg  operator of the electromagnetic
current and $~{\bf D}^{(\gamma)}_{ret}(x,y)~$ is the retarded propagator of the
photon.) The correlator of the  currents is the source of the field which has
scattered the electron. Here, both  photon propagators are  retarded and
indicate the causal order of the process.  The measurement analyzes all those
fluctuation in the hadron, which have developed before the scattering of the
electron and created the virtual photon probed by the electron.

The tensor $\pi_{10}^{\mu\nu}(x,y)$ is naturally expressed via the two
quark correlators,
\begin{equation}
 \pi_{10}^{\mu\nu}(x,y)=
{\rm Tr}\gamma^\mu{\bf G}_{10}(x,y)\gamma^\nu{\bf G}_{01}(y,x)~~.
\label{eq:T.8}\end{equation} 
The correlators ${\bf G}_{10}$ and ${\bf G}_{01}$ are evolved from
the earlier times via equations (3.26) and (3.27) of Ref~\cite{QGD}. The field 
correlator generated via the source of the previous step of evolution is
\begin{eqnarray}
 {\bf G}_{01} \to {\bf G}^{\ast}_{01} = 
 -{\bf G}_{ret} \Sigma^{\ast}_{01}{\bf G}_{adv}~.           
\label{eq:T.9}
\end{eqnarray}
The newly created correlation ${\bf G}_{10}$ in the state of free propagation 
is connected with the  density of the final states,
\begin{eqnarray}
{\bf G}_{10} \to {\bf G}^{\#}_{10} = G^{\#}_{10}
-{\bf G}^{\#}_{ret}\Sigma^{\#}_{10}{\bf G}^{\#}_{adv}~.           
\label{eq:T.10}
\end{eqnarray}
Similarly, the self-energies $\Sigma^{\ast}_{01}$ are defined via the 
correlators ${\bf G}_{01}$ and ${\bf D}_{10}$ which can also be causally
evolved from  previous stages of the evolution. In this way, the causal ladder
of the fluctuations is built even without the concept of partons.

Early discussions of the role of the light cone distances in  high
energy collisions \cite{GIP} resulted in Gribov's idea of the two-step
treatment of  inelastic processes \cite{Gribov2};  the gamma-quantum {\em
first} decays into virtual hadrons and {\em later} these hadrons interact
with the nuclear target. This idea looks very attractive since it 
explains the origin of the two leading jets in electro-production events,
corresponding to  the target and the projectile (photon) fragmentation. It
also provides a reasonable explanation of the plateau in the  rapidity
distribution of the hadrons. However, this elegant qualitative
picture contains a disturbing element, {\em i.e.}, the way  the words
``first'' and ``later'' are used.  

\begin{figure}[htb]
\begin{center}
\mbox{ 
\psfig{file=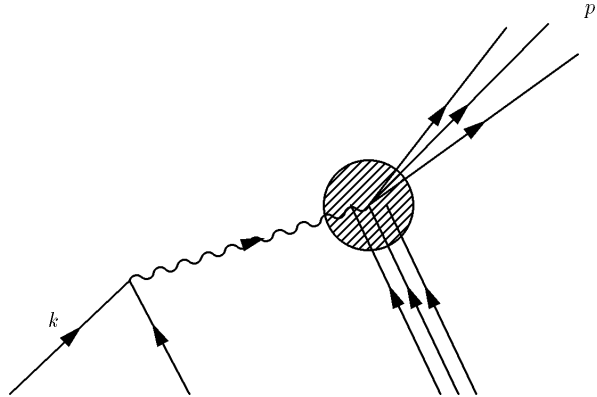,height=2.5in,bb=120 470 430 685}
}
\end{center}
\caption{Fluctuation which is ``active'' in the inclusive measurement
of jet distribution in DIS.}
\label{fig:fig2}
\end{figure}

In order to understand how Gribov's process may be observed practically,
let us change the observable in the same deep inelastic
process initiated by the $ep$-interaction. Let us  trigger events on the
high-$p_t$ quark or gluon jet in the final state regardless of the
momentum of the electron. This means  that all states that are included into
the data set are of the form,
$\alpha_{\lambda}({\bf p}) |X\rangle$, where
$\alpha_{\lambda}({\bf p})$ is the Fock  operator for the final state
quark.  The corresponding observable is the number of the final-state quarks.
To the lowest order of perturbation theory, this is a process of
the type $2\to 1$.  Its inclusive probability is   
\begin{eqnarray}
{dN_q\over d{\bf p}}=
\langle P|a_{\sigma}({\bf k})~S^{\dag}\alpha^{\dag}_{\lambda}({\bf p})
\alpha_{\lambda}({\bf p})~S~ a^{\dag}_{\sigma}({\bf k})|P\rangle~. 
\label{eq:T.11}
\end{eqnarray}
After commutation of the Fock operators with  $S$ and $S^{\dag}$ we arrive
at the expression,
\begin{eqnarray}
{ d N_q\over d {\bf p}}={1\over 2} \sum_{\sigma\lambda}
\int dxdx'dydy'{\overline\psi}_{{\bf k}\sigma}^{(+)}(x)
{\overline q}_{{\bf p}\lambda}^{(+)}(x')
\langle P| {\delta^2 \over \delta{\overline\Psi}(x) \delta \Psi(y)}
\bigg( {\delta S^{\dag} \over \delta q(y')}
{\delta S \over \delta {\overline q}(x')}\bigg) |P\rangle 
\psi_{{\bf k}\sigma}^{(+)}(y)q_{{\bf p}\lambda}^{(+)}(y')~.
\label{eq:T.12}    
\end{eqnarray}
With reference to the Bogolyubov's form of the micro-causality principle
\cite{Bogol},  which (in its simplest form) reads as
\begin{eqnarray}
{\delta \over \delta \phi(x)} \bigg(  S^{\dag} 
{\delta S \over \delta \phi(y)}\bigg) = 0 ,
~~~{\rm unless}~~~(x-y)^2>0~~~{\rm and}~~~x^0>y^0~,
\label{eq:T.13}    
\end{eqnarray}
we may argue that the space-time points $x'$ and $y'$ (where the quark jet 
is created) are inside the forward light cone of the possible points
$x$ and $y$ of the electron scattering.  To the lowest order, the graph for
this process is depicted at Fig.~\ref{fig:fig2}. 

The photon propagator is retarded. To produce the quark,  the hadronic target
has  to be hit either by the photon coming from the first step of the virtual 
fragmentation of the electron or by one of the partons formed by the further
virtual  fragmentation. By causality, both the electron and the hadronic target
must develop appropriate fluctuations {\em before} the moment of the creation 
of the final-state quark. Thus, the change of the trigger  drastically affects
the information read out of the data. Triggering on the high-$p_t$ jet in the
same process allows one  to filter out fluctuations corresponding to Gribov's
picture. 

\section{The evolution ladder to the lowest order}
\label{sec:SNE1}

The way to derive evolution equations in the causal picture of QFK has been
shown in Refs.~\cite{QFK,QGD}. In the case of pure glue-dynamics, we start
with the Eq.~(\ref{eq:Q1.7}) for the tensor $\Pi_{01\choose 10}$, and 
iterate this equation with the aid of the Eqs.~(\ref{eq:Q1.31}). To the lowest
order in $g^2$, the vertices are considered to be bare (without the loop
corrections).  The iteration process explicitly accounts for the two types of
fields corresponding either to the core of the proton (these are not yet touched
by the evolution) or to the fields of anomalous fluctuations (which eventually
become the field of radiation.) Therefore, every off-diagonal correlator, 
${\bf D}_{01}$ or ${\bf D}_{10}$, corresponds either to the initial field of
colliding composite particles, or to the final state of the gluon field. In the
first case, we use equations (\ref{eq:Q1.31}) and express the field  correlators
via their sources formed during the preceding evolution,
\begin{eqnarray}
 {\bf D}_{{\stackrel{10}{\scriptscriptstyle 01}}} \to
 {\bf D}^{\ast}_{{\stackrel{10}{\scriptscriptstyle 01}}} = 
-{\bf D}_{ret} \Pi^{\ast}_{{\stackrel{10}{\scriptscriptstyle 01}}}
{\bf D}_{adv}~.           
\label{eq:E1.1}
\end{eqnarray}
The first term in Eq.~(\ref{eq:Q1.31}) is dropped, which corresponds to the main
physical assumption that the proton or nucleus is a fundamental mode of QCD and
that there is no {\em free} gluon fields defined by the initial data.   In the
second case, we replace  ${\bf D}_{{\stackrel{10}{\scriptscriptstyle 01}}}$ by
the densities of the final states,
\begin{eqnarray}
{\bf D}_{{\stackrel{10}{\scriptscriptstyle 01}}} \to
 {\bf D}^{\#}_{{\stackrel{10}{\scriptscriptstyle 01}}}  = 
D^{\#}_{{\stackrel{10}{\scriptscriptstyle 01}}}
-{\bf D}^{\#}_{ret}\Pi^{\#}_{{\stackrel{10}{\scriptscriptstyle 01}}}
 {\bf D}^{\#}_{adv}~.           
\label{eq:E1.2}
\end{eqnarray}
Here, the first term corresponds to the immediate creation of the final state
on-mass shell gluons (therefore, their propagators have no radiative
corrections). By keeping this term intact, we implicitly recognize that the
{\em states} of freely propagating color fields do exist. This is both the 
main assumption and the corner-stone of   perturbative QCD. Though this step
allows one to approach the exclusive  high-energy processes from the ``safe''
side of the QED-like perturbation theory, it completely disregards the
existence of the hadronic scale and creates artificial problems for collinear
processes.  Further analysis indicates that  in the picture  with 
realistic final states, this term has either to be modified by the redefinition
of the normal modes ({\em e.g.}, in dense matter), or be abandoned (if the
final states are hadronic jets).  The second term corresponds to multiparticle
emission (via the intermediate off-mass-shell gluon field) and is more
physical. 

For now, we limit ourselves by the condition that only one
two-point correlator of this expansion can originate from the QCD evolution
of the initial state and that the emitted gluon is on-mass-shell,
\begin{eqnarray}
  \Pi^{\mu\nu}_{01}(p)=-{i\over 2}\int  {d^4 k \over (2\pi)^4}
  V^{\mu\alpha\nu}_{acf}(p,k-p,-k)
\left[{\bf D}_{ret}(k) \Pi_{01}(k){\bf D}_{adv}(k)\right]^{\alpha\beta}_{cc'}
 V^{\nu\beta\sigma}_{bc'f'}(-p,p-k,k) 
 D_{10, f'f}^{\#\lambda\sigma}(k-p) \}~. 
\label{eq:E1.3}\end{eqnarray} 
This is the full tensor form of the evolution equations of pure 
glue-dynamics to order $g^2$. They have to be projected onto the normal
modes of the gluon field and be rewritten in the form of scalar equations
for the invariant functions.

Our next goal is to find the evolution equations for the  invariants, $w_1$
and $w_2$ of the gluon field.  Using the Eqs.~(\ref{eq:E1.6}) and 
(\ref{eq:E1.9}) we  extract evolution equations for the various invariants
from the tensor evolution equations (\ref{eq:E1.3}).  We obtain
two equations for the invariants of the gluon source:
\begin{eqnarray}
{p^2~w_{1}^{(01)}(p)\over {\cal W}^{R}_{1}(p){\cal W}^{A}_{1}(p)}
={ N_c \alpha_s\over \pi^2}
\int d^4 k  \delta_+[(k-p)^2] 
\bigg[ {-p^2/z+k^2 \over {\cal W}^{R}_{1}(p){\cal W}^{A}_{1}(p)}P_{gg}(z)
{k^2 w_{1}^{(01)}(k)\over  
{\cal W}^{R}_{1}(k){\cal W}^{A}_{1}(k)} - \nonumber\\
- {(z-1/2)^2\over {\cal W}^{R}_{1}(p){\cal W}^{A}_{1}(p)}
~ {w_{2}^{(01)}(k) \over
{\cal W}^{R}_{2}(k){\cal W}^{A}_{2}(k)} \bigg]~~,
\label{eq:E1.10}\end{eqnarray}    
\begin{eqnarray}
{w_{2}^{(01)}(p)\over {\cal W}^{R}_{2}(p){\cal W}^{A}_{2}(p)}
={2 N_c \alpha_s\over \pi^2}
\int d^4 k \delta_+[(k-p)^2]  ~{(1/z - 1/ 2)^2
 \over {\cal W}^{R}_{2}(p){\cal W}^{A}_{2}(p)}
{k^2 w_{1}^{(01)}(k)\over {\cal W}^{R}_{1}(k){\cal W}^{A}_{1}(k)} ~,
\label{eq:E1.11}\end{eqnarray} 
where $P_{gg}(z)$ is a well-known splitting kernel,
\begin{eqnarray}
P_{gg}(z)={1-z\over z}+{z\over 1-z}+z(1-z)~. 
\label{eq:E1.12}\end{eqnarray}
Within the accuracy of this approximation, {\em viz}, when 
$^{(2)}\Pi_{01}\sim g^2$ is used as the basis for the iteration procedure,
the propagators  ${\cal W}^{R}(p)$
and ${\cal W}^{A}(p)$ on the right-hand side do not include loop
corrections. 

The first term on the right side of Eq.(\ref{eq:E1.10}) has a transverse
source in it, which describes a step of evolution when the initially
transverse field pattern remains transverse after the emission. In what
follows, we shall refer to this type of transition as the TT-mode.  The second
term in this equation contains a longitudinal (static) source and after one
act of real  emission the new field  pattern becomes transverse. This field
can be, {\em e.g.}, the proper field of the static source, which  became a
propagating  transverse field after the static source has been accelerated
(TL-transition mode). The second equation describes the process of creation of
the new static field of the source after the acceleration has been terminated
(LT-transition mode).  There is no LL-transition mode since any rearrangement 
of a charged system between two different static configurations requires at
least two emissions. The first emission extinguishes the old static field, the
second emission creates a new one.

Connections between the invariants $w_{j}^{01}(p)$ and the structure functions
follow from the description of measurement, 
\begin{eqnarray}
c~\int dp^-{ip^+p^2w_{1}(p)\over {\cal W}^{R}_{1}(p){\cal W}^{A}_{1}(p)}= 
{d G (x,p_{t}^{2})\over d p_{t}^{2}}~,~~~
c~\int dp^-{ip^+~w_{2}(p)\over {\cal W}^{R}_{2}(p){\cal W}^{A}_{2}(p)} = 
{\cal G} (x,p_{t}^{2})~,
\label{eq:E1.13}\end{eqnarray}   
where $c$ is some common normalization constant. In terms of these
functions the evolution equations acquire a habitual form,
\begin{eqnarray}
{d G(p^+,p_{t}^{2})\over d p_{t}^{2}} 
= - {N_c \alpha_s\over \pi^2}\int {d k^+ d {\vec k_t} \over 2 }
 \bigg[ \int dp^-{\delta_+[(k-p)^2] \over p^2}
P_{gg}(z) {d G (k^+,k_{t}^{2})\over d k_{t}^{2}} + \nonumber\\
+\int dp^- {\delta_+[(k-p)^2] \over [p^2]^2}~\big(z-{1\over 2}\big)^2~
{\cal G} (k^+,k_{t}^{2}) \bigg] ~~,
\label{eq:E1.14}\end{eqnarray}  
\begin{eqnarray}
{\cal G} (p^+,p_{t}^{2}) 
= - {N_c \alpha_s\over \pi^2}\int {d k^+ d {\vec k_t} \over 2 }
 \int dp^-\delta_+[(k-p)^2] ~\big({1\over z}-{1\over 2}\big)^2~
 {d G (k^+,k_{t}^{2})\over d k_{t}^{2}} ~~.
\label{eq:E1.15}\end{eqnarray}
We must account for the static components in course of the QCD evolution  as
long as we wish to describe this process as developing in time, and as  long as
the fields are considered not only in the far zone, but also in the near zone,
where they continuously interact with the sources.

None of the integrations $dp^-$ in Eq.~(\ref{eq:E1.14}) is singular at the
point $k^+=p^+$. Indeed, assuming the strong ordering of the emission,
$p_t\gg k_t$, we obtain,
\begin{eqnarray}
{d G(p^+,p_{t}^{2})\over d p_{t}^{2}} 
={N_c \alpha_s\over \pi^2}\int_{p^+}^{P^+} {d k^+  \over 2k^+ }
\bigg[{1\over p_t^2}P_{gg}(z) 
\int d {\vec k_t}{d G (k^+,k_{t}^{2})\over d k_{t}^{2}} + 
{1-z\over p_t^4}~\big(z-{1\over 2}\big)^2~
\int d {\vec k_t}{\cal G} (k^+,k_{t}^{2}) \bigg] ~~.
\label{eq:E1.14a}\end{eqnarray}  

\begin{figure}[htb]
\begin{center}
\mbox{ 
\psfig{file=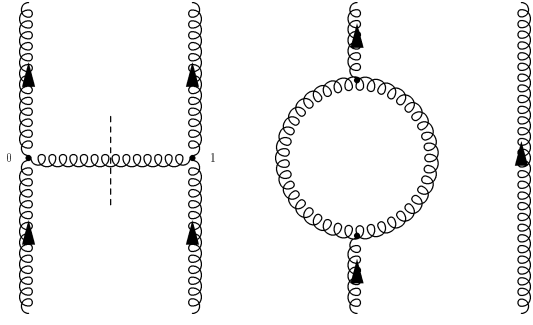,height=2.in,bb=100 359 545 655}
}
\end{center}
\caption{Two diagrams used to form the $(+)$-prescription.}
\label{fig:fig3}
\end{figure}

It is common to regulate the singularity of the splitting kernel $P_{gg}(z)$ at
$z=1$ with the aid of (+)-prescription.  In Ref.~\cite{Qui}, this  prescription
to order $g^2$  was obtained using two diagrams  depicted in
Fig.\ref{fig:fig3}. The singularity  in the one-loop  correction to the free 
propagation compensates for the divergence due to the collinear emission. 

In the case of  emission (that converts transverse field into the
longitudinal one), we encounter the same type of end-point singularity. 
Integration $dp^-$ in Eq.~(\ref{eq:E1.15}) is  singular and leads to the pole
$(1-z)^{-1}$ at $k^+=p^+$, {\em i.e.},
\begin{eqnarray}
{\cal G}(x,p_{t}^{2}) =-{N_c\alpha_s\over 2\pi}
N_c \int_{p^+}^{P^+} {d k^+ \over k^+-p^+}~{(2p^+-k^+)^2\over 4 p^+k^+}
G (k^+) ~,
\label{eq:E1.16}\end{eqnarray}       
where 
\begin{eqnarray}
G(p^+)=\int_{Q_0^2}^{\infty} d p_t^2 {d G(p^+,p_{t}^{2})\over d p_{t}^{2}}~,
\label{eq:E1.17}\end{eqnarray} 
is the $x$-fraction of the glue converted into  radiation, integrated over all
transverse momenta starting from some low (confinement ?) scale $Q_0^2$. 

Though the  pole in Eq.~(\ref{eq:E1.16}) is similar to the one in $P_{gg}(z)$
no self-energy correction to this process exists,  since the two-point
function cannot convert a T-mode into an L-mode. The only hope for the possible
removal of this divergence is connected to the radiative corrections of the
emission process.  Besides, unlike (transverse) partons the longitudinal
fields are beyond the factorization scheme and the RG for  the S-matrix. It is
not at all obvious what kind of running coupling accompanies the splitting
kernel. Therefore, we have to consider the evolution with the elementary cell
at least to the order $g^4$.

\section{The evolution equations to the order  $\alpha_s^2$}
\label{sec:SNE2}

There are two kinds of problems which cannot be resolved within the
approach of a simple ladder. First, we encounter collinear singularities which
may be (at least partially) compensated by loop corrections like  
${\cal W}^{R,A}_{i}(p)$. Second, the loop corrections and  the
additional real emission processes are required to assemble the running
coupling.

Let us start with the full expression for $^{(4)}\Pi_{01}(p)$, which emerges
after the vertex of the $g^3$-order (\ref{eq:Q1.11}) is substituted into 
the Eq.(\ref{eq:Q1.7}),
\begin{eqnarray}
 ^{(4)}\Pi^{\alpha\beta,ab}_{01}(x,y)= -{1\over 2} 
 \int d z_{1} d z_{2} \sum_{R,S=0}^{1}(-1)^{R+S}
V^{\alpha_1\alpha\alpha_2}_{a_1aa_2}(x)
{\bf D}_{0R}^{a_2r_2,\alpha_2\rho_2}(x,z_1)
V^{\rho_2\rho\rho_1}_{r_2rr_1}(z_1)
{\bf D}_{R1}^{r_1b_1,\rho_1\beta_1}(z_1,y)\times \nonumber\\
\times V^{\beta_1\beta\beta_2}_{b_1bb_2}(y) 
{\bf D}_{1S}^{b_2s_2,\beta_1\sigma_2}(y,z_2)
 V^{\sigma_2\sigma\sigma_1}_{s_2ss_1}(z_2) 
{\bf D}_{S0}^{s_1a_1,\sigma_2\alpha_1}(z_2,x)
{\bf D}_{RS}^{rs,\sigma\rho}(z_1,z_2)~~.
\label{eq:E2.1}
\end{eqnarray}
In this equation, we still keep all the gluon correlators dressed. We decompose
it using the following guidelines. First, we divide terms into two major groups
corresponding to $R=S$ and $R\neq S$. The former gives rise to the virtual
vertex corrections, while the latter describes all possible real emission
processes. The loops of virtual corrections are formed by the diagonal
propagators. They are ultraviolet-divergent and  require renormalization. 
Inside each group, every off-diagonal correlator, ${\bf D}_{01}$ or ${\bf
D}_{10}$ corresponds either to the initial field of the colliding composite
particles, or to the final state of the gluon field.  Thus, we again use Eqs.
(\ref{eq:E1.1}) and (\ref{eq:E1.2}) along with the previous agreement  
that only one two-point correlator of this expansion can originate from the
QCD evolution of the initial state.

In this approximation, two types of radiative corrections appear in the 
evolution equations. First, we have to calculate the basic polarization tensor
$\Pi_{01}$ to order $g^4$ by accounting for the vertex corrections. Second, we
should account for the loop corrections in all propagators. The formal counting
of the order comes from the following decomposition,
\begin{eqnarray}
{\bf D}_{01}=-{\bf D}_{ret}\Pi_{01}{\bf D}_{adv}=
-D_{ret}~[1+\Pi_{ret}D_{ret}]~\Pi_{01}~[1+D_{adv}\Pi_{adv}]~D_{adv} \nonumber\\
\approx -D_{ret}[^{(4)}\Pi_{01} + {}^{(2)}\Pi_{ret}D_{ret}{}^{(2)}\Pi_{01}
+{}^{(2)}\Pi_{01}D_{adv}{}^{(2)}\Pi_{adv}]D_{adv},          
\label{eq:E2.4}
\end{eqnarray}
where $~^{(n)}\Pi\sim g^n$. 
After separation of the transverse and longitudinal modes, we obtain,
\begin{eqnarray}
{p^2~w_{1}^{01}(p)\over {\cal W}^{R}_{1}(p){\cal W}^{A}_{1}(p)}=
{p^2~^{(4)}w_{1}^{01}(p)\over [p^2]^2} +
2p^2{^{(2)}w_{1}^{s}(p) ^{(2)}w_{1}^{01}(p)\over [p^2]^2}~, 
\label{eq:E2.5}\end{eqnarray}
\begin{eqnarray}
{p^2~w_{2}^{01}(p)\over {\cal W}^{R}_{2}(p){\cal W}^{A}_{2}(p)}=
~^{(4)}w_{2}^{01}(p)+2{}^{(2)}w_{2}^{s}(p){} ~^{(2)}w_{2}^{01}(p)~.
\label{eq:E2.6}\end{eqnarray}

\subsection{The gluon self-energy corrections}
\label{sbsec:SBE2a} 

Various gluon self-energies $^{(2)}\Pi_{AB}(p)$ of the order $g^2$ are  needed
to compute the elementary cell of the QCD ladder to order $g^4$. Though the
results are well-known, we consider it instructive to go through their
derivation in some details.

The simplest of  $^{(2)}\Pi_{AB}$, the self-energy $^{(2)}\Pi_{10}^{\#}(p-k)$, 
is the correction to the emission process due to the ``mass'' of the decoupled
gluon which splits further into two massless gluons.  These two gluons belong
to the space of the final freely-propagating states. Therefore,
\begin{eqnarray}
  \Pi^{\alpha\beta}_{10}(p)={ig^{2}\over 2(2\pi)^2}\int d^4 q
 \theta(q^+) \theta(p^+-q^+)\delta(q^+q^- -q_t^2)
 \delta[(p^+-q^+)(p^--q^-)-(\vec{p_t}-\vec{q_t})^2]\nonumber\\
 \times   V^{\mu\alpha\rho}_{acf}(q-p,p,-q)\bar{d}_{\rho\sigma}(q)
 V^{\sigma\beta\nu}(q,-p,p-q) \bar{d}_{\nu\mu}(q-p)~. 
\label{eq:E2.7}\end{eqnarray}
After integration over $q^-$ with the aid of the first $\delta$-function
we obtain the following expressions for the transverse and longitudinal 
invariants $w_{1}^{\# 10}$ and $w_{2}^{\# 10}$:
\begin{eqnarray}
\left[ \begin{array}{c}  w_{1}^{\# 10}(p) \\ w_{2}^{\# 10}(p)\end{array}\right]
=-{ig^2N_c\over (2\pi)^2} \int_{0}^{p^+}{dq^+\over q^+}
\left[ \begin{array} {c} -p^2 P_{gg}({q^+\over p^+}) \\ 
                 2\bigg( {1\over 2}-{q^+\over p^+}\bigg)^2 \end{array}\right]
                 \int d^2 \vec{q_t} 
\delta\big[(p^+-q^+)\big(p^--{q_t^2\over q^+}\big)-
(\vec{p_t}-\vec{q_t})^2 \big]~.
\label{eq:E2.8}\end{eqnarray}
The divergence of this integral can be viewed in different ways. First, let us
notice that the integration $d^2 \vec{q_t}$ involves only a $\delta$-function
and the result of this integration is $\pi p^+/q^+$. Thus, the divergence
becomes connected  with the $dq^+$ integration, and requires both lower, and
upper cut-offs in the remaining integral for $w_{j}^{\# 10}(p)$,
\begin{eqnarray}
 w_{j}^{\# 10}(p) 
=-{ig^2N_c\over 8\pi} \theta (p^2) \theta (p^+)\beta_{j}(p^+, \epsilon)~,
\label{eq:E2.9}\end{eqnarray}
where we have introduced the functions
\begin{eqnarray}
 \beta_{1}(p^+, \epsilon)=
-2 N_c \int_{\epsilon}^{p^+-\epsilon}{dq^+\over p^+}
P_{gg}({q^+\over p^+})~\to~
N_c \bigg({11\over 3}-
4\ln {p^+\over\epsilon}\bigg),~~{\rm when}~~\epsilon\to 0, 
\label{eq:E2.10}\end{eqnarray}
\begin{eqnarray}
\beta_{2}(p^+, \epsilon)=
N_c \int_{\epsilon}^{p^+-\epsilon}{dq^+\over p^+}
\bigg( 1-2{q^+\over p^+}\bigg)^2 ~\to~{N_c\over 3},~~{\rm when}~~\epsilon\to 0~.
\label{eq:E2.11}\end{eqnarray}

Being introduced in this way, the cut-off may be thought of as a regulator for
the spurious poles in the gluon correlators $D_{10}(q)$ and $D_{01}(q-p)$ which
form the loop. However, such an interpretation is excluded by the fact that
these correlators are on-mass-shell and include only transverse fields.
Therefore, the poles of the polarization sums $\bar{d}(q)$ and $\bar{d}(q-p)$
are due to the transverse momenta of the on-mass-shell loop gluons and the
singularity of the remaining integration is a physical collinear (mass)
singularity. To remove this singularity, one has to eliminate the  collinear
region by a fiat, {\em viz.}, by introducing a physical parameter of resolution
for the emission process. The value of this parameter has to be extracted from
the data, {\em e.g.}, from the shape of jets in DIS. In the case of  heavy-ion
collisions, the cut off will naturally come from the finite density of the
final-state particles. Indeed, in the coordinate space, a collinear singularity
is caused by the infinite time-of-interaction between the radiation and its
massless source. A single interaction with the ``third body'' suffices to
interrupt the emission process and to allow the field of  radiation to decouple
from its source. 

If the cut-off parameter is sufficiently small, than the  integration results
in the well known numbers associated with the Gell-Mann-Low  beta-function. 
However, one should keep in mind that  if the $\ln\epsilon$-term is not
miraculously canceled due to some interference process and has to be kept
finite, then the magic ``$(11N_c/3)$'' becomes an approximate number. In fact,
the cut-off $\epsilon$ can be re-expressed via the lower limit of the invariant
mass of the two final-state gluons and it has to be even kept in the finite
quantity $\beta_2$.

To the lowest order, the retarded self-energy of a gluon can be computed via 
its imaginary part. This statement is not trivial. Indeed, the retarded
self-energy is as follows,
\begin{eqnarray}
 \Pi^{\alpha\beta,ab}_{{ret\choose adv}}(p)= -{i\over 4} 
 \int {d^4 q\over (2\pi)^4}
 [ V^{\mu\alpha\rho}_{fac}(q-p,p,-q)
{\bf D}_{{ret\choose adv}}^{cc',\rho\sigma}(q)
 V^{\rho\beta\nu}_{c'bf'}(q,-p,p-q) 
{\bf D}_{1}^{f'f,\nu\mu}(q-p) +\nonumber\\
+ V^{\mu\alpha\rho}_{fac}(q-p,p,-q)
{\bf D}_{1}^{cc',\rho\sigma}(q)
 V^{\rho\beta\nu}_{c'bf'}(q,-p,p-q) 
{\bf D}_{{adv\choose ret}}^{f'f,\nu\mu}(q-p)]~.
\label{eq:E2.12}
\end{eqnarray}
Though in both terms we encounter the retarded propagators, only their
transverse parts are truly causal. Thus, if we wish to use the dispersion
relations to calculate the real part of $\Pi_{ret}$  (which is responsible 
for the phase shifts in the propagation of the gluon field)
via the imaginary part (which is directly connected with the real processes),
we have to separate causal and non-causal parts in Eq.~(\ref{eq:E2.12}),
\begin{eqnarray}
D_{ret}(q)= D_{ret}^{(T)}(q)+D^{(L)}(q),~~
D_{adv}(q)= D_{adv}^{(T)}(q)+D^{(L)}(q) ~.
\label{eq:E2.13}
\end{eqnarray} 
Thus, we obtain the dispersive part of the retarded (advanced) self-energy,
\begin{eqnarray}
 \Pi^{(D)}_{{ret\choose adv}}(p)=-{i\over 4} 
 \int {d^4 q\over (2\pi)^4}
 [ V(q-p,p,-q) D_{{ret\choose adv}}^{(T)}(q)V(q,-p,p-q)  D_{1}(q-p) +\nonumber\\
+ V(q-p,p,-q) D_{1}(q) V(q,-p,p-q) D_{{adv\choose ret}}^{(T)}(q-p)]~,
\label{eq:E2.14}
\end{eqnarray}
and the non-dispersive part,
\begin{eqnarray}
 \Pi^{(ND)}_{{ret\choose adv}}(p)=-{i\over 4} 
 \int {d^4 q\over (2\pi)^4}
 [ V(q-p,p,-q) D^{(L)}(q)V(q,-p,p-q)  D_{1}(q-p) +\nonumber\\
+ V(q-p,p,-q)
D_{1}(q) V(q,-p,p-q) 
D^{(L)}(q-p)]~.
\label{eq:E2.15}
\end{eqnarray}
The difference,
\begin{eqnarray}
\Pi_{0}= \Pi_{ret}- \Pi_{adv}\equiv 
\Pi^{(D)}_{ret}- \Pi^{(D)}_{adv}=\Pi_{01}- \Pi_{10}~,
\label{eq:E2.16}
\end{eqnarray}
is the imaginary part of the retarded self-energy, which has been already
computed. Projecting onto the two normal modes, we obtain the real part of the
invariants as the dispersion integral of the imaginary part with one
subtraction at  some value $p^-=-\Omega$ ,
\begin{eqnarray}
{\rm Re}w_{j}^{ret}(p)\equiv w_{j}^{s}(p)={1\over \pi}
\int d \omega^- \bigg( {1\over \omega^- - p^-}
- {1\over \omega^- +\Omega}\bigg) {\rm Im}w_{j}^{ret}(\omega)~,
\label{eq:E2.17}
\end{eqnarray}
where
\begin{eqnarray}
{\rm Im} w_{j}^{ret}(\omega) 
={g^2N_c\over 16\pi} \theta (p^+\omega^- - p_t^2) {\rm sign} (p^+) 
\beta_{j}(p^+, \epsilon)~.
\label{eq:E2.18}\end{eqnarray}
An explicit calculation of the dispersion integral results in the renormalized
retarded self-energy,
\begin{eqnarray}
{\rm Re} w_{j}^{ret}(p) 
=-{g^2N_c\over 16\pi^2}  
\beta_{j}(p^+, \epsilon)\ln {\mu^2\over -p^2}~,
\label{eq:E2.19}\end{eqnarray}
where the subtraction point  $p^-=-\Omega$ is translated into
$p^2=-\mu^2=-p^+\Omega -p_t^2$.

The non-dispersive part identically vanishes for the real part of 
$w_{2}^{ret}$ (because of the polarization properties of the three-gluon
vertex). For the non-renormalized non-dispersive function we obtain a divergent
integral,
\begin{eqnarray}
{\rm Re}~[p^2 w_{1}^{(ND)}(p)] 
={-g^2 N_c\over 2(2\pi)^3}  \int dq^+ ~d^2\vec{q_t}
\bigg\{ {1 \over |q^+|}~
\bigg({q^+ + p^+\over q^+ - p^+}\bigg)^2 +{1 \over |q^+-p^+|}~
\bigg({q^+ - 2 p^+\over q^+ }\bigg)^2 \bigg\}~,
\label{eq:E2.20}\end{eqnarray}
which is independent of $p^-$ and is, therefore, a quasi-local function
proportional to $\delta(x^+)$. It identically vanishes after one subtraction at
any value of $p^-$. Thus, after the renormalization, the retarded self-energy 
is totally causal both for the transverse and the longitudinal modes.
An explicit choice of the renormalization point will be done later (together
with the virtual vertex corrections).

\subsection{General properties of the virtual vertex corrections}
\label{sbsec:SBE2b} 

Computing this type of correction is most important for our goals because the
vertex strongly depends on the renormalization condition and is affected by 
infrared singularities. It is a very sensitive tool for the physical analysis
of various renormalization prescriptions and from this calculation we draw our
major conclusions. The part of the gluon polarization density of the order
$g^4$ with the virtual vertex corrections has the following form,
\begin{eqnarray}
 \bigg[\Pi^{\alpha\beta,ab}_{01}(p)\bigg]_{VV}=
 2\delta^{ab}{- N_c^2\over 4} 
 \int {d^4k d^4q\over (2\pi)^8 }
V^{\alpha_1\alpha\alpha_2}(q-p,p,-q) 
V^{\rho_2\rho\rho_1}(q,k-q,-k)    \times \nonumber\\
\times [-{\bf D}_{ret}\Pi_{01}(k){\bf D}_{adv}]^{\rho_1\beta_1}
 V^{\beta_1\beta\beta_2}(k,-p,p-k) 
{\bf D}_{10}^{\#,\beta_1\sigma_2}(k-p)
 V^{\sigma_2\sigma\sigma_1}(k-p,q-k,p-q)    \times \nonumber\\
\times [D_{00}^{\alpha_2\rho_2}(q){\bf D}_{00}^{\sigma_2\alpha_1}(q-p)
{\bf D}_{00}^{\sigma\rho}(q-k)+
D_{11}^{\alpha_2\rho_2}(q){\bf D}_{11}^{\sigma_2\alpha_1}(q-p)
{\bf D}_{11}^{\sigma\rho}(q-k)]~~.
\label{eq:E2.23}
\end{eqnarray}
By virtue of the relations, $D_{00}=D_s + D_1/2$  and $D_{11}= -D_s + D_1/2$,
the string,
$D_{00}(q) D_{00}(q-p) D_{00}(q-k) + D_{11}(q) D_{11}(q-p) D_{11}(q-k)$,
which consists of propagators that form the loop of the virtual vertex, 
can be rearranged into 
\begin{eqnarray}
D_{1}(q) D_{s}(k-p) D_{s}(q-k) + D_{s}(q) D_{1}(k-p) D_{s}(q-k) + 
D_{s}(q)D_{s}(k-p) D_{1}(q-k).
\label{eq:E2.25}\end{eqnarray}  

\begin{figure}[htb]
\begin{center}
\mbox{ 
\psfig{file=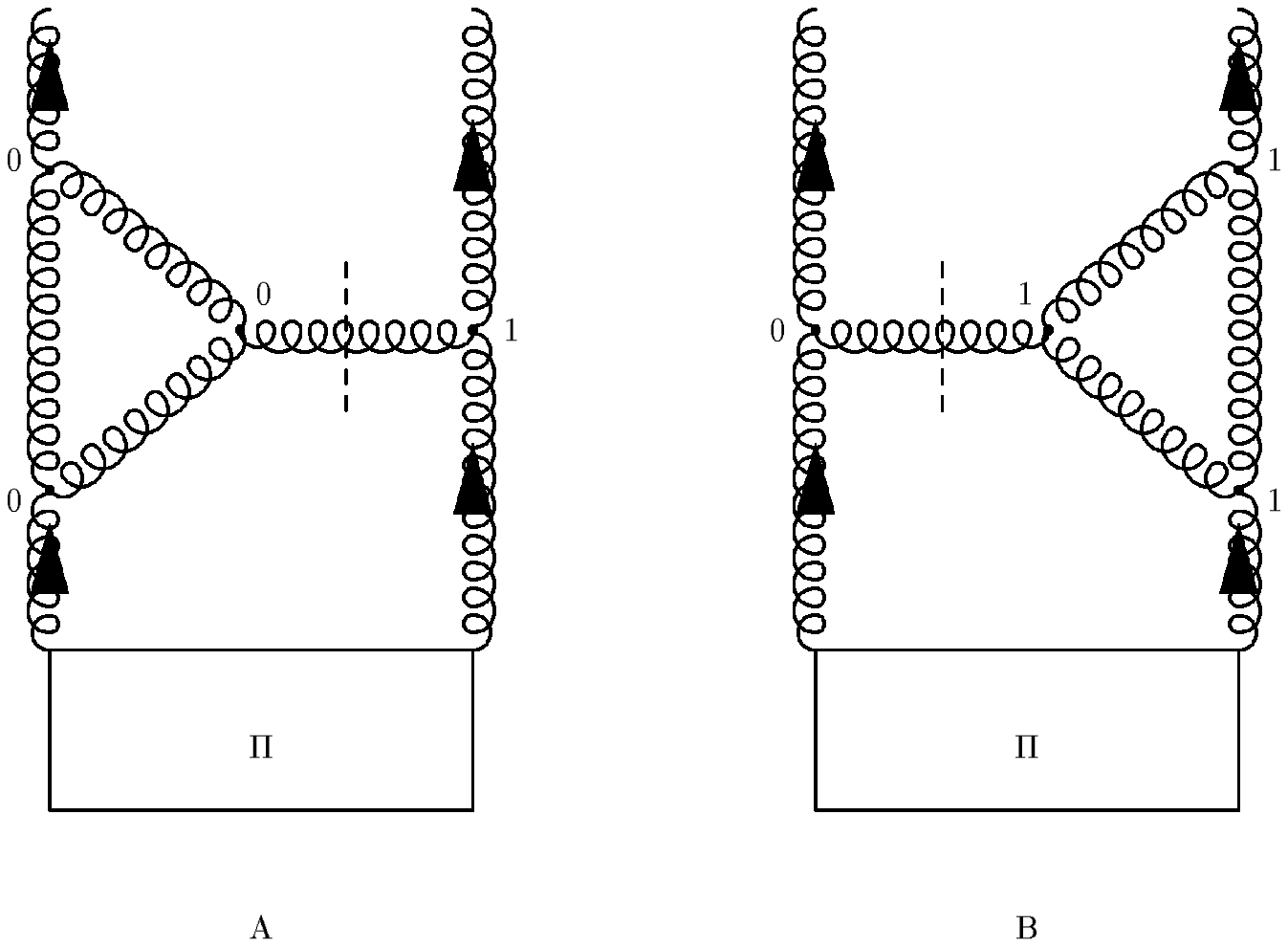,height=2.2in,bb=100 340 520 665}
\hspace{1.cm}
\psfig{file=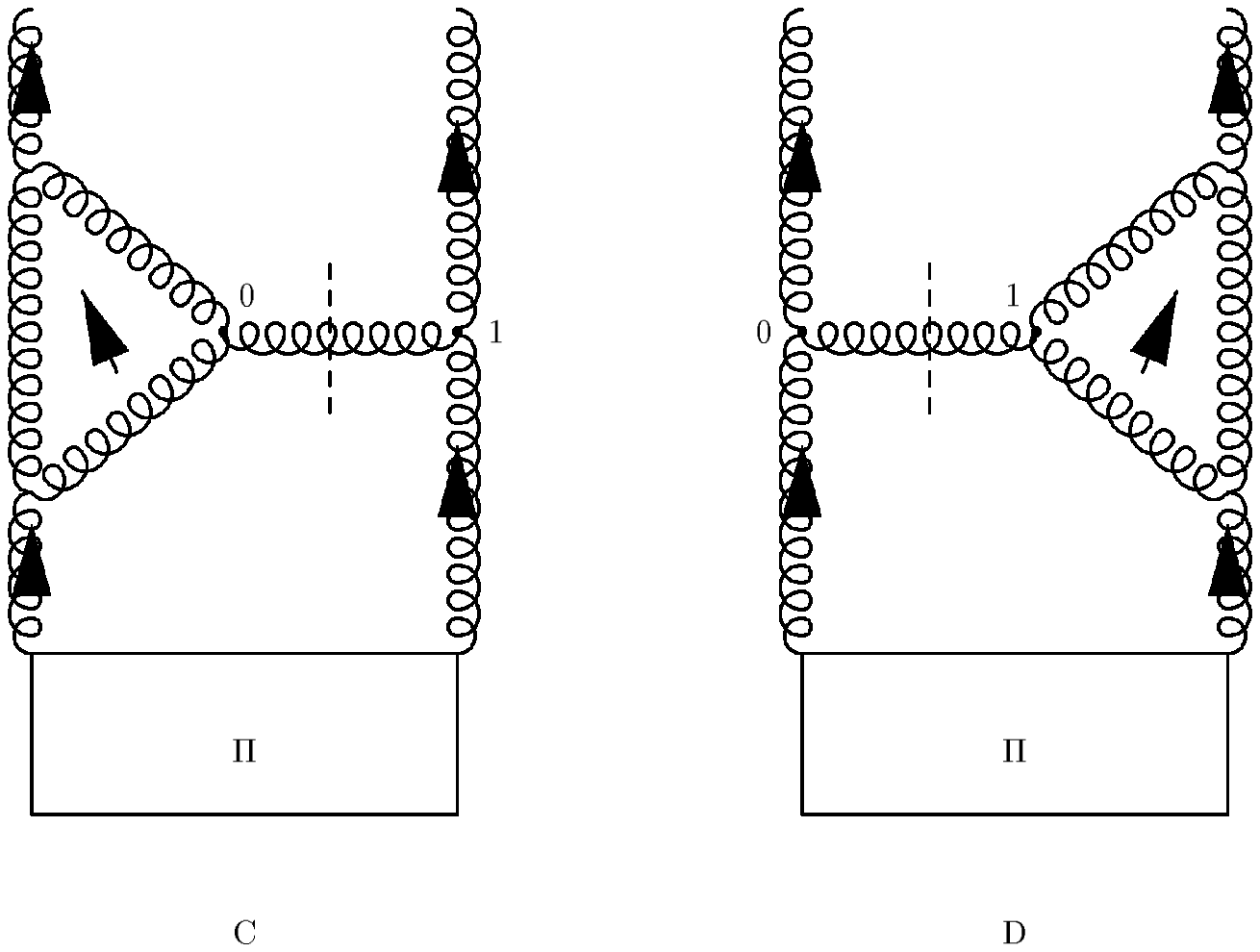,height=2.2in,bb=100 340 520 665}
}
\end{center}
\caption{Virtual vertex corrections. The original sum $V_{000}+V_{111}$
can be transformed into $V_{ret}+V_{adv}$, thus providing the causal
behavior of the ladder with the radiative corrections in the order $\alpha_s^2$}
\label{fig:fig4}
\end{figure}
In Appendix~1,  we show that the vertex function inside (\ref{eq:E2.23}),
rewritten in the form (\ref{eq:E2.25}) is, in fact, the real part of the
retarded vertex, {\em viz.} a triangle graph in which the space-time point $x$
(corresponding to the final momentum $p$) is the latest point on a real time
scale.  Therefore, one can safely replace the original $V_{000}$ by $V_{ret}$
and $V_{111}$ (of the conjugated graph) by $V_{adv}$,  since the
non-causal remainders do not contribute to the virtual vertex
corrections.  Thus, we proved that the virtual loop
corrections do not break the causal picture of the  QCD evolution, which has
been established above, at the tree level. Two retarded virtual vertex
loops (depicted at Figs.\ref{fig:fig4}c  and \ref{fig:fig4}d lead to the
following analytic expression
\begin{eqnarray}
 \bigg[\Pi^{\alpha\beta,ab}_{01}(p)\bigg]_{VV}=
 \delta^{ab}[g^{\alpha\beta}p^2w_{1}^{01}(p) 
 +p^{\alpha}p^{\beta}w_{2}^{01}(p)] 
 =\delta^{ab}{-iN_c^2\over 2} 
 \int {d^4k\over (2\pi)^4 } 2{\rm Re}
^{(3)}{\bf V}_{ret}^{\sigma_1\alpha\rho_2}(k-p,p,-k)\times\nonumber\\
\times  \bigg[ \bar{d}_{\rho_2\beta_2}
{k^2 w_{1}^{(01)}(k)\over {\cal W}^{R}_{1}(k){\cal W}^{A}_{1}(k)}+
n_{\rho_2} n_{\sigma_1}{w_{2}^{(01)}(k) \over
{\cal W}^{R}_{2}(k){\cal W}^{A}_{2}(k)} \bigg]
V^{\beta_2\beta\beta_1}(k,-p,p-k) 
{\bf D}_{10}^{\#,\beta_1\sigma_1}(k-p)~~.
\label{eq:E2.26}
\end{eqnarray}

Since the QCD evolution fully develops {\em before} the measurement, the
physical system of the emitted fields must be coherent. These fields represent,
at most, a virtual decomposition of the hadron in terms of a set of modes which
should emerge as propagating fields only after the interaction happens. This
understanding of the QCD evolution provides us with the natural condition for
the renormalization of the fields in the QCD evolution process. Namely,
radiative corrections to the propagation of the fields should not cause any
phase shifts along the line of propagation of the hadron. If the hadron is
assumed to move along the light cone, the real parts of the vertices and the
self-energies of the retarded propagators should vanish at $p^-=0$. Different
subtractions are also explored below.

The retarded vertex function  ${\bf V}_{ret}(p,k-p,-k)$ is proved to be
analytic in the lower half-plane of the complex light-cone energy, $p^-$.
Equivalently, the advanced function ${\bf V}_{adv}(p,k-p,-k)$ is analytic in 
the upper half-plane of the complex light-cone energy, $p^-$.  According to the
Schwartz symmetry principle, they form a single analytic function in the entire
complex plane with the cuts along the segments of the real axis where the
imaginary  part of this function does not vanish. These segments are 
$p_t^2/p^+ < p^- <\infty$ and  
$-\infty < p^- < -(\vec{p_t}-\vec{k_t})^2/(k^+-p^+)$. 

\begin{figure}[htb]
\begin{center}
\mbox{ 
\psfig{file=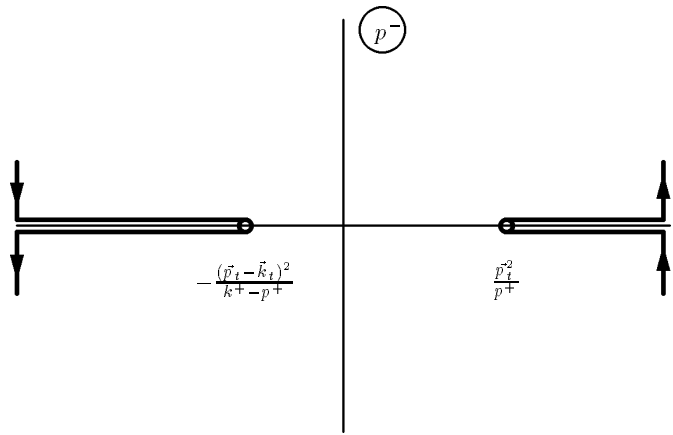,height=2.5in,bb=142 480 425 677}
}
\end{center}
\caption{Contour in the plane of complex $p^-$ used in the proof of the
dispersion relations for the vertex function. It is closed by two arcs 
at the infinity. }
\label{fig:fig5}
\end{figure}

Applying the Cauchy integral formula to  the contour depicted 
at Fig.~\ref{fig:fig5} one can prove the dispersion 
relation (with the subtraction at the point $p^-=-\Omega$),
\begin{eqnarray}
{\rm Re}^{(3)}{\bf V}_{ret}^{(D)}(p,-k,k-p) =
{1\over\pi}\int_{-\infty}^{+\infty} d \omega^- \bigg( {1\over \omega^- - p^-}
- {1\over \omega^- +\Omega}\bigg)
{\rm Im}^{(3)}{\bf V}_{ret}^{(D)}(\omega,-k,k-\omega)~,
\label{eq:E2.27}\end{eqnarray} 
where a four-vector  $\omega^\mu=(p^+,\omega^-,\vec{p_t})$ was introduced for 
brevity. The subtraction in the dispersion relation (\ref{eq:E2.27}) performs
an ultraviolet renormalization of the vertex function. The subtracted term is
quasi-local and does not affect the causal properties of the vertex.  The
subtraction can only be real  and can be performed only at a point where the
imaginary part is zero. This limits the position of the subtraction point to
the segment 
\begin{eqnarray}
{-(\vec{p_t}-\vec{k_t})^2\over k^+-p^+}< p^- < {p_t^2\over p^+ }~  
\label{eq:E2.28}\end{eqnarray} 
of the real axis, where ${\rm Im}^{(3)}{\bf V}_{ret}={\rm Im}^{(3)}{\bf
V}_{adv}=0$. Position $p^-= -\Omega$ of the subtraction point can be
translated into the value of $p^2$,~
$p^2=-\mu^2=-p^+\Omega -p_t^2$.  (Since we discuss the vertex function in its
natural environment (which includes the mass-sell delta-function of the
emitted gluon, $\delta[(k-p)^2]$)~ $p^2$ should be replaced by $-k^+
(\vec{p_t}-\vec{k_t})^2/(k^+-p^+)$.) Under an assumption of strong
ordering of the transverse momenta, $p_t>>k_t$, the allowed values of 
$\mu^2$ are
\begin{eqnarray}
              0< \mu^2 < p_t^2. 
\label{eq:E2.29}\end{eqnarray} 

This inequality may look abnormal since it does not allow one to choose the
renormalization point for the vertex function (corresponding to the currently
resolved momentum, $p$), {\em e.g.},  at $p^2 = Q^2$ (the much higher momentum 
which is measured at the end point of the QCD evolution).  Such a choice would
be almost obvious and even preferable in the S-matrix approach to the
QCD-evolution which does not require any space-time ordering in the evolution
process. Nevertheless, in our approach, the upper boundary in the inequality 
(\ref{eq:E2.29}) is a strict consequence of causality. Indeed, the momentum
transfer $Q$ will be detected later on, and it cannot affect the earlier loops
which correspond to the currently resolved momentum, $p$. {\em Causality
forbids one to renormalize what is not yet resolved.} However, we are free to
choose for the renormalization any of the already resolved lower scales. The
limits prescribed by the inequalities (\ref{eq:E2.28}) and (\ref{eq:E2.29}) are
obtained from the dispersive approach which expresses the real parts of the
loops (regulating the phase shifts) via the imaginary parts (which are related
to the real emission processes and are causally ordered in space and time).  
The renormalized real parts of the propagators and vertices will eventually be
combined into the effective coupling. Thus, the running coupling itself becomes
a dynamically-defined quantity that experiences evolution not only in terms of
the momenta, but also in terms of the space-time variables. This is  manifestly
in contrast with the S-matrix theory which achieves a running regime of the
coupling constant by means of the renormalization group method and assigns the
same value of coupling to the entire scattering process.

\section{Retarded vertex in the TT-mode}
\label{sec:SNE3} 

The dispersive part of the ${\bf V}_{ret}$ yields the following term in the 
evolution,
\begin{eqnarray}
\bigg[{dG(p^+,p_t^2)\over d p_t^2}\bigg]^{VV,(D)}_{TT}
={g^4 N_c^2\over 4(2\pi)^6}
\int {dk^+d{\vec k_t} \over 2 k^+} {dG(k^+,k_t^2)\over d k_t^2} 
[p^+ \Phi_{1}(p,k) + p^+ \Phi_{2}(p,k)]~~,
\label{eq:E3.1}\end{eqnarray}    
where the kernels $\Phi_{1}(p,k)$ and $\Phi_{2}(p,k)$  come from the dispersion
integrals along the right and left cuts, respectively.
\begin{eqnarray}
\Phi_{1}(p,k) = \int d p^-{\delta_{+}[(k-p)^2] \over [p^2]^2}
\int {(p^- +\Omega)d\omega^- \over (\omega^- -p^-)(\omega^-+\Omega)}
\int d^2 \vec{q_t}dq^-\int_{0}^{p^+} dq^+
{\delta_{+}[q^2]\delta_{+}[(q-\omega)^2]\over (q-k)^2}\times\nonumber \\
\times 8\{ -2 C_1 p^2(\omega^- -p^-) - 2 C_2 (k-q)^2(\omega^- -p^-) 
-C_3(\omega^- -p^-)^2 +C_4[(q-k)^2]^2 + C_5[p^2]^2 -2 C_6 p^2(k-q)^2\}~,
\label{eq:E3.2}\end{eqnarray} 
\begin{eqnarray}
\Phi_{2}(p,k) = -\int d p^-{\delta_{+}[(k-p)^2] \over [p^2]^2}
\int {(p^- +\Omega)d\omega^- \over (\omega^- -p^-)(\omega^-+\Omega)}
\int d^2 \vec{q_t}dq^-\int_{p^+}^{k^+} dq^+
{\delta_{+}[(k-q)^2]\delta_{+}[(q-\omega)^2]\over q^2}\times\nonumber \\
\times\{ -16 H_1 p^2q^2 + 16 H_2 q^2(\omega^- -p^-) 
-16 H_3(\omega^- -p^-)p^2 -8H_4(\omega^- -p^-)^2 + 8H_5[p^2]^2 
+8 H_6 [q^2]^2\}~,
\label{eq:E3.3}\end{eqnarray} 
where the expressions in curly brackets are the result of computing the trace
of the product of many Lorentz  matrices. The functions $C_j=C_j(k^+,p^+,q^+)$
and $H_j=H_j(k^+,p^+,q^+)$ are lengthy rational functions. They were obtained
and further handled with the aid of {\em Mathematica} and {\em FeynCalc}. The
terms proportional to $k_t^2$ are omitted because of the assumption of  strong
ordering of the transverse momenta along the ladder, $k_t^2<<p_t^2$.   Here we
give only an explanation of the main steps. As is clearly seen from
(\ref{eq:E3.1}) and (\ref{eq:E3.2}), (\ref{eq:E3.3}), the real part of the
retarded vertex is not calculated and renormalized separately. It is put (in
its tensor form) into the  environment of the ladder  in the form of the
dispersion integral with one subtraction,  and only after that is the trace
computed. We had to take this approach, because  it turned out to be very
difficult to find the tensor decomposition  of the third rank vertex tensor
(similar to  Eq.~(\ref{eq:E1.4}) for the self-energies).  Technically, the
calculations are conducted as follows. We start with integrating out the
variables $q^-$, $\omega^-$, and $p^-$, with the aid of three delta functions.
Then, we integrate over the angle in the plane of the vector $\vec{q_t}$. All
these integrations are finite. The last step is integration over
$q_t=|\vec{q_t}|$. Some of the  integrals happen to be divergent at lower or
upper limits of integration, or even at both of them. However, due to
miraculous relations between different coefficients $C_j$ and $H_j$, the sums
of the integrals are finite. They are as follows:
\begin{eqnarray}
p^+\Phi_{1}(p,k) = {8\pi\over p_t^2}\int_{0}^{p^+} {dq^+\over k^+}
\bigg\{  {k^+-q^+\over p^+} C_4 \ln\bigg(1+{k^+(p^+-q^+)\over p^+(k^+-q^+)}
\big(-1+{(k^+-p^+)\mu^2\over k^+ p_t^2}\big)\bigg) \nonumber \\ 
+{C_3 \over k^+(p^+-q^+)} \bigg[ \ln {(k^+-p^+)\mu^2\over k^+ p_t^2}
- \bigg({(k^+-p^+)\mu^2\over k^+ p_t^2}-1\bigg)
\ln\bigg(1+{q^+p_t^2 \over (p^+-q^+)\mu^2}\bigg) \bigg] \bigg\}~,
\label{eq:E3.4}\end{eqnarray} 
\begin{eqnarray}
p^+ \Phi_{2}(p,k) = {8\pi\over p_t^2}\int_{p^+}^{k^+} {dq^+\over k^+}
\bigg\{{k^+-q^+\over p^+} C_4 \ln\bigg(1+{k^+(p^+-q^+)\over p^+(k^+-q^+)}
\big(-1+{(k^+-p^+)\mu^2\over k^+ p_t^2}\big)\bigg) \nonumber \\
-{ C_3 \over k^+(p^+-q^+)}\bigg({(k^+-p^+)\mu^2\over k^+ p_t^2}-1\bigg)
\ln\bigg(1+{p^+(k^+-q^+)\over k^+(p^+-q^+)}
\big(-1+{(k^+-p^+)\mu^2\over k^+ p_t^2}\big)^{-1}\bigg) \bigg\}~.
\label{eq:E3.5}\end{eqnarray}
Recall that the parameter  $\mu^2=p^+\Omega+p_t^2$ is an equivalent of the
original subtraction point at $p^-=-\Omega$.  The remaining integration over
the light-cone component $q^+$ (of the momentum which flows around the loop) is
singular at the end points. We have carefully checked that no spurious poles
connected with the incomplete fixing of the gauge for the longitudinal part of
the gluon field is involved in the formation of these physical singularities.
They all are collinear (mass) singularities associated with the end points of
the allowed phase space. Moreover, these singularities already appear in the
calculation of the imaginary part of the vertex function thus being connected
to real processes in the complementary domain of the external momenta. Hence,
we are allowed to introduce a cut-off parameter $\epsilon$ (to shield the
singular behavior),  which defines an actual resolution of the process and
serves as a definition of a part of the newly created gluon field, that 
{\em should not} be treated as real emission.

Before the  integration in Eqs.~(\ref{eq:E3.4}) and (\ref{eq:E3.5}), 
it is expedient to select a large
parameter that defines major contribution to the final answer.  The  logarithms
~$\ln (p_t^2/\mu^2)$~  are good candidates since they can  be large if the
ratio $p_t^2/\mu^2$ is either very large or very small. Traditionally, one
would wish to associate $\mu^2$ with the highest momentum transfer $Q^2$ in the
process. Unfortunately, inequality (\ref{eq:E2.28}) deprives us this kind of
luxury. Moreover,  equations (\ref{eq:E3.4}) and (\ref{eq:E3.5}) contain
$\mu^2$ in the  first power, which also makes $\mu^2=Q^2$ very unlikely. Thus,
if we wish to hunt for large logarithms of $p_t^2/\mu^2$ formally, we are
forced to choose the renormalization point somewhere near the left tip of the
right cut in the plane of complex $p^-$. This corresponds to the on-mass-shell
(with respect to the momentum $p^\mu$ ) renormalization of the vertex function,
that is, in the domain which is most unphysical in the context of the QCD
evolution.  We even start the perturbative expansion with coupling typical for
the non-perturbative domain. Regardless to these reservations, we formally
obtain $p_t^2/\mu^2\gg1$  as a large parameter and can continue expecting to
arrive at the leading logarithmic approximation (LLA) with the largest term 
~$\sim\ln (p_t^2/\mu^2)$. The explicit calculation yields
\begin{eqnarray}
\bigg[{dG(p^+,p_t^2)\over d p_t^2}\bigg]^{VV,(D)}_{TT}
={g^4 N_c^2\over (2\pi)^6}~{2\pi\over p_t^2}
\int {dk^+d{\vec k_t} \over 2 k^+} {dG(k^+,k_t^2)\over d k_t^2} 
\bigg\{ {1-2z \over 6} +{\cal P}_{gg}(z)\bigg[{\pi^2\over 3}+
\ln^2{p_t^2\over\mu^2}-
2\ln{p^+\over\epsilon}\ln{ p_t^2 \over (1-z)\mu^2}\nonumber \\
+\ln^2{p^+\over\epsilon}
-{11\over 3}\ln{1-z\over z} +2\ln{p^+\over\epsilon}\ln{1-z\over z^2}
-4\ln z\ln{1-z\over z}\bigg]\bigg\}~~,
\label{eq:E3.6}\end{eqnarray}    
where $z=p^+/k^+$. The answer occurs to be doubly logarithmic which, may
invalidate the perturbative expansion, because no terms  $\sim\ln^2 p_t^2$ 
appear in other diagrams which could compensate it in the vertex.

There is no other way to proceed except for changing the  renormalization
point.   Let us make the new choice relying on the causal picture of the
process of the QCD evolution, which has been discussed previously, and make a
subtraction at the point $p^-=-\Omega=0$ which corresponds to $\mu^2=p_t^2$.
Then the parameter $\mu^2/p_t^2$ becomes unity, and equations (\ref{eq:E3.4})
and (\ref{eq:E3.5}) acquire the following form:
\begin{eqnarray}
p^+\Phi_{1}(p,k) = {8\pi\over p_t^2}\int_{0}^{p^+} {dq^+\over k^+}
\bigg\{  {C_3 \over (k^+)^2(p^+-q^+)} \bigg[ \ln \bigg(1-{p^+\over k^+}\bigg)
- {p^+\over k^+}
\ln\bigg(1-{q^+\over p^+}\bigg) \bigg]
- {k^+-q^+\over k^+p^+}~C_4~ \ln{k^+-q^+\over k^+-p^+} \bigg\}~,
\label{eq:E3.7}\end{eqnarray} 
\begin{eqnarray}
p^+ \Phi_{2}(p,k) = {8\pi\over p_t^2}\int_{p^+}^{k^+} {dq^+\over k^+}
\bigg\{{- p^+ C_3 \over (k^+)^3(p^+-q^+)}
\ln{k^+-p^+\over q^+-p^+}- {k^+-q^+\over k^+p^+} ~C_4~
\ln{k^+-q^+\over k^+-p^+} \bigg\}~.
\label{eq:E3.8}\end{eqnarray} 
The remaining integration is straightforward and results in the expression
\begin{eqnarray}
\bigg[{dG(p^+,p_t^2)\over d p_t^2}\bigg]^{VV,(D)}_{TT}
={g^4 N_c^2\over (2\pi)^6}~{2\pi\over p_t^2}
\int {dk^+d{\vec k_t} \over 2 k^+} {dG(k^+,k_t^2)\over d k_t^2} 
\bigg\{ -{2-5z+11z^2-6 z^3\over 36z(1-z)}+\nonumber \\
+{\cal P}_{gg}(z)\bigg[\pi^2{2+z\over 6}-{11\over 3} +(1-2z){\rm Li}_2(z)
+\ln^2z +({3\over 2}-z)\ln^2(1-z)-{z\over 2}\ln^2{1-z\over z}+
\ln z\ln(1-z) - \nonumber \\
-{11\over 3}\ln(1-z) 
+\bigg(6\ln(1-z)-2\ln z -2z\ln{1-z\over z}\bigg)\ln{p^+\over\epsilon}
+\ln^2{p^+\over\epsilon}\bigg]\bigg\}~.
\label{eq:E3.9}\end{eqnarray}  
It is interesting to trace the origin of the $log^2(p^+/\epsilon)$ in this
formula. Originally, the calculation leads to three doubly-logarithmic terms,
\begin{eqnarray}
z~\ln^2{p^+\over\epsilon} -z~\ln^2{k^+\over\epsilon}
+ \ln^2{p^+\over\epsilon},\nonumber
\end{eqnarray}
of which the first term comes from the vertex cut through the momentum $p$,
while the second and the third terms come from the cut through the momentum
$k-p$. Even though there is partial cancelation  between the first and the
second terms (which removes two of the double logs), this is not a systematic
effect. Indeed,  since the kinematic regions of two cuts do not intersect, this
cancelation cannot be an effect of destructive interference .

\begin{figure}[htb]
\begin{center}
\mbox{ 
\psfig{file=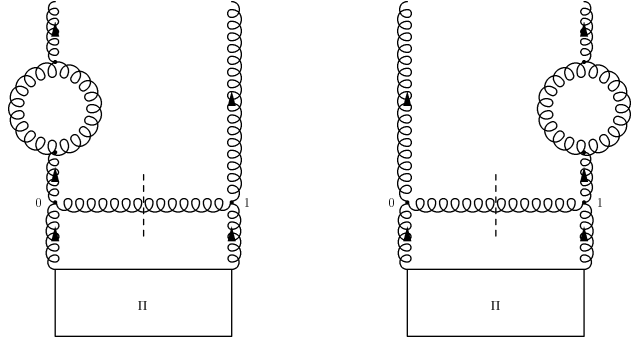,height=2in,bb=100 390 520 665}
}
\end{center}
\caption{Virtual corrections due to the retarded self-energies of order
$\alpha_s^2$.}
\label{fig:fig6}
\end{figure}

Finally, we have a correction due to the retarded self-energies ${\bf
D}_{{ret\choose adv}}(p)$. They are depicted in Fig.~\ref{fig:fig6}.
The result with the arbitrary subtraction point is as follows,
\begin{eqnarray}
\bigg[{dG(p^+,p_t^2)\over d p_t^2}\bigg]^{SE}_{TT}
=-{g^4 N_c\over (2\pi)^6} {2\pi\over p_t^2}\int {dk^+d{\vec k_t} \over 2 k^+} 
{dG(k^+,k_t^2)\over d k_t^2}{\cal P}_{gg}(z)\beta_1(p^+,\epsilon)
\ln\bigg[{p_t^2\over \mu^2(1-z)} \bigg]~.
\label{eq:E3.10}\end{eqnarray} 
Since the renormalization of the self-energy of the propagator is subject
to the same condition as the vertex function, we have to take $\mu^2=p_t^2$
and thus gain no large $\ln p_t^2$ coming from the ultraviolate renormalization.

Several remarks are in order:  

{\em 1}. The dispersive part of the retarded vertex function, renormalized in
accordance with the requirement of causality, does not contain any
$p_t^2$-dependent large numbers coming from the UV cut-off.  The non-dispersive
part of the retarded vertex ${\rm Re}~^{(3)}{\bf V}_{ret}^{(ND)}$  may
undermine the causal picture of the evolution, but it naturally combines with
the residue $\Delta{\rm Re}~{}^{(3)}{\bf V}_{ret}^{(D)}$, so that in the
TT-mode of the evolution equations they together produce the $p^-$-independent
quasi-local function, which identically vanishes after renormalization. This is 
demonstrated by an explicit calculation in Appendix~2. The same is true for
the terms originating from the four-gluon interaction. They prove to be
quasi-local as well, and vanish after subtraction at any value of $p^-$. The
proof is also given in Appendix~2.

{\em 2}.  The retarded self-energy with momentum $p^\mu$ also must be
renormalized by the subtraction at the point $p^-=0$, and thus carries no
logarithmic dependence on $p_t^2$. Therefore, we come to the conclusion, that
in the causal picture of QCD evolution, the renormalization of virtual loops
does not produce large logarithms of $p_t^2$. Hence, the only source of these
logarithms may be the real processes.

{\em 3}. Numerous collinear cut-offs in Eqs.~(\ref{eq:E3.9}) and
(\ref{eq:E3.10}) 
appear already at the level of the calculation of the imaginary parts of the
retarded vertex and retarded self-energy (unitary cuts) and thus are
exclusively due to  real processes. These singularities are intimately
connected with the definition of  emission as a physical process and 
require the parameter of resolution.

{\em 4}.  The expected  large logarithms of $p_t^2$ which are known to drive
the QCD evolution in the DGLAP equations cannot be an artifact of the
renormalization prescription.  Hence, they have to be outside of the
jurisdiction of the renormalization-dependent virtual loops and, therefore,
must originate from the real-emission processes. The renormalization
prescription itself has to be motivated physically.

\section{Real emission  in the TT-mode}
\label{sec:SNE4}  
 
\begin{figure}[htb]
\begin{center}
\mbox{ 
\psfig{file=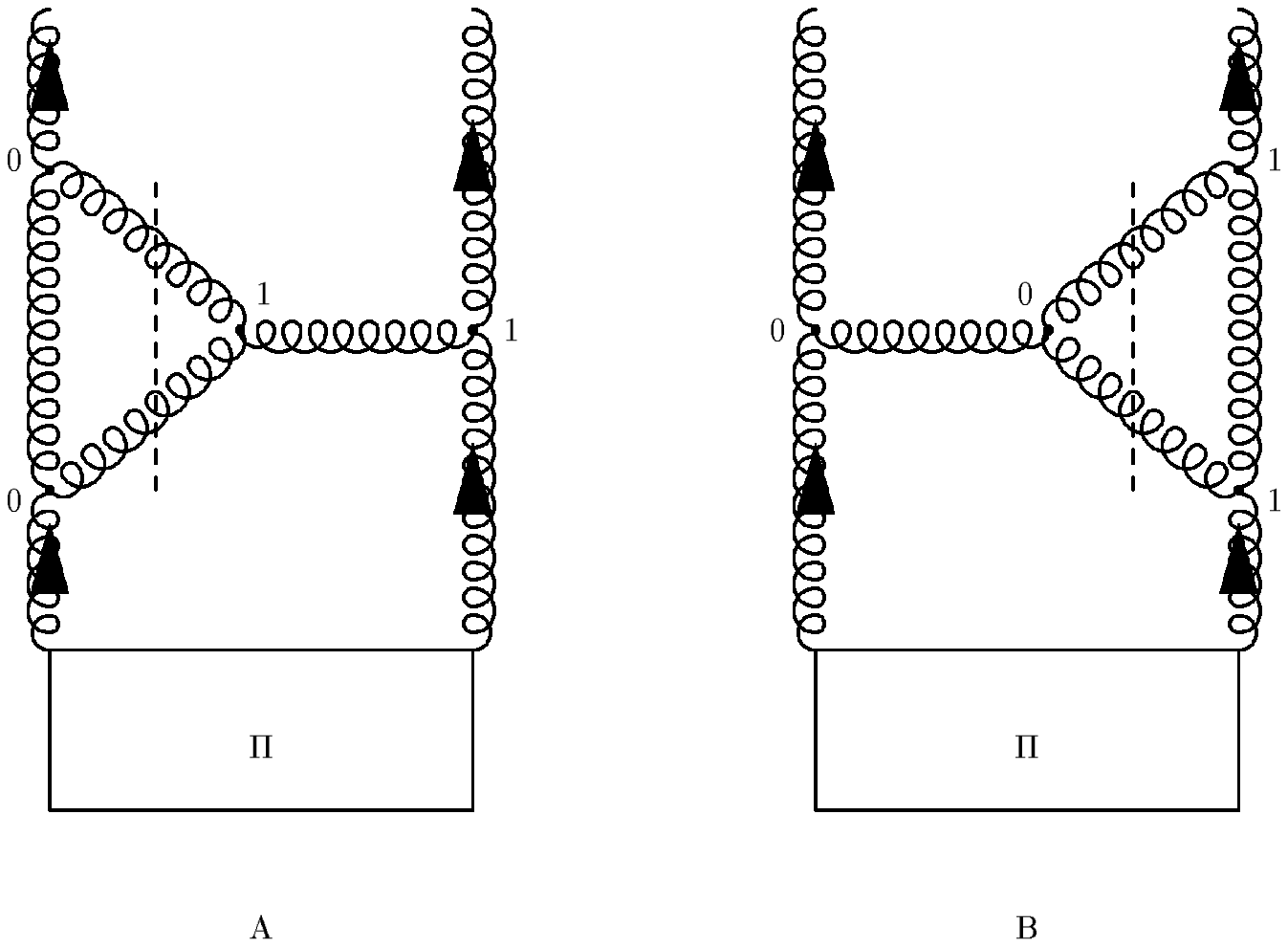,height=2.2in,bb=100 340 520 665}
\hspace{0.8cm}
\psfig{file=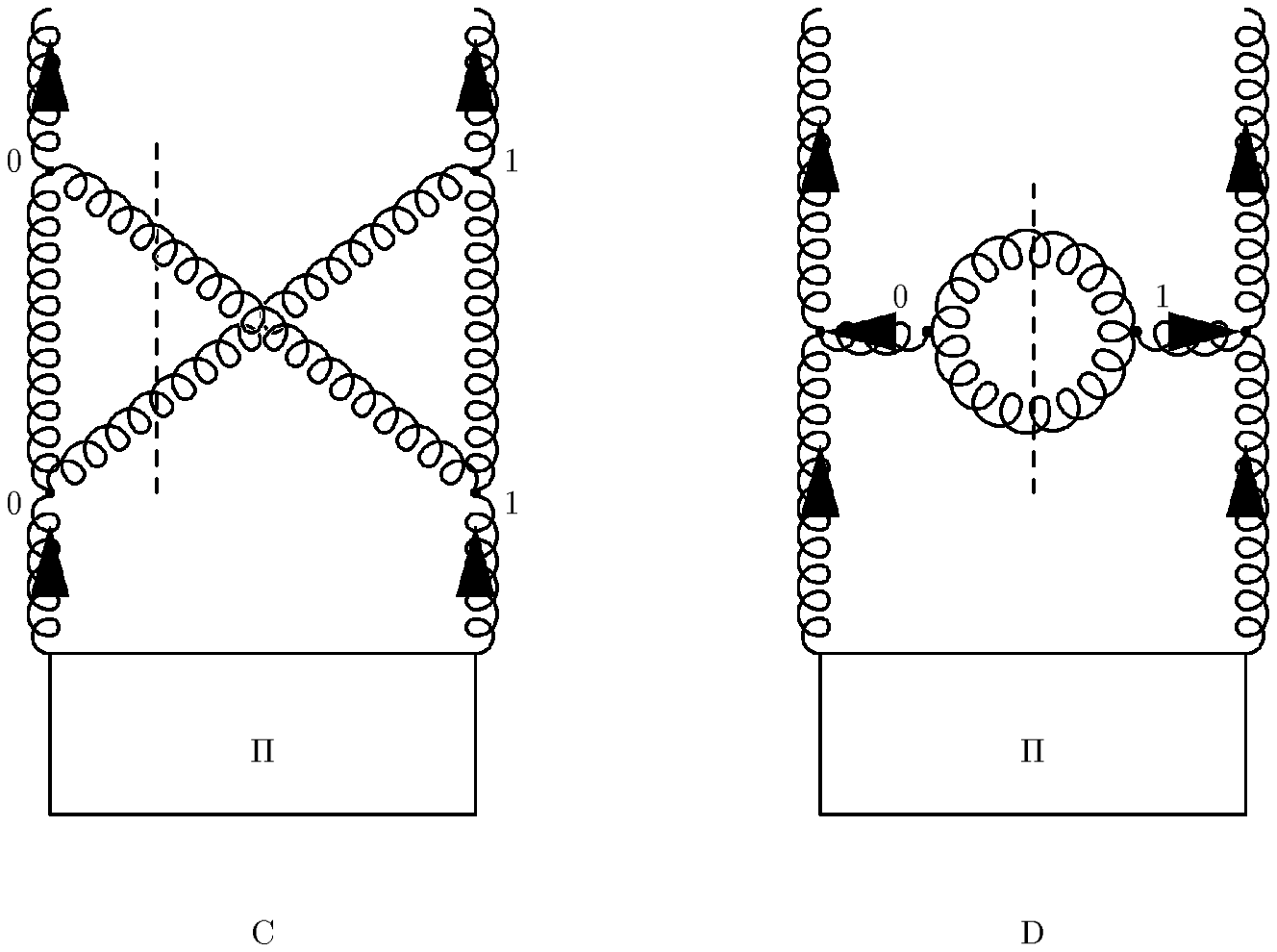,height=2.2in,bb=100 340 520 665}
}
\end{center}
\caption{Real  processes in the order $\alpha_s^2$.}
\label{fig:fig7}
\end{figure}

There are four diagrams which contribute to  real emission. Three of them
originate from the vertex correction (\ref{eq:Q1.7}) with $R\neq S$
(Fig.\ref{fig:fig7}a,b,c).  They differ by the place where the initial state
density $D^{\ast}_{01}$ is built into the diagram. If $D^{\ast}_{01}$ is put on
one of the lines which enters the external vertex of  ${}^{(4)}\Pi_{01}$  we
obtain the diagrams which include the cut vertex  (Fig.\ref{fig:fig7}a,b). If 
$D^{\ast}_{01}$ replaces the internal line $D_{RS}$, then we obtain a fragment
of the ladder with the crossed rungs (Fig.~\ref{fig:fig7}c, the interference
term for the emission of two off-spring gluons). The fourth diagram
(Fig.~\ref{fig:fig7}d) is a  direct descendent of the simple ladder diagram and
its order is enhanced due to the further splitting of the decoupled gluon. 

The two cut vertex diagrams, added together, have a fragment
\begin{eqnarray}
D_{00}(q) D_{11}(k-p) + D_{11}(q) D_{00}(k-p)=
{1\over 2}D_{1}(q) D_{1}(k-p) - 2 D_{s}(q) D_{s}(k-p).
\label{eq:E4.1}\end{eqnarray} 
The first singular term on the right side contributes nothing, while both
propagators of the second term acquire a principal value prescription. 
The analytic expression for the two cut vertex diagrams is as follows,
\begin{eqnarray}
\bigg[{dG(p^+,p_t^2)\over d p_t^2}\bigg]^{RV}_{TT}
={g^4 N_c^2\over (2\pi)^6} \int {dk^+d{\vec k_t} \over 2 k^+} 
{dG(k^+,k_t^2)\over d k_t^2}\int {d p^-\over [(k-p)^2]_{P.V.}} \int d^4 q
{\delta_{+}[(k-q)^2]\delta_{+}[(q-p)^2]\over [q^2]_{P.V.}}\times\nonumber \\
\times\{ R_1 [(p-k)^2]^2 + R_2 [q^2]^2 + R_3 q^2(p-k)^2+R_4[p^2]^2 + 
R_5p^2(p-k)^2 + R_6 p^2 q^2\}~,
\label{eq:E4.2}\end{eqnarray} 
where $R_j=R_j(k^+,p^+,q^+)$ are lengthy rational functions. The general line
of calculations is approximately the same as in the case of the virtual vertex.
First, we integrate out the variables $q^-$ and $p^-$ using the two delta
functions. Then, we integrate over the angle in the plane of the vector
$\vec{q_t}$. As previously, these integrations are finite. The last step is the
integration over $q_t=|\vec{q_t}|$. Some of the  integrals are divergent but,
due to  relations between different coefficients $R_j$ the sum of the integrals
is  finite. It already contains $\ln(k^+-q^+)$; therefore, the final
integration $dq^+$  between $p^+$ and $k^+$ is double-logarithmic. The full
answer is,
\begin{eqnarray}
\bigg[{dG(p^+,p_t^2)\over d p_t^2}\bigg]^{RV}_{TT}
={g^4 N_c^2\over (2\pi)^6}{4\pi\over p_t^2} \int {dk^+d{\vec k_t} \over 2 k^+} 
{dG(k^+,k_t^2)\over d k_t^2}
\bigg\{-{3+17z+2z^2\over 6(1-z)}
+{44-43z+66z^2-45z^3\over 12z(1-z)}\ln z +\nonumber \\
+{\cal P}_{gg}(z)\bigg({\pi^2\over 6}+{34\over 3} +4\ln^2z-4\ln z\ln(1-z)+
2\ln{1-z\over z^2}\ln{p^+\over\epsilon}-4\ln{k^+-p^+\over\epsilon}+
\ln^2{p^+\over\epsilon}\bigg)\bigg\}~.
\label{eq:E4.3}\end{eqnarray} 
Here, the term $\ln^2(p^+/\epsilon)$ is a partner of the similar term in the
virtual vertex (\ref{eq:E3.9}) which originates from the dispersion cut due to
the same real process. If a cancelation between these terms had taken place,
one could have considered it as a result of the interference (which is possible
at least owing to the same geometry of the emission.)  However, these terms 
not only differ by a factor of two, but also have the same sign.

The interference diagram (Fig.~\ref{fig:fig7}c) is calculated in a similar 
way to the cut-vertex diagram. The initial analytic expression is
\begin{eqnarray}
\bigg[{dG(p^+,p_t^2)\over d p_t^2}\bigg]^{INT}_{TT}
={2g^4 N_c^2\over (2\pi)^6} \int {dk^+d{\vec k_t} \over 2 k^+} 
{dG(k^+,k_t^2)\over d k_t^2}\int {d p^-\over [p^2]^{2}_{P.V.}} \int d^4 q
{\delta_{+}[(k-q)^2]\delta_{+}[(q-p)^2]
\over [q^2]_{P.V.} [\kappa^2]_{P.V.}}\times\nonumber \\
\times\{ U_1 [\kappa^2]^2 + U_2 q^2\kappa^2+ U_3 [q^2]^2 +U_4[p^2]^2 + 
U_5 p^2\kappa^2 + U_6 p^2 q^2\}~,
\label{eq:E4.4}\end{eqnarray} 
where $\kappa=p+k-q$. The final answer can be cast in a form
\begin{eqnarray}
\bigg[{dG(p^+,p_t^2)\over d p_t^2}\bigg]^{INT}_{TT}
={g^4 N_c^2\over (2\pi)^6}{8\pi\over p_t^2} \int {dk^+d{\vec k_t} \over 2 k^+} 
{dG(k^+,k_t^2)\over d k_t^2}
\bigg\{2 {\cal P}_{gg}(-z)\bigg({\pi^2\over 6}+2{\rm Li}_2(-z) 
+\ln(1+z)\ln z \bigg)-\nonumber \\
-{(1-z)(1088+143z+1088z^2)\over 36z}
+{(1+z)(22+29z+22z^2)\over 6 z }\ln z 
-{\cal P}_{gg}(z)\bigg(\ln^2z + 4\ln{k^+-p^+\over\epsilon}\bigg)\bigg\}~.
\label{eq:E4.5}\end{eqnarray} 
Neither the ``cut vertex'', nor the ``interference term'' has  
logarithms  $\ln p_t^2$, but both include 
logarithms  $\ln (1/x)$, which are not small  at low $x$.

The last of the three diagrams related to real processes is due to the 
decoupled heavy cluster as an emission field in the final state. To the lowest
possible order, one can mimic it by the diagram in Fig.~\ref{fig:fig7}d, 
where the ``heavy 
off-spring gluon'' decays into two massless gluons. In the lowest order, we
have to keep the propagator of the heavy gluon bare, which results in the
following expression,
\begin{eqnarray}
\bigg[{dG(p^+,p_t^2)\over d p_t^2}\bigg]^{HG}_{TT}
={ig^2 N_c\over (2\pi)^4} \int {dk^+d{\vec k_t} \over 2 k^+}  
{dG(k^+,k_t^2)\over d k_t^2}
\int_{-\infty}^{-{(\vec{k_t}-\vec{p_t})^2+m^2\over k^+-p^+}} d p^- \nonumber \\
\times\bigg\{ 8 {\cal P}_{gg}(z)
\bigg[{1\over p^2}+{z\over 1-z}{(k-p)^2\over [p^2]^{2}}\bigg]
{w_{1}^{\# 01}(k-p)\over (k-p)^2} + 
2z\bigg({1+z \over 1-z}\bigg)^2 {w_{2}^{\# 01}(k-p) \over [p^2]^{2}}\bigg\}~,
\label{eq:E4.6}\end{eqnarray}
where the cut-off $m^2$, is  defined by the properties of the real emission
processes, of which the most important is the existence of the minimal mass $m$
of the emitted jet, which is of the same order of magnitude  as the hadronic
mass. The two integrations $dp^-$ are straightforward:
\begin{eqnarray}
\int_{-\infty}^{-{(\vec{k_t}-\vec{p_t})^2+m^2\over k^+-p^+}} 
{d p^- \over p^2~~(k-p)^2}=
{1\over k^+p_t^2}\ln{m^2p^+\over p_t^2 k^+},~~~~~~
\int_{-\infty}^{-{(\vec{k_t}-\vec{p_t})^2\over k^+-p^+}} 
{d p^- \over [p^2]^2}= {1-z\over p^+p_t^2}~.
\label{eq:E4.7}\end{eqnarray}
Thus, we have obtained the first and the only logarithm of $p_t^2$ which is
scaled by a hadronic mass. This scale does not coincide with an artificial
renormalization scale $\mu^2$. This is a real emission process, which produces
large logarithms, not  virtual loops! The final answer becomes
\begin{eqnarray}
\bigg[{dG(p^+,p_t^2)\over d p_t^2}\bigg]^{HG}_{TT}
=-{g^4 N_c^2\over (2\pi)^6} ~{2\pi\over p_t^2} ~
\int {dk^+d{\vec k_t} \over 2 k^+}  
{dG(k^+,k_t^2)\over d k_t^2}\times\nonumber\\
\times\bigg\{ {\cal P}_{gg}(z)
\bigg[-{11\over 6}+2\ln{k^+-p^+\over\epsilon}\bigg]
\bigg[1+\ln{m^2 p^+\over p_t^2 k^+}\bigg] 
-{1\over 24}{(k^++p^+)^2\over k^+(k^+-p^+)}\bigg\}~.
\label{eq:E4.7a}\end{eqnarray}
Since we have isolated the mechanism responsible for the large $log$'s,  it is
natural to try to elevate its status. This does not need special effort,
since we have already started from the dressed form of the correlator, which
corresponds to the emission rung,
$~{\bf D}^{\#}_{ret}(k-p)\Pi^{\#}_{10}(k-p){\bf D}^{\#}_{adv}(k-p)$.~ 
This string is nothing but the imaginary part of the full propagator 
${\bf D}^{\#}_{ret}(k-p)$ in the kinematic region $k^+-p^+>0$. Indeed, since 
the polarization loop $\Pi_{01}^{\#}(q)$ vanishes at positive $q^+$, we can
write a chain of transformations, 
\begin{eqnarray}
-\big[{\bf D}_{ret}(p)\Pi_{10}(p){\bf D}_{adv}(p)\big]^{\mu\nu}
=-{\bf D}_{ret}(\Pi_{10}-\Pi_{01}){\bf D}_{adv}
={\bf D}_{ret}(\Pi_{ret}-\Pi_{adv}){\bf D}_{adv}=\nonumber \\
=({\bf D}_{ret}{\stackrel{\leftarrow}{D_{(0)}^{-1}}}-1){\bf D}_{adv}-
{\bf D}_{ret}({\stackrel{\rightarrow}{D_{(0)}^{-1}}}{\bf D}_{adv}-1)
={\bf D}_{ret}-{\bf D}_{adv}+{\bf D}_{ret}
({\stackrel{\leftarrow}{D_{(0)}^{-1}}}-
{\stackrel{\rightarrow}{D_{(0)}^{-1}}}){\bf D}_{adv}=\nonumber\\
={\bf D}_{ret}-{\bf D}_{adv}=2i~{\rm Im}~{\bf D}_{ret}
=2i~{\rm Im}\bigg[{\bar{d^{\mu\nu}}(p)\over p^2(1-w_{1}^{ret}(p))}
+{n^\mu n^\nu \over (p^+)^2(1-w_{2}^{ret}(p))}\bigg]~.
\label{eq:E4.8}\end{eqnarray} 
Substituting the last form of this string for the final-state
density of the radiated glue, we arrive at
\begin{eqnarray}
\bigg[{dG(p^+,p_t^2)\over d p_t^2}\bigg]^{HG}_{TT}
={g^4 N_c^2\over (2\pi)^6} 
\int {dk^+d{\vec k_t} \over 2 k^+}~{dG(k^+,k_t^2)\over d k_t^2}
\int_{-\infty}^{-{(\vec{k_t}-\vec{p_t})^2+m^2\over k^+-p^+}} d p^- 
                           \times \nonumber \\
\times\bigg\{ 8 {\cal P}_{gg}(z)
\bigg[{1\over p^2}+{z\over 1-z}{(p-k)^2\over [p^2]^{2}}\bigg]
{\rm Im}{1\over (k-p)^2(1-w_{1}^{ret}(k-p))} + 
2z\bigg({1+z \over 1-z}\bigg)^2 {1 \over [p^2]^{2}}
{\rm Im} {1 \over (1-w_{2}^{ret}(k-p))}\bigg\}~.
\label{eq:E4.9}\end{eqnarray}
To compute the integrals $dp^-$ in this equation we may use the Schwartz 
symmetry principle. Indeed,  basic analytic properties of the retarded and
advanced gluon self-energies in the plane of complex $p^-$   stimulate similar
analytic properties of the gluon propagator. Therefore, the real integration
can be  replaced by the complex integration along the path ${\cal C}$, which 
envelopes  the cut in the complex plane of the function
\begin{eqnarray}
\tilde{w}_{j}(k-p)=
\left\{ \begin{array}{l} w_{j}^{ret}(k-p),~~~~ {\rm Im}~p^-<0, \\
w_{j}^{adv}(k-p),~~~~ {\rm Im}~p^->0 \end{array}\right.
\label{eq:E4.10}\end{eqnarray}
and is closed by a big circle at the infinity. Then the integral is calculated
as the residue in the pole of the integrand. For example,
\begin{eqnarray}
\int_{-\infty}^{-{(\vec{k_t}-\vec{p_t})^2\over k^+-p^+}} 
{d p^- \over p^+p^--p_t^2} ~~{\rm Im}{1\over (k-p)^2(1-w_{1}^{ret}(k-p))}=
\int_{{\cal C}}{d p^- \over p^+p^--p_t^2}~~
{1\over (k-p)^2(1-\tilde{w}_{1}(k-p))}=\nonumber\\
=-{\pi\over k^+p_t^2}~~{1\over 1-{\rm Re}w_{1}^{ret}(k-p;~p^-=p_t^2/p^+))}~.
\label{eq:E4.11}\end{eqnarray}
Two properties of the integrand should be noted. First, the pole at the point
$(k-p)^2=0$ (corresponding to the free propagation) is obviously located at
the second sheet and does not contribute to the integral (this pole has even
to be read with the principal value assignment). Second,  at the pole
~$p^-=p_t^2/p^+$~, the function $\tilde{w}(k-p)$ is real. The other two
integrals are computed in the same way, the only difference being, the pole at
$p^2=0$ is of the second order.  In the simplest case, when the final-state
jet is simulated via a two-gluon system, the functions 
${\rm Re}w_{j}^{ret}(k-p)$, which enter the final answer are,
\begin{eqnarray}
{\rm Re} ~w_{j}^{\#ret}(k-p) 
={g^2N_c\over 16\pi^2}  
~\beta_{j}(k^+-p^+,\epsilon)~\ln {\mu^2\over -(k-p)^2}~.
\label{eq:E4.12}\end{eqnarray}
We have to substitute the value of the argument in the pole, 
$~-(k-p)^2=p_t^2/z$, (in the approximation of strong ordering, $p_t\gg k_t$)
into Eq.~(\ref{eq:E4.12}).
Eventually, we express the result in terms of two effective couplings:
the transverse coupling $\alpha_T(p_t^2/z)$ due to the transverse mode of the
emitted heavy gluon, and the longitudinal coupling  $\alpha_L(p_t^2/z)$, due
to the longitudinal mode of the emitted gluon, {\em i.e.}
\begin{eqnarray} 
\bigg[{dG(p^+,p_t^2)\over d p_t^2}\bigg]^{HG}_{TT}
={N_c\over (2\pi)^4} ~{8\pi\over p_t^2}
\int {dk^+d{\vec k_t} \over 2 k^+}
{dG(k^+,k_t^2)\over d k_t^2}
\big\{  {\cal P}_{gg}(z)~(\alpha_T({p_t^2\over z}) -
\beta_T~\alpha_T^2({p_t^2\over z})+ 
{(1+z)^2 \over 1-z}~\beta_L~\alpha_L^2({p_t^2\over z})\big\}~,
\label{eq:E4.13}\end{eqnarray} 
where,
\begin{eqnarray}
\beta_T={g^2\over 16\pi^2}\beta_1(k^+-p^+,\epsilon),~~~~
\beta_L={g^2\over 16\pi^2}\beta_2(k^+-p^+,\epsilon),\nonumber\\
\alpha_T={g^2/4\pi\over 1+\beta_T\ln{p_t^2 k^+\over \mu^2 p^+}},~~~~
\alpha_L={g^2/4\pi\over 1+\beta_L\ln{p_t^2 k^+\over \mu^2 p^+}}~.
\label{eq:E4.14}\end{eqnarray} 
The running coupling with the same argument, $\alpha(p_t^2/z)$, was obtained by
Dokshitzer {\em et al.} \cite{DMW} by means of the  ``dispersive method'' based
on an {\em ad hoc} dispersion formula for the effective running coupling. Our
construction (Eqs.~(\ref{eq:E4.8}) -- (\ref{eq:E4.14}) ) provides a formal
proof to this approach.  In general, we should not  rely on a simple two-body
model for the density of final states used in these equations. Instead, as  is
conceived in  Ref.\cite{DMW},  we can consider  real density of the final state
jets  which are not yet calculable from a theory, but can be extracted from the
data. In our picture, with the proton (or nuclei) kept intact before the
collision we have to renormalize the gluon self-energy $\Pi^{\#}(p)$ at the
point $p^-=0$. Thus, we have to set $\mu^2=p_t^2$ in Eqs.~(\ref{eq:E4.13}) and
(\ref{eq:E4.14}). Then the $p_t$-dependence of the effective coupling may only
come from the the collinear cut-off, $\epsilon$, which accounts for the
properties of the states of emission.

\section{Radiative corrections in the LT-mode}
\label{sec:SNE5}

There are no major differences in the calculations for this mode compared to
the TT-mode,  since we only change a projector.  Therefore, we shall omit
details in the analysis below. The dispersive part of the virtual vertex before
the final $q^+$-integration is
\begin{eqnarray}
\big[{\cal G}(p^+,p_t^2)\big]^{VV,(D)}_{LT}
={g^4 N_c^2\over (2\pi)^5}
\int {dk^+d{\vec k_t} \over 2 k^+} {dG(k^+,k_t^2)\over d k_t^2} 
\bigg\{\int_{0}^{p^+} 
{ \psi(q^+,k^+,p^+)dq^+ \over 
k^+p^+(k^+-p^+)^2(k^+-q^+)(p^+-q^+)}\times\nonumber \\
\times\bigg[\ln\bigg(1+{k^+(p^+-q^+)\over p^+(k^+-q^+)}
\big(-1+{(k^+-p^+)\mu^2\over k^+ p_t^2}\big)\bigg)
-{k^+(p^+-q^+)\over p^+(k^+-q^+)}
\ln {(k^+-p^+)\mu^2\over k^+ p_t^2}\bigg]+\nonumber \\
+\int_{p^+}^{k^+}{ \psi(q^+,k^+,p^+)dq^+ \over
k^+p^+(k^+-p^+)^2(k^+-q^+)(q^+-p^+)}
\ln\bigg(1+{k^+(q^+-p+)\over p^+(k^+-q^+)}
\big(-1+{(k^+-p^+)\mu^2\over k^+ p_t^2}\big)\bigg) \bigg\}~.
\label{eq:E5.1}\end{eqnarray} 
The UV-renormalization is performed by means of one subtraction at the point
$p^-=-\Omega$. The first and the second integrals are obtained via the
dispersion relation from the imaginary parts corresponding to the unitary cuts
of the vertex diagram  near the external lines with  momenta $p$ and $(k-p)$,
respectively.  Only the first term of the integrand is significant for the
final answer. The reason is that the contribution from the cut vertex diagram
corresponding to the real two-gluon emission, although divergent,  exactly
coincides, but with an opposite sign, with the second term in  
Eq.~(\ref{eq:E5.1}) before the renormalization of the latter by means of
subtraction in the dispersion relation. Since the subtracted counter-term by
its definition is quasi-local, the net yield of  these two terms will be a
quasi-local counter-term only. This term is independent of $p^-$ in  momentum
space and proportional to $\delta(x^+)$ in  coordinate space. Hence, it must
be discarded.  With our previous renormalization on the light cone,
$\mu^2=p_t^2$, the answer reads as
\begin{eqnarray}
\big[{\cal G}(p^+,p_t^2))\big]^{VV,(D)}_{LT}
={g^4 N_c^2\over (2\pi)^5}
\int {dk^+d{\vec k_t} \over 2 k^+} {dG(k^+,k_t^2)\over d k_t^2} 
\bigg\{  {(2-z)(132-132z+35z^2)\over 18 (1-z)^2} -
{7(2-z)^2\over 3z(1-z)}\ln(1-z)-\nonumber \\
-{(2-z)(16-14z+z^3)\over 2 z^2(1-z)}\ln^2(1-z)-
{(2-z)^2(2-2z+z^2)\over  z(1-z)^2} {\rm Li}_2(z) \bigg\}~.
\label{eq:E5.2}\end{eqnarray}
Unlike the TT-mode, the non-dispersive part of the vertex
correction is not quasi-local. After renormalization, it looks like
\begin{eqnarray}
\big[{\cal G}(p^+,p_t^2))\big]^{VV,(ND)}_{LT}
={g^4 N_c^2\over (2\pi)^5}
\int {dk^+d{\vec k_t} \over 2 k^+} {dG(k^+,k_t^2)\over d k_t^2} 
~\int_{0}^{p^+} dq^+
 ~\psi_L(q^+,k^+,p^+)~ \ln {(1-z)\mu^2\over p_t^2}~.
\label{eq:E5.3}\end{eqnarray} 
With the same choice  of the renormalization point at $\mu^2=p_t^2$ we obtain
\begin{eqnarray}
\big[{\cal G}(p^+,p_t^2))\big]^{VV,(ND)}_{LT}
={g^4 N_c^2\over (2\pi)^5}
\int {dk^+d{\vec k_t} \over 2 k^+} {dG(k^+,k_t^2)\over d k_t^2} 
\bigg\{  {(2-z)(8-8z+z^2)\over z^2 (1-z)} \ln^2(1-z)
-4{(2-z)^2\over 2 z(1-z)}\ln(1-z)\bigg\}~.
\label{eq:E5.4}\end{eqnarray}
Finally, we have a correction due to the retarded self-energies ${\bf
D}_{{ret\choose adv}}(p)$. Before specifying the subtraction point,
it is given by,
\begin{eqnarray}
\big[{\cal G}(p^+,p_t^2))\big]^{SE}_{LT}
={g^4 N_c^2\over 2(2\pi)^5} \int {dk^+d{\vec k_t} \over 2 k^+} 
{dG(k^+,k_t^2)\over d k_t^2}{(2-z)^2\over z(1-z)}\beta_2(p^+,\epsilon)
\ln\bigg[{\mu^2(1-z)\over p_t^2} \bigg]~.
\label{eq:E5.5}\end{eqnarray} 
Since the renormalization of the self-energy of the propagator is subjected
to the same condition as the vertex function, we have to take $\mu^2=p_t^2$
and thus obtain no large $\ln p_t^2$. Thus, as in the case of the TT-mode 
we can perform the UV-renormalization in such a way that no large $\ln p_t^2$ 
arises.

Real processes have $log$'s of $p_t^2$. One of them, the cut vertex,
is gone. The two remaining diagrams correspond to the emission of a heavy
gluon and the interference between two sequential gluon emissions. 

Let us begin with the heavy gluon. As previously, we start with the lowest
order contribution (with the bare propagator in the intermediate state),
\begin{eqnarray}
\big[{\cal G}(p^+,p_t^2))\big]^{HG}_{LT}
={ig^2 N_c\over (2\pi)^4}  \int {dk^+d{\vec k_t} \over 2 k^+} 
{dG(k^+,k_t^2)\over d k_t^2}~{2(2-z)^2\over z}
\int_{-\infty}^{-{(\vec{k_t}-\vec{p_t})^2+m^2\over k^+-p^+}} d p^- 
 ~{w_{1}^{\# 01}(k-p)\over (k-p)^2} ~.
\label{eq:E5.6}\end{eqnarray}
Since the density of states $w_1^{\# 01}(k-p)$  only establishes the limit
of the integration $dp^-$ (being constant above the threshold), the integral
$dp^-$ requires two cut-offs, the minimal mass $m$ of a gluon jet,
and the maximal mass, $M$, resulting in
\begin{eqnarray}
\big[{\cal G}(p^+,p_t^2))\big]^{HG}_{LT}
=-{g^4 N_c^2\over 2(2\pi)^5} \int {dk^+d{\vec k_t} \over 2 k^+} 
{dG(k^+,k_t^2)\over d k_t^2}~{(2-z)^2\over z(1-z)}~\beta_1(k^+-p^+,\epsilon)
\ln\bigg[{M^2\over m^2} \bigg]~.
\label{eq:E5.8}\end{eqnarray}
No dependence on $p_t^2$ appears here, just a big number connected with the 
end-points of the mass spectrum of the final-state jets. This has to be
compared with the case of the TT-mode (Eqs. (\ref{eq:E4.6}) and
(\ref{eq:E4.7}) ). The difference is easily understood, since the propagator
of the longitudinal field has no causal pole. The next logical step is to
consider the same process with the dressed propagator of the final-state gluon
field, as has been done for the case of the TT- transition mode. This makes
the integral  converge; the cut-off $M$ is unnecessary, and we find,
\begin{eqnarray}
\big[{\cal G}(p^+,p_t^2))\big]^{HG}_{LT}
={ig^2 N_c\over (2\pi)^4} \int {dk^+d{\vec k_t} \over 2 k^+}   
{dG(k^+,k_t^2)\over d k_t^2}~{2(2-z)^2\over z} \times \nonumber\\
\times \int_{-\infty}^{-{(\vec{k_t}-\vec{p_t})^2+m^2\over k^+-p^+}} d p^- 
~{w_{1}^{\# 01}(k-p)
\over (k-p)^2(1-w_{1}^{ret}(k-p))(1-w_{1}^{adv}(k-p))} ~.
\label{eq:E5.9}\end{eqnarray}
In our model, the propagator of the final-state time-like gluon has no 
resonance poles and the integrand of the equation (\ref{eq:E5.9}) lacks the
propagating pole at $p^2=0$. Thus we cannot continue with  the
contour  integration and have to proceed in a straightforward way with the
simple model of the two-body final state. Then, the integral in the previous
equation is calculated as follows,
\begin{eqnarray}
\big[{\cal G}(p^+,p_t^2))\big]^{HG}_{LT}
=-{g^2 N_c\over (2\pi)^3}  \int {dk^+ d{\vec k_t} \over 2 k^+}   
{dG(k^+,k_t^2)\over d k_t^2}~{2(2-z)^2\over z(1-z)} 
{{\rm arctan} \pi \beta_T \over  \pi \beta_T} ~.
\label{eq:E5.10}\end{eqnarray}

The pole at $z=1$, which was a subject of concern in the order $g^2$, is
persistently reproduced in all radiative corrections in the order $g^4$, and
it requires the physical screening.

\section{Screening effects in the QCD evolution}
\label{sec:SS1}

We have shown above that the UV-renormalization of the evolution equations can
be performed in such a way that it does not bring  any scale into the problem.
Therefore, the only scale which is unambiguously present in the equations of
the QCD evolution is the one associated with the properties of the final
states that may regulate mass divergences. In the evolution picture based on
null-plane dynamics, this scale systematically enters via the collinear
cut-off $~\epsilon~$.  Now, we want to find an explicit expression for this
cut-off in terms of the parameters of the collective modes of the final state.
In the simplest case this will be the mass of the transverse plasmon. 

It turns out that the null-plane dynamics, which employs the light-like
direction $x^+$ as the Hamiltonian time, is ill-suited for computing the
screening effects.  In null-plane dynamics, the screening is just absent 
because the geometry of fields of these dynamics is singular and cannot
provide a mass to a plasmon. In what follows, we prove this statement and 
suggest an alternative type of dynamics, which allows one to derive the {\em
evolution equations with screening}. 

To compute the screening effects, one has to rely on some {\em distribution}
of the excited modes. One can consistently introduce the distribution only in
connection with the procedure of its measurement.  Throughout this paper, we
concentrat on  the inclusively measured one-particle distribution and argue
that the full final state of the collision is always prepared as a fluctuation
before the moment of the inclusive measurement (and that the measurement only
excites the system of the allowed final states.) Below,  supporting this idea,
we  construct the basis of states that naturally complies with the Lorentz
contraction of the incoming objects and includes both strongly and weakly
localized field modes. In terms of this basis, a snapshot of a virtual
fluctuation before a collision (with as of, yet, coherently balanced phases)
already looks like a distribution of hard particles (with  high $p_t$) moving
through the slowly-varying fields (with low  $p_t$). The interaction just
freezes this fluctuation with the wrong phases  and converts it into the true
distribution, in which the slow fields are immediately screened by the
interaction with the hard particles. Thus, even at a very early time after a
collision of two nuclei, the emerging final state is rather  a system of
plasmons, than an ensemble of free partons.

\subsection{The absence of screening in null-plane dynamics}
\label{sbsec:SBS1a}

The QCD evolution in high-energy processes is always described in the
null--plane dynamics, which is the only one where the  standard variable of
evolution equations, the Feynman $x$, is unambiguously defined.  It is tempting
to try to compute the screening effects staying within framework of these
dynamics. The procedure seems to be straightforward. One has to compute the
retarded propagator (\ref{eq:E1.8}) for the field of emission,
\begin{eqnarray}
{\bf D}^{\mu\nu}_{ret}(p)=
{{\overline d}^{\mu\nu}(p)\over p^2 + p^2 w_{1}^{R}(p)}+
{1 \over (p^+)^2}~ {n^{\mu}n^{\nu} 
\over  1 - w_{2}^{R}(p)}~, \nonumber
\end{eqnarray} 
accounting for the effects of the final-state interaction (collective nature of
the final states) in the invariants $p^2 w_{1}^{R}(p)$ and  $w_{2}^{R}(p)$ of
the gluon self-energy.  In this formula, the first and the second terms are the
full propagators of  the  transverse and  longitudinal fields {\em of the
null-plane dynamics}, respectively.  The self-energy $p^2 w_{1}^{R}(p)$ of  the
transverse field has a dispersive part and non-dispersive part. The former is
given by the Eq.~(\ref{eq:E2.14}), and it always is zero at $p^2=0$.
Mathematically, this zero appears in the imaginary part of
$\Pi^{(D)}_{ret}(p)$, and survives through the calculation of the dispersion
integral (\ref{eq:E2.17}). Physically, this happens because the propagation of
the field in the dispersive term of  the self-energy  is mediated by the
transverse propagator $D^{(T)}_{ret}$, which maintains the causality of the
$\Pi^{(D)}_{ret}$. This is true for all physical gauges, which  explicitly
separate the transverse and  longitudinal fields.

In the non-dispersive part (\ref{eq:E2.15}) of $\Pi_{ret}(p)$, the field
propagation is mediated by the longitudinal field, which results in
Eq.~(\ref{eq:E2.20}) for  $p^2 w_{1}^{(ND)}(p) $. In the case, when the
states are populated with  density $n(k)=n(k^+,k_t)$
this equation is modified to,
\begin{eqnarray}
{\rm Re}~[p^2 w_{1}^{(ND)}(p)] 
={-g^2 N_c\over 2(2\pi)^3}  \int dq^+ ~d^2\vec{q_t}
\bigg\{ {1+2n(q) \over |q^+|}~
\bigg({q^+ + p^+\over q^+ - p^+}\bigg)^2 +
{1+2n(q-p) \over |q^+-p^+|}~
\bigg({q^+ - 2 p^+\over q^+ }\bigg)^2 \bigg\}~.
\label{eq:S1.1}\end{eqnarray}
This expression has no zero at $p^2=0$ and seems to be a good
candidate to form the mass term in the dispersion equation for the transverse
field. However, as before,  this term is quasi-local in $x^+$ (it is
independent of the light-cone energy $p^-$) and vanishes after one subtraction
at any value of $p^-$. 

The contact interaction in the tadpole part of the gluon self-energy behaves in
the same way, and results in the $p^-$-independent quasi-local term.

Thus, we can draw two conclusions. (i) The mass of a plasmon with momentum $p$
is generated by the  amplitude  of the forward scattering of a gluon on some
distribution of states only, if the gluon field between the two successive
interactions is longitudinal, or if the interaction is contact.  (ii) In the
null-plane dynamics,  these interactions are local in $x^+$.  Therefore, they
are incapable of producing any screening effects, which are genuinely  {\em
non-local}.  In order to incorporate screening effects into the evolution
equations, we have to exchange the null-plane Hamiltonian dynamics for another
dynamics in which the  fields would behave in a less singular way.

 \subsection{Field states in the proper-time dynamics}
\label{sbsec:SBS1b}

It was explicitly demonstrated that it is impossible to adequately describe 
the basic process of  forward scattering, which is responsible for the
generation of the plasmon mass, in the null-plane dynamics. The problem arises
due to the singular behavior of the field pattern which is defined as the
static field with respect to the time $x^+$. This singular behavior shows that
the  choice of the  Hamiltonian dynamics and  proper definition of the  field
states is highly  nontrivial and important. The problem we encounter in
connection with screening is even more general.  Indeed, a search for the QGP
in heavy-ion collisions is, in the first place, a search for evidence of 
entropy production.  Before the collision, the quark and gluon fields are
assembled into  two coherent wave packets (the nuclei), and therefore, the
initial entropy equals zero. The coherence is lost, and  entropy is created due
to the interaction. Though one may wish to rely on the invariant formula 
$S={\rm Tr}\rho\ln\rho$,  which expresses the entropy $S$ via the density 
matrix $\rho$, at least one basis of states should be found explicitly. Thus,
it is imperative to find a  way to describe quarks and gluons of both nuclei 
as well as the products of their interaction using {\it the same Hamiltonian
dynamics}.   An appropriate choice for the gluons is always difficult because
the  gauge  is a global object (as are the Hamiltonian dynamics) and both
nuclei should be described using the same  gauge condition.

Quantum field theory has a strict definition of  {\it dynamics}. This
notion was introduced by Dirac  \cite{Dirac}  at the end of the 1940's in
connection with his attempt to build a quantum theory of the gravitational
field. Every (Hamiltonian) dynamics includes its specific definition of
the quantum mechanical observables on the (arbitrary) space-like surfaces,
as well as the means to describe the evolution of the observables  from
the ``earlier'' space-like surface to the ``later'' one. 

The primary choice of the degrees of freedom  is effective if, even without any
interaction, the dynamics of the normal modes adequately  reflects the main
physical features of the phenomenon. The intuitive physical picture  clearly
indicates that the normal modes  of the fields participating in the collision
of the two nuclei should be compatible with their Lorentz contraction. Unlike
the incoming plane waves of the standard scattering theory, the nuclei have a
well-defined shape and the space-time domain of their intersection is also
well-defined. Of the ten symmetries of the Poincar\'{e} group, only  rotation
around the  collision $z$-axis, boost along it,  and the translations and 
boosts in the transverse $x$ and $y$-directions  survive. The idea of the
collision  of two plane sheets immediately leads us to  the {\em wedge form}; 
the states of quark and gluon fields before  and after the collision
must be confined within the past and future  light cones (wedges) with  the
$xy$-collision plane as the edge. Therefore, it is profitable to choose, in
advance, the set of normal modes which have the symmetry of the localized
interaction and carry quantum numbers adequate to this symmetry. These
quantum numbers are the transverse components of momentum and the  rapidity of
the particle (which replaces the component $p^z$ of its momentum). In this {\it
ad hoc} approach, all the spectral components of the nuclear  wave functions
ought to collapse in the  two-dimensional plane of the interaction,  even if
all the confining interactions of the quarks and gluons in the hadrons  and the
coherence of the hadronic wave functions are neglected. 

In the wedge form of dynamics, the states of  free quark and gluon fields are
defined (normalized) on the space-like hyper-surfaces of the constant proper
time $\tau$, $\tau^2=t^2-z^2$. The main idea of this approach is to study the
dynamical evolution of the interacting fields along the Hamiltonian time
$\tau$. The gauge of the gluon field is fixed by the condition $A^\tau =0$. 
This simple idea solves several problems. On the one hand,   it becomes
possible to treat the two different light-front dynamics  which describe  each
nucleus of the initial state separately, as  two limits of this single
dynamics.  On the other hand, after the collision, this  gauge simulates a 
local (in rapidity) temporal-axial gauge. This feature provides a smooth
transition to the boost-invariant regime of the  created matter expansion (as
a first approximation).  Particularly, addressing the problem of screening,
we will be able to compute the plasmon mass in a uniform fashion, considering
each rapidity interval separately and using the (local) temporal axial gauge.

The feature of the states to collapse at the interaction vertex  is
crucial for understanding the dynamics of the collision. A simple
optical prototype of the wedge dynamics is the {\em camera obscura} (a
dark chamber with the pin-hole in the wall). Amongst the many possible {\em a
priori} ways to decompose the incoming light, the camera selects  only
one. Only  the spherical harmonics centered at the pin-hole can
penetrate  inside the camera.  The spherical waves reveal their angular
dependence  at some distance from the center and build up the image on the
opposite wall. Here, we suggest to view the collision of two nuclei as a
kind of diffraction of the initial wave functions through the ``pin-hole''
$t=0,~z=0$ in $tz$-plane.

The states of the wedge dynamics appear to be almost ideally suited for the
analysis of the processes that are localized at different times $\tau$ and
intervals of rapidity $\eta$ and are characterized by a different transverse
momentum transfer.  With respect to any particular process, these states are 
easily divided into  slowly varying fields and localized particles. In this
way, one may introduce the distribution of particles and study their effect on
the dynamics of the fields.  Thus, we now can calculate  the plasmon mass
as a local (at some scale) effect which agrees with our understanding of its
physical origin.

For the simplest  qualitative  estimates, it is enough to consider the
one-particle wave functions  of the scalar field. Let us rewrite the wave
function $\psi_{\theta,p_\perp}(x)$ in the following form,
\begin{eqnarray}
\psi_{\theta,p_\perp}(x)= {1 \over 4\pi^{3/2}} 
e^{-ip^0t+ip^zz+i{\vec p}_{\perp}{\vec r}_{\perp}}\equiv
\left\{ \begin{array}{l}  4^{-1}\pi^{-3/2} e^{-im_{\perp}\tau\cosh(\eta-\theta)}
e^{i{\vec p}_{\perp}{\vec r}_{\perp}},~~~~\tau^2>0, \\
 4^{-1}\pi^{-3/2}  e^{-im_{\perp}\tau\sinh(\eta-\theta)}
e^{i{\vec p}_{\perp}{\vec r}_{\perp}},~~~~ \tau^2<0~. \end{array}\right.
\label{eq:S1.2}\end{eqnarray} 
where $p^0=m_{\perp}\cosh\theta$, $p^z=m_{\perp}\sinh\theta$ ($\theta$ being 
the rapidity of the particle), and, as usual,
$m_{\perp}^{2}=p_{\perp}^{2}+m^2$. The above form implies that $\tau$ is
positive in the future of the wedge vertex  and negative in its past. Even
though this wave function is obviously a plane wave which occupies the whole
space, it carries the  quantum number $\theta$ (rapidity of the particle)
instead of the momentum $p_z$. A peculiar property of this wave function is
that it may be normalized in two different ways, either on the hypersurface
where $t=const$,
\begin{eqnarray}
\int_{t=const}\psi^{\ast}_{\theta',p'_\perp}(x)
~i{\stackrel{\leftrightarrow}{\partial \over \partial t}}~ 
\psi_{\theta,p_\perp}(x)=
\delta (\theta -\theta')\delta ({\vec p}_{\bot}-{\vec p'}_{\bot})~~,
\label{eq:S1.2a}\end{eqnarray}   
or, equivalently, on the hypersurfaces $\tau=const$ in the 
future- and the past--light wedges of the collision plane, where $\tau^2>0$,
\begin{eqnarray}
\int_{\tau=const}\psi^{\ast}_{\theta',p'_\perp}(x)
~i{\stackrel{\leftrightarrow}{\partial \over \partial \tau}}~ 
\psi_{\theta,p_\perp}(x)=
\delta (\theta -\theta')\delta ({\vec p}_{\bot}-{\vec p'}_{\bot})~.
\label{eq:S1.2b}\end{eqnarray} 
  
The norm of a particle's wave function always corresponds to the conservation
of its charge or  probability to find it. Since the norm given by
Eq.~(\ref{eq:S1.2b}) does not depend on $\tau$, the particle with a given
rapidity $\theta$ (or velocity $v=\tanh\theta=p^z/p^0$), which is
``prepared''  on the surface  $\tau=const$ in the past light wedge, cannot
flow through the light-like wedge boundaries; the particle is
predetermined to penetrate in the future light wedge through its vertex. The
dynamics of the penetration process can be understood in the following way.

At large $~m_{\perp}|\tau|$, the phase of the wave function 
$\psi_{\theta,p_\perp}$ is stationary in a very narrow interval  around
$~\eta=\theta~$ (outside this interval, the function oscillates with
exponentially increasing  frequency); the wave function describes a particle
with rapidity $~\theta$ moving along the classical trajectory.  However, for
$m_{\perp}|\tau|\ll 1$, the phase  is almost constant along the surface
$\tau=const$.  The smaller $\tau~$ is, the more uniformly the domain of 
stationary  phase is stretched out along the light cone. A single particle
with the wave function $\psi_{\theta,p_\perp}$ begins its life as the wave 
with the given rapidity $\theta$ at large negative $\tau$. Later, it becomes
spread out over the boundary of the past light wedge as $\tau\to -0$. Still
being spread, it appears on the boundary of the future light wedge.
Eventually, it again becomes a wave  with  rapidity $\theta$ at large positive
$\tau$.  The size and location of the interval where the phase of the wave
function is stationary plays a central role in all subsequent discussions,
since it is equivalent to the  localization of a particle. Indeed, the
overlapping of the domains of stationary phases in space and time provides the
most  effective interaction of the fields.

The size $\Delta\eta$ of the $\eta$-interval  around the particle rapidity
$\theta$, where the wave function is  stationary, is easily evaluated.
Extracting  the trivial factor $e^{-im_{\perp}\tau}$ which defines the
evolution of the wave  function in the $\tau$-direction, we obtain an
estimate from the exponential of  Eq.~(\ref{eq:S1.2}),
\begin{eqnarray}
 2~m_{\perp}\tau\sinh^2(\Delta\eta/2)\sim 1~~. 
\label{eq:S1.3}\end{eqnarray}      
The two limiting cases are as follows, 
 \begin{eqnarray}
 \delta\eta \sim \sqrt{2\over m_{\perp}\tau},
~~{\rm when}~~ m_{\perp}\tau \gg 1~,~~~{\rm and}~~~
  \Delta\eta \sim \ln {2\over m_{\perp}\tau},
~~{\rm when}~~ m_{\perp}\tau\ll 1~~.
\label{eq:S1.4}\end{eqnarray} 
In the first case, one may boost this interval into the laboratory
reference frame and see that the interval of  stationary phase is Lorentz 
contracted (according to the rapidity $\theta$) in $z$-direction. 

The modes normalized according to (\ref{eq:S1.2b}) play the same role as the
wave functions of bound states in the examples described at the end of
Sec.~\ref{sec:SNo}. Indeed, the packets of the waves with the same rapidity
$\theta$ do not experience dispersion in the longitudinal direction. Therefore,
an expansion in terms of such waves  can be employed to form  moving  objects. 
The amplitudes and phases of the coefficients in this expansion are
balanced in such a  way that, before the collision, the wave packets represent
the finite-sized nuclei with given rapidities.   These wave packets must at
least partially diagonalize  the Hamiltonian  that includes  interactions 
which maintain the shapes of the nuclei.

If two localized objects simultaneously pass through the vertex, then the
partial waves that form their wave functions effectively overlap in the
vicinity of the light wedge. The interval of time  when the partial wave with
transverse momentum $p_\bot$ is spread out, is of the order $\tau\sim
1/m_\bot$.  The high-$p_\bot$ components of the wave function localize around
the world line of the initial rapidity $\eta=\theta$ earlier,  than the
low-$p_\bot$ components and have less time to interact.  If no interaction 
occurs at sufficiently small $\tau$, then the amplitudes and the phases of
these waves remain unchanged and the packet assembles into the initial
finite-sized object.  

If the interaction takes place and the balance of phases becomes broken, then 
the estimates of Eqs.~(\ref{eq:S1.3}) and (\ref{eq:S1.4}) show how to
construct the picture of screening  at sufficiently small $\tau$, even before
the secondary collisions come into the game. The states with $p_t\ll\tau^{-1}$
vary only slowly at the rapidity interval $\Delta\eta\sim\ln (2/p_t\tau)\gg
1$, and cannot be considered to be particles. At the same time, the states
with $k_t\gg\tau^{-1}$  are well localized in the rapidity direction. For
these states,  $\delta\eta \sim \sqrt{2/k_t\tau}\ll 1$ and  we may safely view
them as the particles with the distribution which can be measured inclusively 
at the time $\tau$. The forward scattering of the particles with transverse
momentum $k_t$ on the field with the momentum $p_t\ll k_t$ results in an
adjoint mass of the soft field which screens the possible collinear
singularity in the evolution equations. The process  of evolution naturally
saturates at the proper time $\tau$ when the adjoint mass becomes comparable
to the  transverse momentum $p_t\sim 1/\tau$. This is the end of the
``earliest stage''.  At later times, collisions between the partons-plasmons
take over.

Another important question is how early the ``earliest stage''  begins.  It is
easy to understand that the  shortest time scale that may be seriously
considered in the theory is defined by the full energy of collision,
$\tau_{min}\sim 1/\sqrt{s}$.  (For nuclei collisions, $\tau_{min}\sim
1/A^{1/3}\sqrt{s}$ where $s$ is the invariant mass per  participating nucleon.)
At this time, the stationary phase of a partial wave is stretched over the
widest rapidity interval $\delta\eta\sim\ln(\sqrt{s}/m_{\perp})$. It is not
surprising that this estimate coincides with the well known kinematically
allowed width $2Y$ of the rapidity plateau,  $2Y\approx log(s/m^2_{char})$. We
just have two complementary ways to obtain the same quantity. 

The entire design of pQCD is aimed at getting rid of logarithms like
$log(s/m^2)$; they lead to mass singularities. Their presence invalidates the
power counting in  the calculation of the scattering amplitudes and makes the
dimensional analysis (including the RG methods) impossible.   On the contrary,
in AA--collisions, this logarithms play a major role defining the energy
density in the emerging dense system and, consequently, the screening effects.
However, even in this case, we may find subprocesses (with the largest
$p_t$--transfer), which are insensitive to the collective effects, and where
the two strategies (screening and removing) must be not too far away from each
other.

A full analysis of the QCD evolution and the collective effects in the scope of
the wedge dynamics is a subject of separate study which is now underway. To
treat these problems in a global fashion, one has to derive propagators of
perturbation theory in curvilinear coordinates, separate longitudinal and
transverse gauge fields,  and perform quantization of all fields
\cite{WD1,WDG}.  Here we shall address only the {\em estimate} of two most
important effects:  

(i) The plasmon mass, which may be computed with reference to the local
properties of the wedge dynamics. 

(ii) Screening in the evolution equations at large $p_t$, where the
difference between the wedge and the null-plane dynamics is expected to be
minimal. 

\subsection{Local screening and the mass of a plasmon}
\label{sbsec:SBS1c}

The local parameterization of the dynamics based on the system of the
space-like surfaces $\tau=const$ is easily implemented with the aid of the
vector $u^\mu(\eta)=(\cosh\eta, \vec{0_t}, \sinh\eta)$, which is normal to the
hyper-surface and indicates the local ``time'' direction. As long as we are
interested in the process, which takes place in a limited slice in rapidity, we
may consider $\eta$ as a label of this slice and perform all calculations in
the Lorentz frame which moves with the rapidity $\eta$. In this frame,
$u^\mu_\ast=(1,0,0,0)$, and the global gauge condition, $~u(\eta)A=A^\tau =0~$,
becomes just $A^0_\ast=0$. A dynamical formation of the plasmon mass is exactly
the case when we can proceed in this simplified manner.

The bare propagator in this (local) temporal-axial gauge is of the following
form,
\begin{eqnarray}
 D^{\mu\nu} (u,k)= {1\over k^2} \bigg(-{\rm g}^{\mu\nu} +
{k^\mu u^{\nu}+k^\nu u^{\mu}\over  (k u) }-{k^\mu k^{\nu} \over (k u)^2}\bigg)
={d_{(T)}^{\mu\nu}(k)\over k^2 }+{d_{(L)}^{\mu\nu}(k)\over (ku)^2 }~,
\label{eq:S1.5} 
\end{eqnarray} 
where the first and the second terms in this equation are the propagators
of the transverse and longitudinal fields, respectively. The 
polarization matrices are,    
\begin{eqnarray}
 d_{(T)}^{\mu\nu}(k)= -{\rm g}^{\mu\nu} +u^\mu u^{\nu}-d_{(L)}^{\mu\nu}(k),~~~
d_{(L)}^{\mu\nu}(k)= 
{[k^\mu -u^{\mu}(ku][k^\nu -u^{\nu}(ku)] \over  (k u)^2 - k^2 },
\label{eq:S1.6} 
\end{eqnarray} 
and the gluon self-energy $\Pi^{\mu\nu}$ is  of the following form,
\begin{equation}
\Pi^{\mu\nu}(p) = d_{(T)}^{\mu\nu}(p)~w_T(p) +
{[p^2 u^\mu -(pu)p^\mu][p^2 u^\nu -(pu)p^\nu]\over p^2~[(pu)^2-p^2]} w_L(p).
\label{eq:S1.7}\end{equation}  
Since $\Pi^{\mu\nu}$ always appears in an assembly $D\Pi D$  the terms, 
like $p^{\mu}u^{\nu}+ u^{\mu}p^{\nu}$  or $u^{\mu}u^{\nu}$ , which are 
necessary to provide transversality of $\Pi^{\mu\nu}$, cancel out.  
The invariants $w_T$ and $w_L$ can be found from the two contractions,
\begin{eqnarray}
-g_{\mu\nu}(p) \Pi^{\mu\nu}(p) = 2~ w_T(p) + w_L(p),~~~{\rm and}~~~
u_\mu u_\nu(p) \Pi^{\mu\nu}(p)= {(pu)^2 -p^2 \over p^2} w_L(p)~.
\label{eq:S1.8}\end{eqnarray} 
The solution of the Schwinger-Dyson equation for the retarded propagator can be
cast in the form
\begin{eqnarray}
 {\bf D}_{ret}^{\mu\nu}(p) ={d_{(T)}^{\mu\nu}(p)\over p^2 -w^R_T(p) }
 +{p^2~d_{(L)}^{\mu\nu}(p)\over (pu)^2 [p^2 -w^R_L(p)]}~,
\label{eq:S1.9} 
\end{eqnarray} 
which exhibits an explicit separation of the transverse and longitudinal
propagators.

Our minimal goal is to estimate the mass of the transverse plasmon (the
collective mode)  which is expected to replace the final state of free
propagation in the QCD evolution equations. Therefore, we need to compute the
self-energy $w^R_T(p)$ of the soft transverse mode accounting for the effect
of the hard part of the (already measured) final-state inclusive  gluon
distribution, $dN_g/dk_t^2dy$. To pick up the dominant effect,  we have to
compute $w^R_T(p)$ at the point $p^2=0$ \cite{Shuryak}. Therefore, we can
parameterize the momentum $p$ of the soft mode as $p^\mu
=(p_t\cosh\theta,\vec{p_t}, p_t\sinh\theta)$. The self-energy $w^R_T(p)$ has 
dispersive and non-dispersive parts, which are defined by equations
(\ref{eq:E2.14}) and (\ref{eq:E2.15}), respectively. Even before the loop
integral is computed, it is straightforward to check that the dispersive part
is proportional to $p^2$ and can be disregarded in our simple estimate. The
non-dispersive part of $w^R_T(p)$ is represented by the integral,
\begin{eqnarray}
w^{(ND)}_{T}(p)=
{g^2N_c\over 4}~{p_0^2\over {\bf p}^2} \int {d^4 q \over V^{(3)}(\tau)}
~\bigg\{{\delta (q^2)\over q_0^2}\big[ 1+2n({\bf q}) \big]
\bigg[{({\bf p}-{\bf q})^2\over 4}~+ \hspace{2cm} \nonumber \\
\hspace{2cm} +~{(q_0+p_0)^2(5q_0^2-2q_0p_0+p_0^2)\over 4({\bf p}-{\bf q})^2}-
{q_0^2+p_0^2\over 2}\bigg]~+~(q\to p-q) \bigg\}~+~{\cal O}(p^2),
\label{eq:S1.10}\end{eqnarray}
where the ``normalization volume'', $V^{(3)}(\tau)$, for the hard modes with 
transverse momenta $q_t\gg p_t$ is defined as the volume of the domain on the
hyper-surface $\tau=const$, where the wave function of the soft plasmon mode 
is stationary, and is given by
\begin{eqnarray}
V^{(3)}(\tau)=\pi R^2~\tau~\Delta\eta~,
~~~\Delta\eta\sim\ln{2\over p_t\tau}~,\nonumber
\end{eqnarray}
where $R$ is the radius of the colliding nuclei, $\pi R^2\approx\pi r_0^2
A^{2/3}$, and $r_0$ is the proton radius. The  expression (\ref{eq:S1.10})
includes a vacuum part which is UV-divergent and requires renormalization
(which we neglect for now). Integrating $q_0$ out and setting $p^2=0$  we
arrive at
\begin{eqnarray}
w^{(ND)}_{T}(p)=
{g^2N_c\over 2\pi R^2~\tau~\Delta\eta}~
\int {d^3 {\bf q} \over |{\bf q}|^3}~{d~N_g\over d^3~{\bf q} }
~\bigg[{({\bf p}-{\bf q})^2\over 2}~+~
{(5|{\bf q}|^4+2|{\bf q}|^2|{\bf p}|^2+|{\bf p}|^4)\over 
2({\bf p}-{\bf q})^2}-
{\bf q}^2-{\bf p}^2\bigg]~.
\label{eq:S1.11}\end{eqnarray}
As for the distribution $d~N_g/d^3{\bf q}=d~N_g/q_td^2\vec{q_t}dy~$,  we
shall assume that it is  homogeneous in the rapidity interval $\Delta\eta$
centered at the rapidity $\theta$ of the soft mode, $n=n(q_t,\theta -y)
\approx n(q_t,\theta)$.
Furthermore, we can take $\theta=0$ and consequently, $~p_z\approx 0$, 
$~p_0\approx p_t$. With the same accuracy, we may take $q_z=q_t\sinh y\approx
0$,   $dq_z=q_t\cosh y dy\approx q_t~dy$ and integrate over $q_t$ with the 
condition $q_t\gg p_t$. In this approximation, we obtain
\begin{eqnarray}
w^{(ND)}_{T}(p_t^2,\theta;p^2=0)= {g^2N_c\over  \pi R^2}~ 
\int_{p_t^2}^{\infty} {d q_t^2 \over \tau q_t}
{d~N_{g} (q_t^2,\theta) \over d~q_t^2 }~, 
\label{eq:S1.12}\end{eqnarray}
where the function $~dN_{g}/dq_t^2~$ is the density of the final-state
gluons in the unit interval near rapidity $\theta$. In the same way we compute
the contribution of the tadpole term,
\begin{eqnarray} 
w^{(tdpl)}_{T}(p_t^2,\theta;p^2=0)= {3 g^2N_c\over \pi R^2\tau~\Delta\eta}~
\int {d^3 {\bf q} \over |{\bf q}|}\bigg({1\over 2}+n({\bf q})\bigg)=
{3 g^2N_c\over \pi R^2}
\int_{p_t^2}^{\infty} {d q_t^2 \over \tau q_t}
{d~N_{g} (q_t^2,\theta) \over d~q_t^2 }~, 
\label{eq:S1.13}\end{eqnarray} 
where the divergent vacuum term vanishes after UV-renormalization.

Now, the renormalization can be performed in the usual way by subtracting the
quasilocal terms  {\em after} the integration with a weight function
$n(\theta)$. Indeed, the point of renormalization can be  chosen in the domain
$\theta\to\mp\infty$ ($p^-\to 0$ or $p^+\to 0$ ) where it does not affect the
finite part of $w(p)_{T}$ produced by the physical distribution $n(\theta,p_t)$
which vanishes for modes with infinite rapidity. Assembling
these two contributions, we obtain,
\begin{eqnarray} 
m_{pl}^2(p_t^2,\theta)=w^{(ND)}_{T}(p_t^2,\theta;p^2=0)+
w^{(tdpl)}_{T}(p_t^2,\theta;p^2=0)=
\big(3+1\big) {g^2N_c\over \pi R^2}
\int_{p_t^2}^{\infty} {d q_t^2 \over \tau q_t}
{d~N_{g} (q_t^2,\theta) \over d~q_t^2 }~. 
\label{eq:S1.14}\end{eqnarray} 
This quantity is defined locally in rapidity and it works as a
feed-back that limits  the possible energy of the emission field from 
below and thus screens the mass singularity in the evolution equations. 

Because the product $~p_t\tau~$ in Eq.(\ref{eq:S1.14}) is close to unity,  the
proper time $\tau$ and momentum $p_t$ are the complementary parameters, this
equation can be read in two ways. On the one hand, the smaller $p_t$ is, the
more  particles with higher momentum interacting with the soft mode are
emitted, and the heavier this soft mode is. On the other hand, we know that
the mode with momentum $p_t$ acquires the status of the emission field
only by a time $\tau\sim 1/p_t$. The later this happens, the more that 
particles  which participate in the formation of the plasmon mass are created,
and the larger the mass of plasmon is.  

\section{Evolution equations with final state screening}
\label{sec:SNS2}

\subsection{Screening in the order $\alpha_s$.}
\label{sbsec:SBS2a}

Our last problem is to show that the replacement of the free radiation field
by the collective modes of the true final state screens the collinear
singularities that we encountered in the null-plane dynamics. Previously,
these singularities were coming  either from the poles of the polarization 
sum $d^{\mu\nu}$, or from  the integration $dp^-$ of the  $\delta_+[(k-p)^2]$.
In order to smooth out this behavior,  we consider the evolution equations in
the same dynamics which allows for the effect of local mass generation.  While
the calculation of the plasmon mass could be performed in the co-moving
reference frame, the  evolution must be treated in a more global fashion.  In
order to find the space-time domain where the QCD evolution described by the 
DGLAP equation develops, let us go back to the general four-vector
$u^\mu=(\cosh\eta, \vec{0_t}, \sinh\eta)$, that defines  the local time
direction at the rapidity  $\eta$ and keep the latter as a parameter. The
gauge of the gluon field is fixed by the condition $uA=0$ which may be
conveniently rewritten as
\begin{eqnarray} 
  u(\eta)A= {1\over 2}e^{-\eta} A^+ + {1\over 2}e^{\eta} A^- =0~. 
\label{eq:S2.1}\end{eqnarray}
{}From this, we infer that the gauge condition $A^+=0$ holds as $\eta\to
-\infty$, that is, in the nearest vicinity of the hyper-plane $x^+=0$.  In the
process of {\em e--p} DIS, the Lorentz frame moving with this rapidity is the
rest frame for the electron, and the infinite-momentum frame for the proton.
The momentum of the virtual photon of the DIS  has components $q^+=0$ and
$q^-=2\nu/P^+$. Thus, the DGLAP equations describe the process which is
localized in the space-time domain of the electron fragmentation.

In what follows, we  attempt to answer only two questions: 

(i) Does the plasmon mass shield the singularity in the evolution equations? 

(ii) Does the equation with screening allow for a smooth transition to the
DGLAP equation when the screening mass becomes small? 

For these limited goals, we can  drop all regular terms. To derive the
evolution equation, we start with Eq.~(\ref{eq:E1.3}) and extract the  scalar
equations by means of Eqs.~(\ref{eq:S1.8}).  In this way we obtain an analog
of  the Eqs.~(\ref{eq:E1.10}), (\ref{eq:E1.14}) for the evolution of the
transverse fields,
\begin{eqnarray}
{d G (x,p_{t}^{2})\over d p_{t}^{2}}
={-N_c \alpha_s\over \pi^2}\int dp_0
\int {d k_zd^2\vec{k_t} \over 2\omega} \delta (k_0-p_0-\omega) 
{dG(k^z,k_t^2)\over d~k_t^2}
 {p_0^2 \over {\bf p}^2}P_{gg}({p^0\over k^0})
\bigg[{1\over p^2}- {p^0\over k^0}~{k^2\over [p^2]^2}\bigg]
+{\cal O}({p_t^2\over s},{k_t^2\over s})~~,
\label{eq:S2.2}\end{eqnarray}    
where only the terms which do not vanish at $s\to\infty$   
($\eta\to -\infty$) are retained . In this equation,
\begin{eqnarray}
\omega^2 = ({\bf k_\ast}-{\bf p_\ast})^2 + m^2((\vec{k_t}-\vec{p_t})^2) =
(k_\ast^z-p_\ast^z)^2 +(\vec{k_t}-\vec{p_t})^2+
m^2((\vec{k_t}-\vec{p_t})^2)~, \nonumber
\end{eqnarray}
and the temporal and longitudinal components of momenta are those of
the infinite momentum frame, 
\begin{eqnarray}
k_\ast^0\equiv (ku)={1\over 2}e^{-\eta} k^+ + {1\over 2}e^{\eta} k^- 
\approx k_\ast^+/2, ~~~~
k_\ast^z={1\over 2}e^{-\eta} k^+ - {1\over 2}e^{\eta} k^- \approx k_\ast^+/2,
~~~~\eta\to -\infty~. 
\label{eq:S2.3}\end{eqnarray}
Both are large and scale with the momentum of the proton.(In
Eq.~(\ref{eq:S2.2}), and  in what follows the label $\ast$ is dropped.) The
operational definition of the structure function in the new variables changes
slightly with respect to the previous definition (\ref{eq:E1.13}),
\begin{eqnarray}
\int dp_0{p_0~w_{T}^{(01)}(p)\over (p^2-w^{R}_{T}(p)) (p^2-w^{A}_{T}(p))}= 
{d G (x,p_{t}^{2})\over d p_{t}^{2}}~,
\label{eq:S2.4}\end{eqnarray} 
where by virtue of (\ref{eq:S2.3}), the domain of the integration $dp_0$ 
is strongly localized near the point $p^0=p^z$. The mass
$m^2((\vec{k_t}-\vec{p_t})^2)$ was computed in Sec.~\ref{sbsec:SBS1c} and is
used as the pole mass of the emitted gluon-plasmon. It has the property to
increase with decreasing transverse momentum. Integrating out $p^0$ with
the aid of the delta-function in the Eq.~(\ref{eq:S2.2}), we arrive at
\begin{eqnarray}
{dG(p^z,p_t^2)\over p_t^2}
={- N_c \alpha_s\over \pi^2} \int {d k_zd^2\vec{k_t} \over 2\omega}  
{dG(k^z,k_t^2)\over d~k_t^2}
 { (k_z-\omega)^2 \over {\bf p}^2[(k_z-\omega)^2 -{\bf p}^2] }
\bigg[{k_z \over\omega} + {k_z \over k_z-\omega}-2 + 
{\omega (k_z-\omega)\over k_z^2}\bigg],
 \label{eq:S2.5}\end{eqnarray} 
where the expression in the square brackets is just the splitting kernel
$P_{gg}$ and the strong ordering ($p^2\gg k^2$) is assumed from now on.   
Once again, this equation is written in the first approximation with
respect to the small parameter $p_t^2/k_z^2\sim p_t^2/s$, and the former pole
at $p^0=k^0$  is translated into the non-singular factor $1/\omega$. 

By examination, Eq.~(\ref{eq:S2.5})  has no singularity in the remaining
integration $dk_z$. It is reliably screened by the finite $p_t^2$ regardless
of the presence of the plasmon mass. Such a screening happens because now the
proton has  {\em finite rapidity}; and the dynamics are free from the
artificial singularity of the null-plane. However, if we consider a
non-ordered emission, or just take $p_t$ to be small, then only the mass, 
$m((\vec{k_t}-\vec{p_t})^2)$, would provide the desired protection. Only in
the formal limit $m^2/s\to 0$, the singularity at $k_z=p_z$ becomes a reality,
and we recover the usual DGLAP equation. It is easy to demonstrate, that the
same mechanism of shielding of the mass singularity works for 
Eq.~(\ref{eq:E1.15}), which describes the transition between the longitudinal
and transverse modes of the gluon field. 

With the mass  computed according to (\ref{eq:S1.12}), Eq.~(\ref{eq:S2.5}) can
be considered as the simplest prototype of the evolution equation with 
screening for the fluctuations in the AA-collisions.  One should keep in mind,
that (by virtue of (\ref{eq:S2.3})) the set of limits, $k_z-p_z\to 0$,
$k^+-p^+\to 0$, and $\eta\to -\infty$, which has been employed for its
derivation is intrinsically ambiguous. In order to obtain the evolution
equations that can handle the emission in the central rapidity region we
cannot approach the limit of $\eta\to -\infty$. These equations can be derived
only in the framework of the ``wedge dynamics'' \cite{WD1,WDG}.

\subsection{Suppression of the radiative corrections in the order
 $\alpha_s^2$.}
\label{sbsec:SBS2b}

\begin{figure}[htb]
\begin{center}
\mbox{ 
\psfig{file=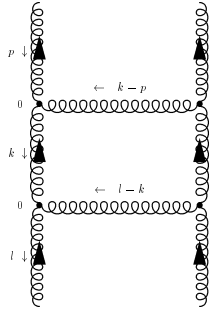,height=2.in,bb=119 381 309 684}
}
\end{center}
\caption{Fragment of the ladder with two rungs of emission.}
\label{fig:fig3a}
\end{figure}

The last thing that we are going to demonstrate with our limited ``local''
approximation is that the radiative corrections of the order $\alpha_s^2$ are
doubly-suppressed by the plasmon masses. Indeed, a review of our previous
calculations in this order shows that all of them are connected with
processes with at least two real emissions.  An example of such a process with
two consecutive emissions is depicted at Fig.~\ref{fig:fig3a}.

In any of these cases, we have to integrate
the expressions with two mass-shell delta functions, 
\begin{eqnarray}
\int d^4 k~\delta_+[(k-p)^2-m_1^2]~\delta_+[(l-k)^2-m_2^2] P_1(p^+/k^+)
P_2(k^+/l^+)~~,
\label{eq:S2.6}\end{eqnarray}
where $P_{1,2}(z)$ may have a pole at $z=1$, 
$m_1=m_1((\vec{k_t}-\vec{p_t})^2)$, $m_2=m_2((\vec{l_t}-\vec{k_t})^2)$, and
the total emitted momentum is $l-p=s$. For example, let the collinear
singularity emerge when the integration $dk^+$ reaches one or both ends of the
interval $p^+<k^+<l^+$.  Introducing the new variables, $f=k-p$, we can
integrate out $f^-$ with the aid of one of the delta-functions and remove the
second one by integrating over the  angle between the transverse vectors
$\vec{f_t}$ and $\vec{s_t}$. The latter integration  defines the limits for
the magnitude of the transverse component  $f_t$, which has to be confined
between the two positive roots of the equation,
\begin{eqnarray} 
 [(s_t+f_t)^2 -(s^+-f^+)(s^--{f_t^2-m_1^2\over f^+})-m_2^2]
 ~[(s^+-f^+)(s^--{f_t^2-m_1^2\over f^+})-m_2^2-(s_t-f_t)^2]=0 ~.
\label{eq:S2.7}\end{eqnarray}
These roots exist (the measure of integration $df_t$ does not vanish)
only if the variable $f^+$ is within the following limits,
\begin{eqnarray} 
 1+{m_1^2-m_2^2\over s^2} -\sqrt{1-2 {m_1^2+m_2^2\over s^2}+
 \bigg( {m_1^2-m_2^2\over s^2}\bigg)^2} 
 ~<~2~{f^+\over s^+}~< \hspace{2cm} \nonumber\\
<~1+{m_1^2-m_2^2\over s^2} + \sqrt{1-2 {m_1^2+m_2^2\over s^2}+
  \bigg( {m_1^2-m_2^2\over s^2}\bigg)^2} ~~.
\label{eq:S2.8}\end{eqnarray}
This inequality shows that the domain of integration over $k^+$ gets smaller
with the  growth of the masses, and it just vanishes when, {\em e.g.}, either
$m_1^2\geq s^2$ or $m_2^2\geq s^2$, and the emission of two gluons becomes
impossible. Hence, the finite plasmon masses not only screen the mass 
singularities, but also strongly reduce the phase-space for the  terms of the
order $\alpha_{s}^{2}$.

\vspace{2cm}

\noindent {\bf ACKNOWLEDGMENTS}

We are grateful to Yu. Dokshitzer,  A. Kovner, L. McLerran, B. Muller, and H.
Weigert for many stimulating discussions.  We wish to thank Edward Shuryak for
helpful conversations at various stages in the development of this work.  We
are grateful to Jianwei Qiu for attracting our attention to his paper
\cite{Tw4}.  We appreciate the help of Scott Payson who critically read the
manuscript.

This work was supported by the U.S. Department of Energy under
Contract  No. DE--FG02--94ER40831.       

\bigskip  

\renewcommand{\theequation}{A1.\arabic{equation}}
\setcounter{equation}{0}

\section{Appendix 1. Causal properties of the evolution ladder}
\label{sec:SA1}

In the course of computing the $g^4$-elementary ladder cell we met various
types of radiative corrections with  retarded order of their temporal
arguments. This circumstance allowed us to make a firm conclusion (which was
verified at the tree-level) about the time ordering in process the QCD
evolution process. It is also fortunate in one more respect. The retarded
(advanced) functions have the unique analytic properties that they are regular
in the upper (lower) half-plane of the complex energy. Since we use
Hamiltonian dynamics with the coordinate $x^+$ chosen as the ``time''
direction, the corresponding conjugate variable is  the light-cone energy,
$p^-$, and we can rely on the analytic properties in the complex plane of
$p^-$.Beyond the tree-level calculations, we include radiative corrections
with loops, self-energies and vertex functions. We must check whether they
maintain the aforementioned causal behavior. Self-energy corrections to the
propagators are of an explicitly  retarded type, and they do not cause any
problems. The loops of the vertices have yet to be examined. The two vertex
functions, $V_{000}$, and $V_{111}$~, which come from the terms with $R=S$ in
Eq.(\ref{eq:E2.23}), have the same surroundings and appear in the combination
$V_{000}+V_{111}$.  This sum forms the following string of propagators which
constitute the loop, 
\begin{eqnarray}
D_{00}(q) D_{00}(k-p) D_{00}(q-k) + D_{11}(q) D_{11}(k-p) D_{11}(q-k).
\label{eq:A1.1}\end{eqnarray}   
Since $D_{00}=D_s + D_1/2$  and $D_{11}= -D_s + D_1/2$, this string can be
identically rewritten as,
\begin{eqnarray}
D_{1}(q) D_{s}(k-p) D_{s}(q-k) + D_{s}(q) D_{1}(k-p) D_{s}(q-k) + 
D_{s}(q)D_{s}(k-p) D_{1}(q-k),
\label{eq:A1.2}\end{eqnarray}
where we omitted the term $D_{1} D_{1} D_{1}/2$, which identically vanishes in
the surroundings of the evolution ladder. We want to show that expression
(\ref{eq:A1.2}) is nothing but the real part of the retarded vertex, which is
an analytic function in the lower half-plane of energy $p^-$.
To prove this statement, we should derive the formula for the retarded vertex 
to order $g^3$. This is more easily done in the coordinate form. 

Let us start with the definition of the retarded vertex
in the form of the functional derivative of the retarded self-energy
with respect to the field ${\cal A}(z)$,
\begin{eqnarray}
 ^{(3)}{\bf V}^{\nu\beta\sigma}_{bcf;ret}(x,y,z)=-
 {{\delta ~ ^{(2)}\Pi^{bc;\nu\beta}_{ret}(x,y) } \over
  {g \delta {\cal A}^{f}_{\sigma}(z_{B}) }  }~.
\label{eq:A1.3}
\end{eqnarray} 
The self-energy $\Pi_{ret}$ is built from the  propagators
$D_{ret}$, $D_{adv}$, and  $D_1$. The simplest way to differentiate them 
is to use the equations  (\ref{eq:Q1.5}) for $D_{ret}$, $D_{adv}$, and  $D_1$ 
in the presence of the classical external field. These equations can be 
written  as follows:
\begin{eqnarray}
{\bf D}_{ret\choose adv}(x,y)= D_{ret\choose adv}(x,y)+
\int dz D_{ret\choose adv}(x,z) V(z){\cal A}(z)D_{ret\choose adv}(z,y)~, 
\label{eq:A1.4}
\end{eqnarray} 
\begin{eqnarray}
{\bf D}_{1}(x,y)= D_{1}(x,y)+
\int dz [ D_{ret}(x,z) V(z){\cal A}(z)D_{1}(z,y) + 
D_{1}(x,z) V(z){\cal A}(z)D_{adv}(z,y)~.
\label{eq:A1.5}
\end{eqnarray} 
It is straightforward to obtain,
\begin{eqnarray}
^{(3)}{\bf V}_{ret}^{\alpha\rho\sigma}(x,y,z)= - {i\over 4}
\big(~ [~ V^{\alpha_1\alpha\alpha_2}(x)D_{ret}^{\alpha_2\sigma_2}(x,z)
V^{\sigma_2\sigma\sigma_1}(z)D_{ret}^{\sigma_1\rho_2}(z,y)
V^{\rho_2\rho\rho_1}(y)D_{1}^{\rho_1\alpha_1}(y,x)  \nonumber \\
+V^{\alpha_1\alpha\alpha_2}(x)D_{ret}^{\alpha_2\sigma_2}(x,z)
V^{\sigma_2\sigma\sigma_1}(z)D_{1}^{\sigma_1\rho_2}(z,y)
V^{\rho_2\rho\rho_1}(y)D_{adv}^{\rho_1\alpha_1}(y,x)  \nonumber \\
+V^{\alpha_1\alpha\alpha_2}(x)D_{1}^{\alpha_2\sigma_2}(x,z)
V^{\sigma_2\sigma\sigma_1}(z)D_{adv}^{\sigma_1\rho_2}(z,y)
V^{\rho_2\rho\rho_1}(y)D_{adv}^{\rho_1\alpha_1}(y,x) ~] \nonumber \\
+ [~ V^{\alpha_1\alpha\alpha_2}(x)D_{ret}^{\alpha_2\rho_2}(x,y)
V^{\rho_2\rho\rho_1}(y) D_{ret}^{\rho_1\sigma_2}(y,z)
V^{\sigma_2\sigma\sigma_1}(z)D_{1}^{\sigma_1\alpha_1}(y,x) \nonumber \\
+V^{\alpha_1\alpha\alpha_2}(x)D_{ret}^{\alpha_2\rho_2}(x,y)
V^{\rho_2\rho\rho_1}(y) D_{1}^{\rho_1\sigma_2}(y,z)
V^{\sigma_2\sigma\sigma_1}(z)D_{adv}^{\sigma_1\alpha_1}(y,x) \nonumber \\
+V^{\alpha_1\alpha\alpha_2}(x)D_{1}^{\alpha_2\rho_2}(x,y)
V^{\rho_2\rho\rho_1}(y) D_{adv}^{\rho_1\sigma_2}(y,z)
V^{\sigma_2\sigma\sigma_1}(z)D_{adv}^{\sigma_1\alpha_1}(y,x) ~] ~\big)~.
\label{eq:A1.6}
\end{eqnarray} 
Here, the groups of the first three,  and of the last three terms 
differ only by the direction in which the arguments are
going around the loop. Therefore, these two groups are identical.
In the momentum representation, we obtain
\begin{eqnarray}
^{(3)}{\bf V}_{ret}(p,-k,k-p)=-{i\over 2}\int dxdydze^{ip(x-z)+ik(y-z)}
\big[~ V(x)\underline{D_{ret}(x-y)V(y)D_{ret}(y-z)}V(z)D_{1}(z-x)\nonumber \\
+ V(x)D_{ret}(x-y)V(y)D_{1}(y-z)V(z)\underline{D_{adv}(z-x)}+
V(x)D_{1}(x-y)V(y)D_{adv}(y-z)V(z)\underline{D_{adv}(z-x)} ~\big].
\label{eq:A1.7}
\end{eqnarray} 

\begin{figure}[htb]
\begin{center}
\mbox{ 
\psfig{file=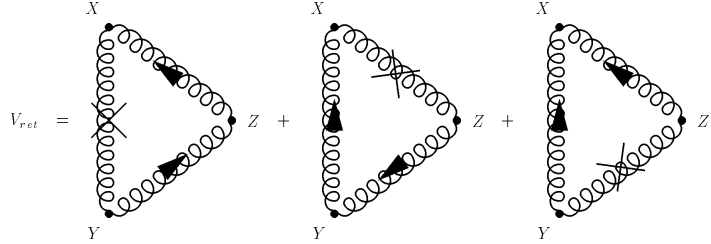,height=1.5in,bb=110 505 580 670}
}
\end{center}
\caption{Three diagrams contributing to the retarded vertex.
Lines with arrows correspond to the propagators $D_{ret}$ and point
$x$ is the latest. The crossed lines correspond to the correlators
$D_1$, the densities of real states.}
\label{fig:fig9}
\end{figure}

The elements which are responsible for the analytic properties of the retarded
vertex with respect to  momentum $p$ are underlined in the formula and 
marked by the time-directed arrows in Fig.~\ref{fig:fig9}. 
In the same way, we may introduce an advanced vertex function and put
it in a form similar to (\ref{eq:A1.7}),
\begin{eqnarray}
^{(3)}{\bf V}_{adv}(p,-k,k-p)=-{i\over 2}\int dxdydze^{ip(x-z)+ik(y-z)}
\big[~ V(x)\underline{D_{adv}(x-y)V(y)D_{adv}(y-z)}V(z)D_{1}(z-x)\nonumber \\
+V(x)D_{adv}(x-y)V(y)D_{1}(y-z)V(z)\underline{D_{ret}(z-x)}+
V(x)D_{1}(x-y)V(y)D_{ret}(y-z)V(z)\underline{D_{ret}(z-x)} ~\big]~,
\label{eq:A1.8}
\end{eqnarray} 
which shows that at the real light-cone energy $p^-$, $V_{adv}$ is the complex
conjugate of $V_{ret}$. This is exactly the sum  
$V_{ret}+V_{adv}=V_{000}+V_{111}$ (twice the real part of $V_{ret}$) that 
defines the vertex loop correction to
the elementary cell of the ladder. Indeed, since
\begin{eqnarray}
D_{ret}=D_{s}+{1\over 2} D_0,~~~ D_{adv}=D_{s}-{1\over 2} D_0,
\label{eq:A1.9}
\end{eqnarray} 
we find that
\begin{eqnarray}
{\rm Re}~{}^{(3)}{\bf V}_{ret}^{(D)}(p,-k,k-p)=
-{i\over 4}\int {d^4 q \over (2\pi)^4}
\big[~V_x D_{1}(q) V_y D_{s}(q-k)V_z  D_{s}(q-p) \nonumber \\
+ V_x D_{s}(q) V_y D_{1}(q-k) V_z D_{s}(q-p) + 
V_x D_{s}(q) V_y D_{s}(q-k) V_z D_{1}(q-p) ~\big] ~.
\label{eq:A1.10}
\end{eqnarray} 
Thus, it is natural to use the dispersion relation (\ref{eq:E2.27}) to compute
the real part of ${\bf V}_{ret}$. However, one should remember that each of
the retarded propagators of the gauge field includes a part that corresponds
to longitudinal fields, which are not driven by the Hamiltonian equations of
motion. Only the transverse fields  propagate and obey causality. Therefore,
those retarded and advanced functions, which maintain the causal properties of
the vertex (and are underlined in Eq.(\ref{eq:A1.7}) ) have to be split  into 
longitudinal and transverse parts, 
\begin{eqnarray}
D_{ret}(q)= D_{ret}^{(T)}(q)+D^{(L)}(q),~~~~
D_{adv}(q)= D_{adv}^{(T)}(q)+D^{(L)}(q)~,
\label{eq:A1.11}
\end{eqnarray} 
and only the causal transverse part has the proper analytic behavior, which
results in  dispersion relations linking the real and imaginary parts
of the vertex function. Thus we encounter a necessity of separating
the dispersive part of the vertex, $V^{(D)}$, from the non-dispersive
part, $V^{(ND)}$.  

For our practical goals, we need only the real part of the dispersive term 
of the  vertex function, which is,
\begin{eqnarray}
^{(3)}{\bf V}_{ret}^{(D)}(p,-k,k-p)=-{i\over 4}
\int {d^4 q \over (2\pi)^4}
\big[~ V_xD_{ret}^{(T)}(q)V_yD_{ret}^{(T)}(q-k)V_zD_{1}(q-p)+\nonumber \\
 +V_xD_{ret}(q)V_yD_{1}(q-k)V_zD_{adv}^{(T)}(q-p)+
V_xD_{1}(q)V_yD_{adv}(q-k)V_zD_{adv}^{(T)}(q-p) ~\big].
\label{eq:A1.12}
\end{eqnarray} 
It can be computed by means of dispersion relation via the imaginary part,
which consists of two terms. The main term, is symmetric with respect to the
singular functions  (commutators $D_{0}$ and densities of states $D_{1}$),
and is given by,
\begin{eqnarray}
{\rm Im}~{}^{(3)}{\bf V}_{ret}^{(D)}(\omega,-k,k-\omega)=
-{i\over 4}\int {d^4 q \over (2\pi)^4} \times \hspace{2in} \nonumber \\
\times\bigg(~  [~V_x D_{0}(q)V_y  D_{s}^{(T)}(q-k) V_z D_{1}(q-\omega) -
V_x D_{1}(q) V_y D_{s}^{(T)}(q-k) V_z D_{0}(q-\omega)~]+ \nonumber \\
 + [~V_x D_{s}^{(T)}(q) V_y D_{0}(q-k)V_z  D_{1}(q-\omega) -
V_x D_{s}^{(T)}(q) V_y D_{1}(q-k)V_z  D_{0}(q-\omega)~]+\nonumber \\
+[~V_x D_{0}(q) V_y D_{1}(q-k) V_z D_{s}^{(T)}(q-\omega)-
V_x D_{1}(q) V_y D_{0}(q-k) V_z D_{s}^{(T)}(q-\omega) ~]~ \bigg). 
\label{eq:A1.13}
\end{eqnarray}
Symmetrization has been achieved by an extra splitting  
of the two retarded propagators which do not depend on the momentum $p$ and
do not participate in maintaining the analytic properties with respect to the 
(light-cone) energy $p^-$. Therefore, we have an additional residue of the
splitting procedure which contains the longitudinal propagators $D^{(L)}$,
{\em i.e.},
\begin{eqnarray}
\Delta{\rm Im}~{}^{(3)}{\bf V}_{ret}^{(D)}(\omega,-k,k-\omega)=
-{i\over 4}\int {d^4 q \over (2\pi)^4} \times\hspace{2in}\nonumber \\
\times\big[ - V_x D_{s}^{(L)}(q)V_y  D_{1}(q-k) V_z D_{0}(q-\omega) -
V_x D_{1}(q) V_y D_{s}^{(L)}(q-k) V_z D_{0}(q-\omega)  \big]. 
\label{eq:A1.14}
\end{eqnarray}
In the last two equations, we have introduced the four-vector 
$\omega^\mu=(p^+,\omega^-,\vec{p_t})$. The first two terms in
Eq.~(\ref{eq:A1.13}) correspond to the unitary cut of the diagram for
$V_{ret}$ near the external line with  momentum $p$. The next two terms
correspond to the unitary cut near the external line with momentum $p-k$. At
real $\omega^-$, the imaginary part due to these terms differs from zero  for
$\omega^->p_t^2/p^+$ and for $\omega^- < -(\vec{p_t}-\vec{k_t})^2/(k^+-p^+)$,
respectively. Therefore, the subtraction point in the dispersion relations
with respect to the light-cone energy $p^-$ is confined to the {\em finite
interval} of $p^-$ between the tips of the two cuts. The last two terms in
Eq.(\ref{eq:A1.13}) would have corresponded to the unitary cut near the
external line with  momentum $k$, if the momentum $k$  were time-like. As long
as dispersion relations do not involve the momentum $k$, these terms vanishe
in the ladder kinematics, since we always have $k^2<0$.

The non-dispersive part of the vertex is not causal and can only be calculated
in a straightforward manner, 
\begin{eqnarray}
 ^{(3)}{\bf V}_{ret}^{(ND)}(p,-k,k-p)=
-{i\over 2}\int {d^4 q \over (2\pi)^4}
\big[ V_x D_{s}^{(T)}(q)V_y D^{(L)}(q-k)V_z D_{1}(q-p)+\nonumber \\
+V_x D^{(L)}(q)V_y D_{s}^{(T)}(q-k)V_z D_{1}(q-p)+
 V_x D_{s}(q)V_y D_{1}(q-k)V_z D^{(L)}(q-p)+\nonumber \\
+ V_x D_{1}(q)V_y D_{s}(q-k)V_z D^{(L)}(q-p) \big].
\label{eq:A1.15}
\end{eqnarray} 
We use the vertex function only when it is embedded in the ladder
diagram, and therefore, we do not need to compute it separately.

\renewcommand{\theequation}{A2.\arabic{equation}}
\setcounter{equation}{0}

\section{Appendix 2. The vanishing contributions}
\label{sec:SA2}
There are several types of terms that do not contribute to the evolution
equations for various reasons.

\subsection{The terms with the four-gluon vertices}
\label{subsec:SA2a} 

\begin{figure}[htb]
\begin{center}
\mbox{
\psfig{file=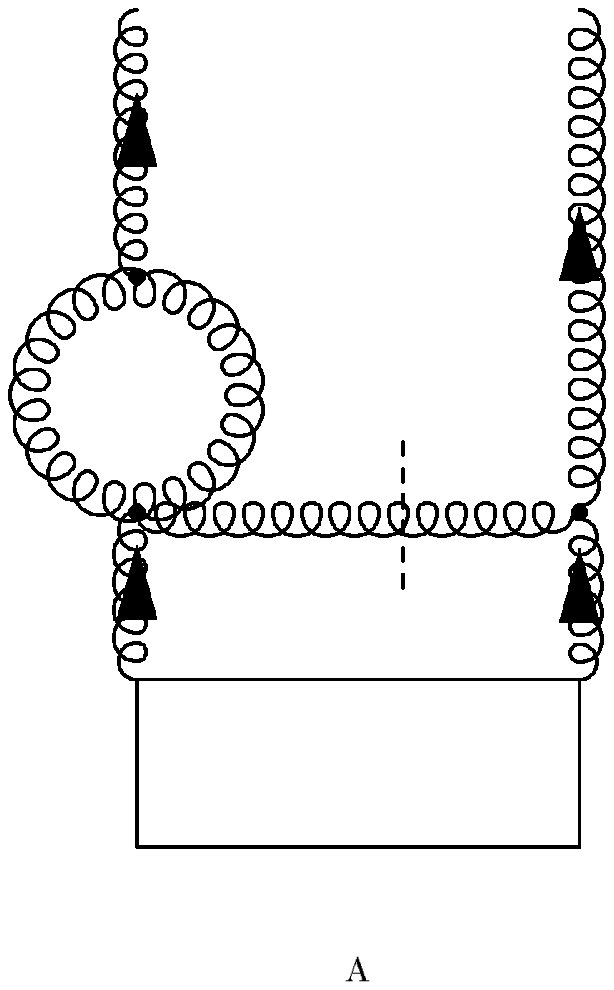,height=1.74in,bb=100 370 293 675}
\psfig{file=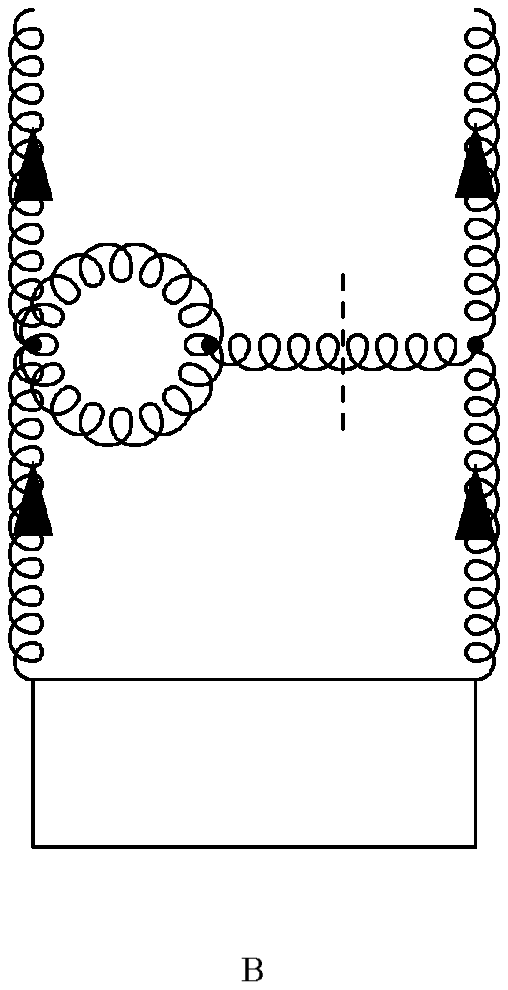,height=1.74in,bb=100 370 293 675}
\psfig{file=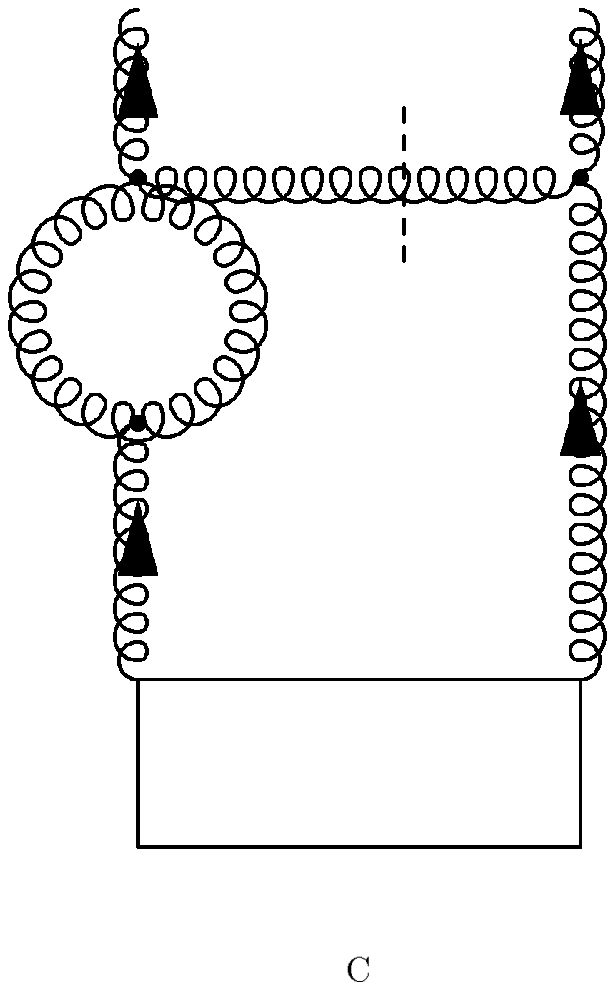,height=1.74in,bb=100 370 293 675}
\psfig{file=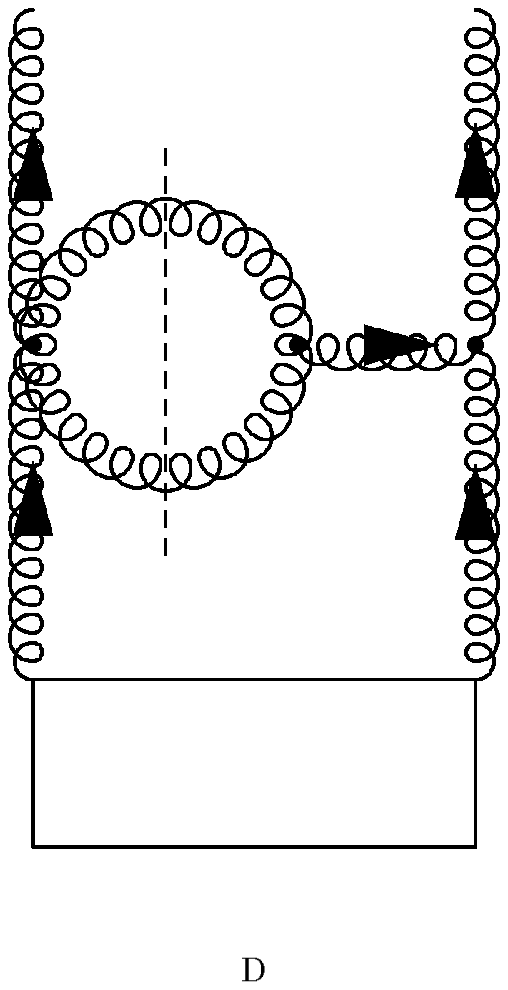,height=1.74in,bb=100 370 293 675}
\psfig{file=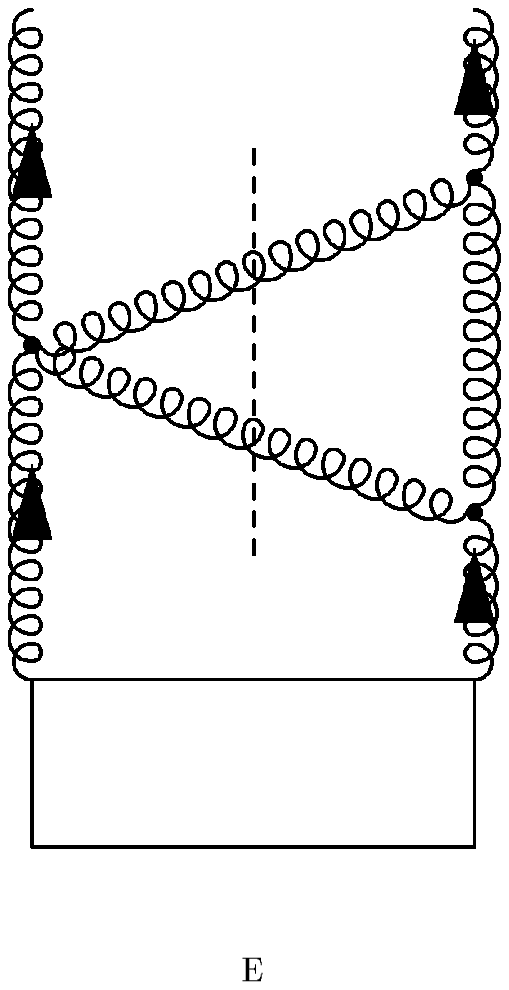,height=1.74in,bb=100 370 293 675}
\psfig{file=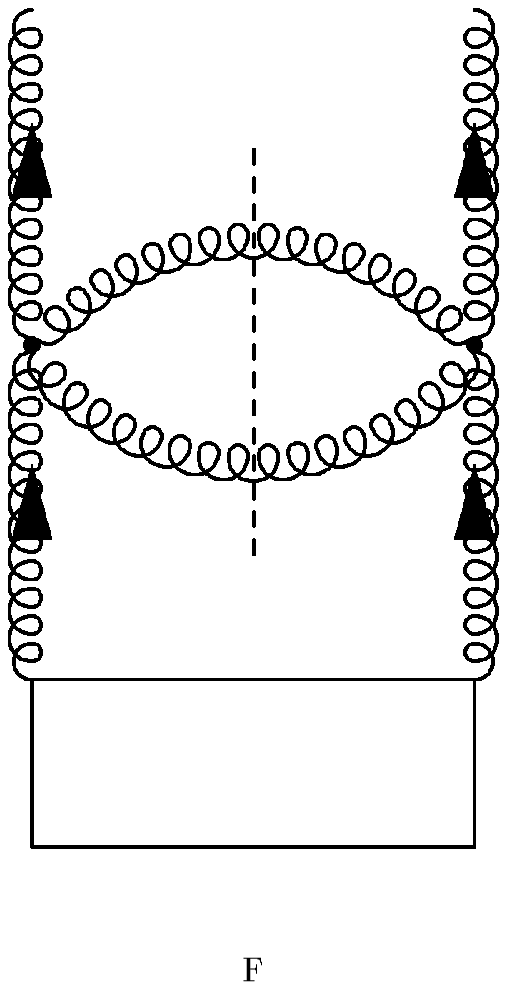,height=1.74in,bb=100 370 293 675}
}
\end{center}
\caption{Diagrams with four-gluon vertices.}
\label{fig:fig10}
\end{figure}

In the order $g^4$, we have eleven diagrams with  four-gluon vertices. They
can appear only in the TT-transition mode.  
Ten of them turn out to be zero. These are diagrams ({\em a--e}) in 
Fig.~\ref{fig:fig10} and their ``mirror reflections''.  
Diagrams  {\em a--c} include the virtual loops that are the real parts of the 
retarded functions. Hence, they can be computed via the dispersion relations. 
Even though the dispersion integral is found to be UV-finite, it still
requires renormalization. Therefore, the dispersion integral has to be 
rewritten
with the subtraction at some point $p^-=-\Omega$. For the diagram {\em (a)}
we get,
\begin{eqnarray}
\bigg[{d G(p^+,p_t^2)\over d p_t^2}\bigg]_{(a)}=
-{ig^2N_c\over (2\pi)^2}
\int {dk^+d{\vec k_t} \over 2 k^+}~{dG(k^+,k_t^2)\over d k_t^2}~
\delta \big[(k-p)^2\big] \int dp^-
 \int \bigg({d\omega^-\over \omega^- - p^-}-{d\omega^-\over
  \omega^- + \Omega} \bigg) \times \nonumber\\
\times\int {dq^+ d^2 \vec{q_t} \over 2}
{\delta_+\big[(\omega-q)^2\big] \delta_+\big[q^2\big]\over  [p^2\big]^2 }
\bigg\{{(\omega^- -p^-)\over p^+}+{(k-q)^2\over
   k^+(p^+-q^+)}-{p^2(2k^+q^+-k^+p^+-q^+p^+)\over
k^+p^{+2}(p^+-q^+)  }  \bigg\}~=~0.
\label{eq:g4.20}\end{eqnarray}
This virtual correction identically vanishes after subtraction at any value
of the light-cone energy. Therefore, in the coordinate space this term is
quasi-local ( $\sim \delta(x^+)$ ) and does not contribute to the evolution.

Diagram~{\em (b)} behaves similarly: it is finite and we
calculate it using the dispersion relation over $p^-$ making a
subtraction at some point $p^-=-\Omega$ to renormalize it,
\begin{eqnarray}
\bigg[{d G(p^+,p_t^2)\over d p_t^2}\bigg]_{(b)}=
-{ig^2N_c\over (2\pi)^2} 
\int {dk^+d{\vec k_t} \over 2 k^+}~{dG(k^+,k_t^2)\over d k_t^2}
    ~\delta \big[(k-p)^2\big] dp^- \nonumber\\
    \times \int \bigg({d\omega^-\over \omega^- - p^-}-{d\omega^-\over
  \omega^- + \Omega} \bigg) 
\int{dq^+d^2 \vec{q_t}\over 2}
{\delta_+\big[(q-\omega)^2\big]
\delta_+\big[(k-q)^2\big]\over [p^2\big]^2 }\nonumber\\
\times \bigg\{{(\omega^- -p^-)\over (k^+-q^+)(k^+-p^+)}-
{q^2\over k^+(q^+-p^+)(k^+-q^+)}+{p^2 \over (k^+-p^+)k^+(q^+-p^+) }\bigg\}~=~0.
\label{eq:g4.25}\end{eqnarray}
Without renormalization, these two terms would produce finite corrections
comparable to those coming from the three-gluon vertex (they include both 
the Coulomb logarithm, and the collinear cut-off).

Diagram~{\em (c)} is of the form,
\begin{eqnarray}
\bigg[{d G(p^+,p_t^2)\over d p_t^2}\bigg]_{(c)}=
-{ig^2N_c\over (2\pi)^2}
\int  {dk^+d{\vec k_t} \over 2 k^+}~{dG(k^+,k_t^2)\over d k_t^2}~
\int dp^-{\delta \big[(k-p)^2\big]\over [p^2\big]^2}
\int{dq^+d^2 \vec{q_t}\over 2}
\left[ {\delta\big[(k-q)^2\big]\over q^2}+
{\delta\big[q^2\big]\over (k-q)^2}\right] \nonumber\\
\times\bigg\{{(k-q)^2(\omega -p^-)\over p^+}+{\big[(k-q)^2\big]^2\over
   k^+(p^+-q^+)}-{p^2(k-q)^2(2k^+q^+-k^+p^+-q^+p^+)\over
k^+p^{+2}(p^+-q^+)  }  \bigg\}~=~0.
\label{eq:g4.27}\end{eqnarray}
It proves to be zero  after delta-functions are used to integrate out $p^-$
and $q^-$. (The expression in the curly brackets is  zero.)

The diagrams {\em (d)} and {\em (e)}  correspond to real processes.  They
vanish identically, which  is seen only after the integrations $dp^-$, $dq^-$,
and $d{\vec q_t}$ are carried out.

The only term which has a four-gluon vertex that survives, is
given by diagram~(f). It corresponds to the real emission
\begin{eqnarray}
\bigg[{d G(p^+,p_t^2)\over d p_t^2}\bigg]_{(f)}=
-{24 i g^4 N_c^2 \over (2\pi)^2}
\int {dk^+d{\vec k_t} \over 2 k^+}~{dG(k^+,k_t^2)\over d k_t^2} dp^-
\int_{p^+}^{k^+}{dq^+d^2 \vec{q_t}\over 2}                 
{\delta\big[(q-p)^2\big] \delta\big[(k-q)^2\big]
\over [p^2\big]^2} \times \nonumber\\ 
= -{g^4 N_c^2\over (2\pi)^2}
\int {dk^+d{\vec k_t} \over 2 k^+}~{dG(k^+,k_t^2)\over d k_t^2} 
~ {6(k^+ -p^+)\over
    k^+p^+p_t^2} .~~~~~~~
\label{eq:g4.40}\end{eqnarray}

\subsection{The non-dispersive and relative terms in three gluon vertex.}
\label{subsec:SA2b} 

The two terms in the three-gluon vertex in the TT-transition mode, which 
were not discussed in the body of the paper, are
given by Eqs.~(\ref{eq:A1.14}) and (\ref{eq:A1.15}). 
To compute the contribution of the longitudinal (non-causal) fields in the
dispersive part we have to 
substitute (\ref{eq:A1.14}) into Eq.~(\ref{eq:E2.27}) and then 
proceed with Eq.~(\ref{eq:E2.26}). These steps result in
\begin{eqnarray}
\bigg[{dG(p^+,p_t^2)\over d p_t^2}\bigg]^{VV,(\Delta L)}_{TT}
={g^4 N_c^2\over 4(2\pi)^6}
\int {dk^+d{\vec k_t} \over 2 k^+}~ 
{dG(k^+,k_t^2)\over d k_t^2}\int dp^- ~\delta \big[(k-p)^2\big]~
[p^+ \Phi_{\Delta L1}(p,k) + p^+ \Phi_{\Delta L2}(p,k)]~,
\label{eq:nd.5}\end{eqnarray}
where $\Phi_{L1}$ and $\Phi_{L2}$ are given by 
\begin{eqnarray}
\Phi_{\Delta L1}=-
 \int \bigg({d\omega^-\over \omega^- - p^-}-{d\omega^-\over
  \omega^- + \Omega} \bigg)
\int {dq^+dq^-\over 2}{\rm sign}(q^+-p^+)
                 \int d^2 \vec{q_t} 
{\delta\big[(q-\omega)^2\big]\delta\big[q^2\big]\over 
[p^2\big]^2}\times\nonumber\\
\times\bigg\{{(\omega^- -p^-)\over p^+}+{(k-q)^2\over
   k^+(p^+-q^+)}-{p^2(2k^+q^+-k^+p^+-q^+p^+)\over  
k^+p^{+2}(p^+-q^+)  }  \bigg\} \Psi_1(q^+)~
\label{eq:nd.20}\end{eqnarray}
and
\begin{eqnarray}
\Phi_{\Delta L2}= \int \bigg({d\omega^-\over \omega^- - p^-}-
             {d\omega^-\over \omega^- + \Omega} \bigg)
\int{dq^+dq^-\over 2}{\rm sign}(q^+-p^+) \int d^2 \vec{q_t} 
{ \delta\big[(q-\omega)^2\big]
            \delta\big[(k-q)^2\big]\over [p^2\big]^2 }\times\nonumber\\
\times \bigg\{ {(\omega -p^-)\over (k^+-q^+)(k^+-p^+)}-
{q^2\over k^+(q^+-p^+)(k^+-q^+)}+
{p^2\over  (k^+-p^+)k^+(q^+-p^+) }  \bigg\} \Psi_2(q^+)~,
\label{eq:nd.25}\end{eqnarray}
where $\Psi_{1,2}(q^+)$ are rational function.
After the integration over $q^-$, $\omega^-$, and $\vec q_t$, both
$\Phi_{\Delta L1}$ and  $\Phi_{\Delta L2}$ vanish for an arbitrary subtraction
point $\Omega$. Therefore, the longitudinal fields contribute 
only a quasi--local term to the vertex function.
The non-dispersive part, given by Eq.~(\ref{eq:A1.15}), being substituted 
into Eq.~(\ref{eq:E2.26}), results in 
\begin{eqnarray}
\bigg[{dG(p^+,p_t^2)\over d p_t^2}\bigg]^{VV,(ND)}_{TT}
={g^4 N_c^2\over 4(2\pi)^6}\int {dk^+d{\vec k_t} \over 2 k^+}~
{dG(k^+,k_t^2)\over d k_t^2}\int dp^-
[p^+ \Phi_{ND1}(p,k) + p^+ \Phi_{ND2}(p,k)]~\delta\big[(k-p)^2\big]~~, 
\label{eq:nd.35}\end{eqnarray}
with $\Phi_{ND1}$ and $\Phi_{ND2}$ given by
\begin{eqnarray}
\Phi_{ND1}= 
\int d^4q
~{\delta\big[(q-p)^2\big]
\delta\big[q^2\big]\over [p^2]^2}
\bigg\{p^2(2k^+q^+-k^+p^+-q^+p^+)-(k-q)^2 (p^+)^2)\bigg\}\tilde\Psi_1(q^+)~
\label{eq:nd.50}\end{eqnarray}
and
\begin{eqnarray}
\Phi_{ND2}= 
\int d^4q
{\delta\big[(q-p)^2\big]
\delta\big[(k-q)^2\big]\over [p^2\big]^2 }
\bigg\{p^2(k^+-q^+)- q^2 (k^+-p^+) \bigg\}\tilde\Psi_2(q^+)~.
\label{eq:nd.55}\end{eqnarray}
As with all other constituents of the vertex function, the non-dispersive
part, even being finite, requires renormalization. We make a subtraction  at
$p^-=-\Omega$. This results in both functions, $\Phi_{ND2}$ and $\Phi_{ND2}$,
separately being zero. Thus, all terms associated with the longitudinal fields
in the vertex function in the TT-transition mode are  quasi--local.


\begin{references}
\bibitem{HPQCD} G.Sterman et al., {\em Handbook of perturbative QCD}, 
                 Rev.Mod.Phys. {\bf 67}(1995)157.
\bibitem{Shuryak} E.V. Shuryak, Sov. Phys. JETP {\bf 47} (1978) 212.
\bibitem{Gribov} V.N. Gribov, {\em Space-time description of hadron
              interactions at high energies}, in Proceedings of the 8-th
              Leningrad Nuclear Physics Winter School, February 16-27, 1973.
\bibitem{BPQCD}  Yu.L. Dokshitzer et al., {\em Basics of perturbative QCD},
                 Editions Frontieres, 1991.    
\bibitem{Brodsky} G.P. Lepage and S.J. Brodsky,  Phys.Rev. {\bf D22},
               (1979)2157. 
\bibitem{Larry}  L. McLerran, R. Venugopalan, Phys.Rev. {\bf D49} (1994) 2233;
               {\bf D49} (1994)3352.
\bibitem{Kovner}  J. Jalilian-Marian,  A. Kovner, L.McLerran,
                  H.Weigert, Phys.Rev. {\bf D55} (1997) 5414;  
                 J. Jalilian-Marian,  A. Kovner,  H.Weigert,   The wilson
                renormalization group for low x physics: gluon evolution at
         finite parton density, Preprint TPI-MINN-97-26,1997 ( hep-ph/9709432).
\bibitem{Xiong}L. Xiong, E.V. Shuryak, Nucl. Phys {\bf A590} (1995) 589.
\bibitem{Eskola}  Kari J. Eskola, Berndt Muller, and Xin-Nian Wang,
    Selfscreened parton cascades, Preprint DUKE-TH-96-120 ( nucl-th/9608013).
\bibitem{DGL} L.N. Lipatov, Sov.J.Nucl.Phys. 20 (1975) 94;~~
        V.N. Gribov, L.N. Lipatov, Sov.J.Nucl.Phys. {\bf 15}(1975)438 and 675;
   ~~Yu.L. Dokshitzer, Sov.Phys. JETP {\bf 46} (1977)641.  
\bibitem{AP} G. Altarelli, G. Parisi, Nucl.Phys. {\bf B126} (1977)298;      
\bibitem{DMW}   Yu.L. Dokshitzer, G. Marchesini, B.R. Webber, 
                  Nucl.Phys. {\bf B469} (1996) 93.
\bibitem{DSh}     Yu.L. Dokshitser, D.V. Shirkov  Z.Phys.{\bf C67}(1995) 449. 
\bibitem{DDT} Yu.L. Dokshitzer, D.I. Dyakonov and  S.I. Troyan,  
              Phys.Rep. {\bf 58}(1980)269.  
\bibitem{Keld}  L.V. Keldysh, Sov. Phys. JETP {\bf 20} (1964) 1018; 
               E.M. Lifshits, L.P. Pitaevsky, Physical kinetics, 
             Pergamon Press, Oxford, 1981. 
\bibitem{QFK} A. Makhlin, Phys.Rev. {\bf C 51} (1995) 3454. 
\bibitem{Tw4} Jianwei Qui,  Phys.Rev. {\bf D42}, 30(1990). 
\bibitem{Qui} J.C. Collins and Jianwei Qui,  Phys.Rev. {\bf D39},
               1398(1989). 
\bibitem{Sakharov}  A.D. Sakharov, JETP {\bf 18}, 631  (1948). 
\bibitem{WD1} A. Makhlin, The wedge form of relativistic dynamics,
                  Preprint WSU-NP-96-11, 1996, hep-ph/9608259   
\bibitem{WDG} A. Makhlin, The wedge form of dynamics. II. The gluons.
              Preprint WSU-NP-13, 1996, hep-ph/9608261
\bibitem{Bogol} N.N. Bogolyubov, D.V. Shirkov, Introduction to the
               theory of quantized fields, Interscience, NY,1959 
\bibitem{YF} C.N.Yang, D. Feldman, Phys.Rev. {\bf 79} (1950) 972
\bibitem{GIP} V.N. Gribov, B.L. Ioffe, and I.Ya.Pomeranchuk,
               Sov. J. Nucl. Phys. {\bf 2},  549 (1966).
\bibitem{Gribov2}  V.N. Gribov, Sov. Phys. JETP, {\bf 30},  709 (1970).
\bibitem{Weinberg} S. Weinberg, Phys.Rev. {\bf 130} (1962) 776;
                 Phys.Rev. {\bf 137B} (1964) 672; The Quantum theory of fields,
             Ch.10,  Cambridge Univ. Press, 1995 
\bibitem{QGD} A. Makhlin, Phys.Rev. {\bf C 52} (1995) 995. 
\bibitem{Dirac} P.A.M. Dirac, Rev.Mod. Phys, {\bf 21}, 392  (1949). 

\end{references}
\end{document}